\documentclass[reprint,amsmath,amssymb,aps,nofootinbib,twocolumn]{revtex4-2}
\pdfoutput=1
\usepackage{graphicx,amsmath,amssymb,amsthm}
\usepackage[hidelinks]{hyperref}
\usepackage{xcolor}
\usepackage{ulem}
\usepackage{thmtools, thm-restate}
\usepackage{epsfig}
\usepackage{epstopdf}
\usepackage{graphicx}
\usepackage{booktabs}
\usepackage{bbm}
\usepackage{color}
\usepackage{latexsym}
\usepackage{physics}
\usepackage{tensor}
\usepackage{verbatim}
\usepackage[caption=false]{subfig}
\usepackage{tikz}
\usepackage{ifthen}
\usetikzlibrary{matrix}
\usetikzlibrary{decorations.markings,calc,shapes,decorations.pathmorphing}
\usetikzlibrary{patterns}
\usetikzlibrary{positioning}

\usepackage{tikz}
\usetikzlibrary{external}

\usepackage{hyperref}

\usepackage{xcolor}

\hypersetup{
colorlinks,
linkcolor={blue!80!black},
citecolor={blue!80!black},
urlcolor={blue!80!black},
linktoc=page
}

\newcommand{\normord}[1]{\xcentcolon\mathrel{#1}\xcentcolon}
\newcommand{\xcentcolon}{%
  \mathrel{\vbox{\hbox{$:$}\kern.2ex}}%
}

\newcommand\beq{\begin{equation}}
\newcommand\eeq{\end{equation}}

\newcommand{\be}{\begin{equation}}
\newcommand{\ee}{\end{equation}}

\begin{document}

\title{Structure of Holographic BCFT Correlators from Geodesics}
\author{Jani Kastikainen$^{a,b}$}
\email{jani.kastikainen@helsinki.fi}
\author{Sanjit Shashi$^{c}$}
\email{sshashi@utexas.edu}
\affiliation{$^a$Department of Physics, P.O. Box 64, FIN-00014 University of Helsinki, Finland,}
\affiliation{$^b$Universit\'e de Paris, CNRS, Astroparticule et Cosmologie, F-75013 Paris, France,}
\affiliation{$^c$Theory Group, Department of Physics, University of Texas, Austin, Texas 78712, USA.}

\preprint{\today}
\begin{abstract}
\noindent We compute correlation functions, specifically 1-point and 2-point functions, in holographic boundary conformal field theory (BCFT) using geodesic approximation. The holographic model consists of a massive scalar field coupled to a Karch-Randall brane---a rigid boundary in the bulk AdS space. Geodesic approximation requires the inclusion of paths reflecting off of this brane, which we show in detail. For the 1-point function, we find agreement between geodesic approximation and the harder $\Delta$-exact calculation, and we give a novel derivation of boundary entropy using the result. For the 2-point function, we find a factorization phase transition and a mysterious set of anomalous boundary-localized BCFT operators. We also discuss some puzzles concerning these operators.

\end{abstract}
\pacs{04.20.Cv,
04.60.Bc,
98.80.Qc
}

\maketitle


\section{Introduction}

Recent work has highlighted the importance of Karch-Randall (KR) braneworlds \cite{Randall:1999vf,Karch:2000ct} to understanding gravitational phenomena. Such models are doubly-holographic---there exist three equivalent pictures \cite{Karch:2000gx,Geng:2020qvw,Neuenfeld:2021wbl}:
\begin{itemize}
\item[(I)] a $d$-dimensional boundary CFT (BCFT) \cite{McAvity:1995zd,Cardy:2004hm},
\item[(II)] a CFT + gravity on an asymptotically AdS$_d$ space, connected by transparent boundary conditions to a nongravitating $d$-dimensional BCFT bath,
\item[(III)] Einstein gravity on an asymptotically AdS$_{d+1}$ space containing an end-of-the-world brane.
\end{itemize}
Much work in these systems has used duality between (II) and (III) to compute the fine-grained entanglement entropy of black hole information in (II) via the island rule \cite{2020JHEP...09..002P,Almheiri:2019hni,Almheiri:2019psy,Chen:2020uac,Almheiri:2020cfm} to get the semiclassical Page curve in (II) from the classical geometry of (III). In this spirit, we examine \textit{geodesic approximation} \cite{Balasubramanian:1999zv,Balasubramanian:2012tu} for correlators of ``heavy" ($\Delta \gg d$) CFT operators by summing exponentiated geodesic lengths between  boundary insertions. Note that geodesics also compute entanglement entropy in $d = 2$---a point emphasized in AdS/BCFT by \cite{Takayanagi:2020njm}.

Specifically, we take inspiration from the island rule in the context of duality between (I) and (III)---the AdS/BCFT correspondence \cite{Takayanagi:2011zk,Fujita:2011fp}. The island rule is the AdS/BCFT version of the Ryu-Takayanagi (RT) \cite{Ryu:2006bv} and Hubeny-Rangamani-Takayanagi (HRT) \cite{Hubeny:2007xt} prescriptions for holographic entanglement entropy of CFT subsystems. These involve minimizing some bulk surface, but in (III) we must include surfaces ending on the brane.

It is natural to ask about other analogs to known entries of the AdS/CFT dictionary, e.g. geodesic approximation for propagators,
\begin{equation}
\expval{\mathcal{O}(X_1) \mathcal{O}(X_2)} = \int \mathcal{D}\mathcal{P}\,e^{-\Delta L(\mathcal{P})},
\end{equation}
where $\mathcal{P}$ is an arbitrary bulk path from $X_1$ to $X_2$ and $L(\mathcal{P})$ is the renormalized length functional. By a saddle-point approximation, we have that around $\Delta \to \infty$,
\begin{equation}
\expval{\mathcal{O}(X_1)\mathcal{O}(X_2)} \sim \sum_{\text{geodesics}} e^{-\Delta L}.
\label{intgeod}
\end{equation}
The main result of this paper is to extend the geodesic approximation \eqref{intgeod} to holographic models with boundaries such as KR braneworlds. This requires taking into account geodesics that reflect off of the boundary, as described originally in \cite{McAvity_1991,mcavity_asymptotic_1991,mcavity_quantum_1993} in a nonholographic context.

We compute both the heavy 1-point and 2-point functions of a scalar BCFT operator at nonzero brane tension for the first time using geodesic approximation. While \cite{Fujita:2011fp} also computes the 1-point function, a more general expression is needed for consistency with geodesic approximation.\footnote{The difference from \cite{Fujita:2011fp} arises from a different AdS\slash BCFT dictionary for which we give a careful treatment in Section \ref{secIIb}.} We compute it explicitly---a much more difficult calculation than simply computing geodesic lengths---and confirm that it matches the approximation. In $ d=2 $, we are able to extract boundary entropy from the 1-point function, ultimately matching the standard result found from the RT formula \cite{Takayanagi:2011zk}.

The 2-point function with a tensionless brane \cite{Alishahiha:2011rg,Almheiri:2018ijj} can also be found using a method of images valid for finite $\Delta$ (Appendix \ref{app:images}), but this fails for nonzero tension. Geodesic approximation provides an approach which can be applied even at nonzero tension, which we confirm reproduces the method of images result at zero tension.

Our connected 2-point function includes a (generically) subleading image term as discussed by \cite{Almheiri:2018ijj}. However, we find a ``factorization" phase transition for any negative tension brane beyond which connected geodesic saddles are lost. This recalls similar behavior in holographic large-$N$ gauge theories \cite{Witten:1998zw,Brandhuber:1998bs}, specifically chiral symmetry breaking \cite{Albash:2006bs,Johnson:2008vna}. We also use the boundary operator product expansion (BOPE) \cite{McAvity:1995zd,Karch:2017fuh} of the 2-point function to get the spectrum of BCFT boundary scalar operators. Nonzero tension yields a mysterious extra set of ``anomalous" operators \cite{Karch:2017fuh}.

\section{Geodesic Approximation in the A{\lowercase{d}}S/BCFT Correspondence}

We present the necessary ingredients to compute heavy BCFT correlation functions by geodesic approximation.

\subsection{Review of AdS/BCFT}

We start by reviewing AdS$_{d+1}$/BCFT$_d$ \cite{Takayanagi:2011zk,Fujita:2011fp}. In the ($d+1$)-dimensional bulk, take the gravitational action,
\begin{equation}
\begin{split}
I_G =\ & \frac{1}{16 \pi G_N} \int_{\mathcal{M}} d^{d+1}X\sqrt{g}\,\left[R + \frac{d(d-1)}{\ell^2}\right]\\
&+ \frac{1}{8\pi G_N}\int_{\mathcal{Q}} d^d \hat{x}\sqrt{h}\,(K-T)
\end{split}
\end{equation}
where $\ell$ is the AdS radius (set to $1$) and $\mathcal{Q}$ is an end-of-the-world Randall-Sundrum (RS) brane \cite{Randall:1999vf} with tension $T$ and extrinsic curvature $K_{\mu\nu}$.\footnote{We use $ X^{a} $ with Latin indices to denote coordinates of $ \mathcal{M} $ and $ \hat{x}^{\mu} $ with Greek indices to denote worldvolume coordinates of $ \mathcal{Q} $.} We impose Neumann boundary conditions on the bulk metric at the brane,\footnote{These are for when $ \mathcal{Q} $ is approached from the interior of $ \mathcal{M} $.}
\begin{equation}
K_{\mu\nu} = (K-T)h_{\mu\nu}.\label{krNeumann}
\end{equation}
We are interested in the Karch-Randall (KR) branes, for which the tension is subcritical---$|T| < d-1$---and the induced geometry is (asymptotically) AdS$_d$ \cite{Karch:2000ct,Karch:2000gx}.

\begin{figure}

\begin{tikzpicture}[scale=0.75]
\draw[->] (0,0) to (2,0);
\draw[->] (0,0) to (0,2);

\node at (2,0.3) {$y$};
\node at (-0.2,2) {$z$};

\draw[-,very thin] (0.5,0) arc (0:135:0.5);

\node at (0.5,0.5) {$\theta$};

\draw[-,thick,red] (0,0) to (-2,2);

\node[red,rotate=-45] at (-1.2,0.8) {$y = z\cot\theta$};

\node[red] at (0,0) {$\bullet$};

\end{tikzpicture}

\caption{A fixed $\vec{x}$ slice of AdS$_{d+1}$ with a planar KR brane \eqref{planarKR}. $\theta$ is the angle of the brane with the $y$-axis on which the BCFT lives. The bulk region to the left of the brane ($y < z\cot\theta$) is excised in the KR braneworld.}

\label{figs:planarBrane}
\end{figure}
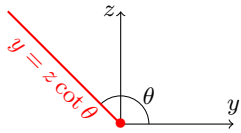

In the Euclidean sector, one bulk solution is AdS$_{d+1}$,
\begin{equation}
ds^2 = \frac{1}{z^2}(dz^2 + dy^2 + d\vec{x}^2),\label{met}
\end{equation}
where $z > 0$, $y \in \mathbb{R}$, and $\vec{x} \in \mathbb{R}^{d-1}$. There then exists a particularly simple class of planar KR branes parametrized by a brane angle $\theta$,
\begin{equation}
y = z\cot\theta.\label{planarKR}
\end{equation}
The bulk geometry is shown in Figure \ref{figs:planarBrane}. The dual Euclidean BCFT$_d$ lives on a half-space ($z = 0$, $y > 0$).

The tension $T$ and induced length scale $\bar{\ell}$ of \eqref{planarKR} are given by,
\begin{equation}
T = -(d-1)\cos\theta,\ \ \bar{\ell}^2 = \csc^2\theta.
\end{equation}
$T$ counts the dual BCFT$_d$ state's boundary degrees of freedom, represented by either a $g$-function describing the overlap between the vacuum state and the boundary state $\bra{0}\ket{b}$ or a boundary entropy $S_{bdy} = \log{\langle 0 \lvert b \rangle}$ \cite{Affleck:1991tk,Azeyanagi:2007qj}. For example for $d = 2$ \cite{Takayanagi:2011zk},
\begin{equation}
S_{bdy} = \frac{\text{tanh}^{-1}{(T)}}{4G_{N}} = -\frac{1}{4G_{N}}\log\cot\frac{\theta}{2}.\label{bdryEntropy}
\end{equation}
Negative tensions are \textit{a priori} possible in this construction, but they only make physical sense if KR branes are treated as nonfluctuating, with a nondynamical radion in the effective action. See \cite{Karch:2020flx} for a discussion and an example of negative tension in an explicit BCFT dual.

As we are interested in scalar operators in BCFT$_d$, we also include a Euclidean bulk scalar field action,
\begin{equation}
\begin{split}
I_S[\Phi] =\ &\frac{1}{2}\int_{\mathcal{M}} d^{d+1}X \sqrt{g} \left(\nabla^a \Phi \nabla_a \Phi + m^2 \Phi^2\right)\\
&- \int_{\mathcal{Q}} d^d \hat{x}\sqrt{h}\,V(\Phi).\label{scalarAct}
\end{split}
\end{equation}
$V(\Phi)$ is a (usually polynomial) coupling of $\Phi$ to the brane.\footnote{We use a $-$ sign in the action so that our scalar equations of motion match those of \cite{Takayanagi:2011zk,Fujita:2011fp}. Their convention is different because they excise the $y > z\cot\theta$ part of the bulk.} Varying $\Phi$ yields the Klein-Gordon equation,\footnote{We neglect backreaction of the field on the metric for simplicity by working in the regime $m^{d-1} G_N \to 0$.}
\begin{equation}
(\square + m^2)\Phi = 0,\ \ \square\Phi = -\nabla^a \nabla_a \Phi,\label{kgEq}
\end{equation}
with Dirichlet or general boundary conditions for $\Phi$,
\begin{align}
&\text{Dirichlet:} &&\delta\Phi|_{\mathcal{Q}} = 0,\\
&\text{General:} &&[n^a \partial_a \Phi + V'(\Phi)]|_{\mathcal{Q}} = 0,\label{gen}
\end{align}
where $(n^{z},n^{y}) = (-z\cos{\theta},z\sin{\theta})$ is the inward-directed unit normal of $\mathcal{Q}$. The general boundary condition reduces to a Neumann boundary condition if $V$ is a constant in $\Phi$---i.e. if $\Phi$ is not directly coupled to the brane. It becomes a Robin boundary condition if $V$ is quadratic.


The AdS/CFT dictionary \cite{Freedman:1999gp,Aharony:1999ti} relates $m$ to the conformal dimension $\Delta$ of a dual scalar operator $\mathcal{O}$ by,
\begin{equation}
m^2 = \Delta(\Delta - d),\label{massCD}
\end{equation}
so ``heavy" scalar operators are dual to heavy (large-$m$) scalar fields. We have the same duality in AdS/BCFT. Throughout the paper, we work in standard quantization and assume that $ \Delta > d\slash 2 $ \cite{Klebanov:1999tb}.

Our goal is a geometric computation of the 1-point and 2-point functions, so we briefly discuss their structure from conformal symmetry induced by the SO$(d-1,2)$ isometries of the AdS$_d$ brane.

Starting with 1-point functions, for a single insertion point $(y,\vec{x})$ and BCFT boundary $y = 0$ \cite{Cardy:1984bb,DeWolfe:2001pq},
\begin{equation}
\expval{\mathcal{O}(y,\vec{x})} = \frac{a_\mathcal{O}}{(2y)^\Delta}.\label{opGeneric}
\end{equation}
As well as being a $\Delta,\theta$-dependent normalization, $a_\mathcal{O}$ also depends on the coupling of the scalar field to the brane, which acts like a ``source." $a_\mathcal{O} = 0$ if $V$ is constant.

For 2-point functions, we define a cross-ratio of insertion points $(y_1,\vec{x}_1)$ and $(y_2,\vec{x}_2)$,
\begin{equation}
\xi = \frac{(y_2 - y_1)^2 + |\vec{x}_2 - \vec{x}_1|^2}{4y_1 y_2} \in [0,\infty).\label{crossRatio}
\end{equation}
Then, by symmetry we have \cite{Liendo:2012hy,Mazac:2018biw},
\begin{equation}
\expval{\mathcal{O}(y_1,\vec{x}_1)\mathcal{O}(y_2,\vec{x}_2)} = \frac{1}{(4y_1 y_2)^\Delta} \mathcal{F}(\xi),\label{tpGeneric}
\end{equation}
where $ \mathcal{F}(\xi) $ is an arbitrary function of the cross-ratio. Unlike \eqref{opGeneric}, 2-point functions do not require a nontrivial coupling of $\Phi$ to the brane to be nonzero.

In our holographic analysis, we obtain the forms \eqref{opGeneric} and \eqref{tpGeneric}, including tension dependence, simply by computing geodesic lengths---a much more straightforward procedure than doing field theory on curved backgrounds.

\subsection{Boundary Correlators from Bulk Fields}\label{secIIb}

We now relate the BCFT operator correlators to bulk field propagators, being particularly careful about normalization. With $X = (z,y,\vec{x})$, define,
\begin{align}
Z_S[J] = \int \mathcal{D}\Phi\,e^{-I_S[\Phi]},
\label{partfunction}\\
J(y,\vec{x}) = \lim_{z\rightarrow 0}z^{\Delta-d} \Phi(X),
\end{align}
as respectively the bulk scalar partition function with the fields $\Phi$ satisfying $\eqref{gen}$ (Dirichlet also works) and the source at $z = 0$.\footnote{We also impose regularity at the Poincaré horizon $ z = \infty $.} Denoting boundary insertions by $\zeta = (y,\vec{x})$, BCFT correlators are functional derivatives,
\begin{equation}
\langle \mathcal{O}(\zeta_1)\cdots \mathcal{O}(\zeta_n)\rangle =  \frac{\delta^{n}\log{Z_S[J]}}{\delta J(\zeta_1)\cdots \delta J(\zeta_n)}\bigg\lvert_{J = 0}.
\label{correlators}
\end{equation}
To compute the partition function, we use the \textit{background field technique} \cite{mcavity_quantum_1993}; writing,
\begin{equation}
\Phi(X) = \phi_B(X) + \phi(X),\ \ \overline{I_S[\Phi]} = I_S[\Phi] - I_S[\phi_B],
\end{equation}
this ``shifted" action can be expanded in $\phi$,
\begin{equation}
\begin{split}
\overline{I_S[\Phi]}
=\ &\frac{1}{2}\int_{\mathcal{M}} d^{d+1}X \sqrt{g}\left(\nabla^a \phi \nabla_a \phi + m^2\phi^2\right)\\
&- \int_{\mathcal{Q}} d^d \hat{x}\sqrt{h}\left[\frac{V''(\phi_B)}{2}\phi^2 + O(\phi^3)\right]\\
&+\int_{\mathcal{M}} d^{d+1} X\sqrt{g}\,\phi(\square + m^2)\phi_B\\
&- \int_\mathcal{Q} d^d\hat{x}\sqrt{h}\,\phi\left[n^a \partial_a \phi_B + V'(\phi_B)\right]\\
&- \int_{z\to 0} d^{d}\zeta\,\frac{1}{z^{d-1}}\phi\partial_z\phi_B,
\end{split}\label{actionShift}
\end{equation}
where $ \zeta $-integrations are over half-space $ y>0 $, and we have integrated the $\nabla^a \phi \nabla_a \phi_B$ term by parts.

Our goal is to use a saddle-point approximation on \eqref{partfunction}, so we need the on-shell value of the action. To get this, assume $\phi_B$ classically solves the Klein-Gordon equation, obeys \eqref{gen}, and is normalizable. Around $z = 0$,
\begin{equation}
\phi_B(X) = z^\Delta f_B(\zeta) + \cdots.\label{modeBack}
\end{equation}
If $\Phi$ is also taken to be on-shell, $\phi$ further solves the Klein-Gordon equation with the following boundary conditions,
\begin{equation}
\left.\left[n^a \partial_a \phi + V''(\phi_B)\phi + \frac{V'''(\phi_B)}{2}\phi^2 + O(\phi^3)\right]\right|_{\mathcal{Q}} = 0,\label{bcFieldA}
\end{equation}
and has the non-normalizable mode. We thus write,
\begin{align}
\phi(X) &= z^{d-\Delta}\left[J(\zeta) + \cdots\right] + z^\Delta \left[A(\zeta) + \cdots\right],\label{fieldAsympt}\\
A(\zeta) &= \int_{y>0} d^d \zeta'\,J(\zeta')H(\zeta,\zeta'),
\end{align}
where $H$ is a linear response function; it is the $ z\rightarrow 0 $ limit of the bulk-to-boundary propagator that obeys required boundary conditions at the brane. 

We now assume a quadratic brane coupling ($\lambda_{1,2} \in \mathbb{R}$),
\begin{equation}
V(\Phi) = \lambda_1 \Phi + \frac{1}{2}\lambda_2 \Phi^2,\label{quadp}
\end{equation}
The cubic terms in \eqref{actionShift} and quadratic terms in \eqref{bcFieldA} vanish. Equation \eqref{bcFieldA} becomes a Robin condition \cite{mcavity_quantum_1993}, and the on-shell shifted action is entirely boundary terms,
\begin{equation}
\overline{I_S}^{\text{cl}} = -\int_{z\to 0} d^d\zeta\,\frac{1}{z^{d-1}}\left(\phi\,\partial_z\phi_B + \frac{1}{2}\phi\,\partial_z\phi\right).
\end{equation}
The first term is finite, but the second term diverges for $ \Delta > d\slash 2 $ as in standard AdS\slash CFT with no brane \cite{Klebanov:1999tb}. To cancel the divergence, we use the covariant counterterm,
\begin{equation}
\frac{d-\Delta}{2}\int_{z \rightarrow 0} d^{d}\zeta\sqrt{\gamma}\,\phi(z,\zeta)^{2}, \quad \sqrt{\gamma} = \frac{1}{z^{d}},
\label{countterm}
\end{equation}
where $ \gamma $ is the induced metric of a fixed-$ z $ slice. This counterterm contributes an extra finite piece, so the renormalized on-shell shifted action is,
\begin{equation}
\begin{split}
-\overline{I_S}^{\text{cl,r}}
=\,&\Delta \int d^d \zeta\,J(\zeta)f_B(\zeta)\\
&+ \frac{2\Delta - d}{2}\int d^d \zeta \int d^d\zeta'\,J(\zeta)\,J(\zeta')\,H(\zeta,\zeta'),
\end{split}
\label{splitren}
\end{equation}
with the coefficient of the first term agreeing with \cite{Almheiri:2018ijj} and that of the second term with \cite{Klebanov:1999tb}. Notice that if the counterterm \eqref{countterm} was built from $ \Phi $ instead of $ \phi $, the first term in \eqref{splitren} would be canceled as well, corresponding to a change in the renormalization scheme.

The corresponding 1-point and connected 2-point functions are,
\begin{equation}
\expval{\mathcal{O}(\zeta)} = \Delta f_B(\zeta),\ \ \expval{\mathcal{O}(\zeta)\mathcal{O}(\zeta')} = (2\Delta -d) H(\zeta,\zeta').
\label{holcols}
\end{equation}
We conclude by noting that $f_B(\zeta)$ and, thus, the 1-point function can be written as integrals over the scalar propagator. The boundary value problem,\footnote{We impose vanishing conditions at $ z=0,\infty $.}
\begin{equation}
(\square + m^2)\phi_B = 0,\ \ \left.(n^a \partial_a + \lambda_2) \phi_B\right|_{\mathcal{Q}} = -\lambda_1,
\end{equation}
has the solution \cite{mcavity_heat_1992,mcavity_surface_1993},\footnote{The formula \eqref{osbsol} can be proven using Green's theorem. In \cite{mcavity_heat_1992,mcavity_surface_1993}, only Neumann problem is considered with $ \lambda_2 = 0 $, but \eqref{osbsol} holds also for Robin boundary conditions.}
\begin{equation}
\phi_B(X) = \lambda_1 \int_{\mathcal{Q}} d^d \hat{x}\sqrt{h}\,G(X,Y(\hat{x})),
\label{osbsol}
\end{equation}
where $ G(X,X') $ is the bulk-to-bulk propagator satisfying Robin boundary conditions at the brane,
\begin{equation}
(n^{a}\partial_a + \lambda_2)G(X,X')\lvert_{X\in\mathcal{Q}} = 0,
\label{Gneumann}
\end{equation}
and $Y(\hat{x})$ is the embedding $\mathcal{Q} \to \mathcal{M}$. Then by using \eqref{modeBack} and \eqref{holcols}, we get the 1-point function,
\begin{equation}
\expval{\mathcal{O}(\zeta)} = \lambda_1\Delta  \lim_{z\to 0}\int_{\mathcal{Q}}d^d\hat{x} \sqrt{h}\,z^{-\Delta}G(X,Y(\hat{x})).\label{propOP}
\end{equation}
We see that the 1-point function vanishes if $ \lambda_1 = 0 $. 

The bulk-to-boundary propagator is obtained from the bulk-to-bulk propagator via,
\begin{equation}
\mathcal{K}(X,\zeta') = (2\Delta - d) \lim_{z'\rightarrow 0}(z')^{-\Delta}\,G(X,X'),
\end{equation}
where we assume that the relative normalization is the same as in standard AdS\slash CFT without a brane \cite{giddings_boundary_1999,Klebanov:1999tb}.\footnote{This should be the case because the Robin propagator $ G(X,X') $ behaves like the empty AdS propagator when $ X \sim X' $.} For the 1-point function, we get,
\begin{equation}
\expval{\mathcal{O}(\zeta)} = \frac{\lambda_1\Delta}{2\Delta - d}\int_{\mathcal{Q}}d^d\hat{x} \sqrt{h}\,\mathcal{K}(Y(\hat{x}),\zeta).
\label{1pointK}
\end{equation}
and for the 2-point function,
\begin{equation}
\expval{\mathcal{O}(\zeta)\mathcal{O}(\zeta')} = (2\Delta - d)^{2}\lim_{z,z'\rightarrow 0}(zz')^{-\Delta}\,G(X,X'),
\label{extr}
\end{equation}
with normalization as in \cite{harlow_operator_2011}. In the large-$ \Delta $ limit, the overall normalizations will not play any role.

We have shown that the 1-point and 2-point functions can be computed in terms of bulk field theoretic quantities. Indeed, we can compute the 1-point function by solving for $\phi_B$ directly (Appendix \ref{appB}) or by integrating against the brane (at zero tension). However computing the 2-point function is more difficult. This motivates the use of geodesic approximation for the propagators.

\subsection{Geodesic Approximation with Boundaries}\label{sec2B}

We now realize geodesic approximation of propagators in the presence of a bulk boundary $\mathcal{Q}$ with Dirichlet or Robin boundary conditions corresponding to the quadratic potential \eqref{quadp}. The starting point is to compute the Euclidean bulk-to-bulk propagator $G(X,X')$,\footnote{We use the notation $ \delta_{\mathcal{M}}(X,X') \equiv \delta^{(d+1)}(X-X')\slash \sqrt{g} $.}
\begin{equation}
(\square + m^2)G(X,X') = \delta_{\mathcal{M}}(X,X'),
\end{equation}
with boundary conditions,
\begin{align}
&\text{Dirichlet:} && G(X,X')|_{X \in \mathcal{Q}} = 0,\label{BCs1}\\
&\text{Robin:} && (n^a\partial_a + \lambda_2)G(X,X')|_{X \in \mathcal{Q}} = 0.
\label{BCs2}
\end{align}
The derivative operators here act on $ X $. To solve for the propagator, we consider the auxiliary heat equation,
\begin{equation}
\begin{split}
\left(\partial_\tau +  \square\right)\mathcal{G}(X,X';\tau) &= 0,\\
\mathcal{G}(X,X';0) &= \delta_{\mathcal{M}}(X,X').
\end{split}
\label{heateq}
\end{equation}
whose solution $ \mathcal{G}(X,X';\tau) $, the heat kernel, is also required to obey an appropriate boundary condition for all $ \tau \geq 0 $. The propagator is then a Laplace transform,
\begin{equation}
G(X,X') = \int_0^\infty d\tau\,e^{-m^{2}\tau}\, \mathcal{G}(X,X';\tau),\label{laplace}
\end{equation}
with respect to the ``auxiliary time" variable $ \tau $.

In \cite{McAvity_1991,mcavity_asymptotic_1991,mcavity_quantum_1993}, a generalization of the DeWitt ansatz for the heat kernel on a $D$-dimensional\footnote{In this section, we define $D = d+1$ for convenience.} manifold $\mathcal{M}$ with boundary $\mathcal{Q}$ was proposed. It takes into account the presence of the boundary by formally summing over geodesics $ \gamma_n $ that reflect off the boundary $ n $ times. This ansatz is,
\begin{equation}
\mathcal{G}(X,X';\tau) = \frac{1}{(4\pi\tau)^{D\slash 2}} \sum_{n=0}^{\infty}e^{-\sigma_n\slash (2\tau)}\,\Omega_n(X,X';\tau),
\label{dewitt}
\end{equation}
where $ \Omega_n(X,X;0) = 1 $. $ \sigma_n(X,X') $ is Synge's world function \cite{poisson_motion_2011} for a reflecting geodesic $ \gamma_n(s) $ ($s \in [0,1]$) with endpoints $ \gamma_n(0) = X $ and $ \gamma_n(1) = X' $,
\begin{equation}
\sigma_n(X,X') = \frac{1}{2}\int_0^1 ds\,g_{ab}(\gamma_n(s))\,\dot{\gamma}_n^a(s)\, \dot{\gamma}_n^b(s).
\end{equation}
For $n$ reflections, this depends on additional reflection data $\{\hat{x}_k,p_k\}_{k=1,...,n}$ defined as follows. 

Denote the $k$th reflection point ($k = 1,...,n$) of the geodesic by $\gamma_n(p_k) = X_k(\hat{x}_k) \in \mathcal{Q}$, where $ X_k $ is the reflection point in the coordinates of $ \mathcal{M} $ and $\hat{x}_k$ is the point in the worldvolume coordinates of $\mathcal{Q}$. Also,
\begin{equation}
p_k < p_{k+1},\ \ p_0 = 0,\ \ p_{n+1} = 1.
\end{equation}
The worldline endpoints are $X_0 = X$ and $X_{n+1} = X'$.

Define $\gamma_{n,k}(s_k)$ with $s_k \in [p_k,p_{k+1}]$ as the segment of the worldline between two adjacent reflections. Then,
\begin{equation}
\sigma_n(X,X') = \sum_{k=0}^n \frac{\sigma(X_k,X_{k+1})}{p_{k+1} - p_k},\label{nRefAct}
\end{equation}
where we write Synge's world function for each segment,
\begin{equation}
\sigma(X_k,X_{k+1}) = \frac{1}{2}\int_{0}^{1} ds\,g_{ab}(\gamma_{n,k}(s))\,\dot{\gamma}_{n,k}^{a}(s)\,\dot{\gamma}_{n,k}^{b}(s).
\end{equation}
The reflection data $\{\hat{x}_k,p_k\}_{k=1,...,n}$ is then determined by the extremization conditions,
\begin{equation}
\frac{\partial \sigma_n}{\partial p_k} = 0, \quad \frac{\partial\sigma_n}{\partial\hat{x}_k} = 0 \quad (k=1,...,n).
\label{extremeReflect}
\end{equation}
We more thoroughly analyze these conditions in Appendix \ref{app:extrem}. For now, note the first condition implies that,
\begin{equation}
\sqrt{2\sigma_n} = L_n(X,X') = \sum_{k=0}^{n} L(X_{k},X_{k+1}),
\label{sqrtsigma}
\end{equation}
where $ L_n $ is the full reflecting geodesic length and $ L(X_{k},X_{k+1}) $ is the length of the segment $\gamma_{n,k} $.

After imposing the first condition, the second condition in \eqref{extremeReflect} becomes the \textit{law of reflection} which states that the angles of the incoming and outgoing portions of the geodesic with respect to the normal must match, as depicted in Figure \ref{figs:reflectCartoon} (see Appendix \ref{app:extrem} for details).

\begin{figure}
\centering
\begin{tikzpicture}[scale=0.8]
\draw[-,red,thick] (1.5,0) arc (0:180:1.5);

\draw[-,thick] (2,0) to (1.3,0.75);
\draw[-,thick] (1.3,0.75) .. controls (1,1.73) .. (0,1.5);

\draw[-,thick] (-2,0) to (-1.3,0.75);
\draw[-,thick] (-1.3,0.75) .. controls (-1,1.73) .. (0,1.5);

\draw[-,very thin] (1.516,0.875) arc (30:-45:0.25);
\node at (1.816,0.675) {$\psi_3$};
\draw[-,very thin] (1.624,0.9375) arc (30:106:0.375);
\node at (1.524,1.3375) {$\psi_3$};

\draw[-,very thin] (0,1.75) arc (90:12.5:0.25);
\node at (0.3,1.95) {$\psi_2$};
\draw[-,very thin] (0,1.875) arc (90:180-12.5:0.375);
\node at (-0.3,2.075) {$\psi_2$};

\draw[-,very thin] (-1.516,0.875) arc (180-30:180+45:0.25);
\node at (-1.816,0.675) {$\psi_1$};
\draw[-,very thin] (-1.624,0.9375) arc (180-30:180-106:0.375);
\node at (-1.524,1.3375) {$\psi_1$};

\draw[-,dashed,very thin,red,dash pattern=on 2pt off 1pt] (2.17,1.25) to (0.43,0.25);
\draw[-,dashed,very thin,red,dash pattern=on 2pt off 1pt] (0,0.5) to (0,2.5);
\draw[-,dashed,very thin,red,dash pattern=on 2pt off 1pt] (-2.17,1.25) to (-0.43,0.25);

\node at (1.3,0.75) {\footnotesize$\bullet$};
\node at (-1.3,0.75) {\footnotesize$\bullet$};
\node at (0,1.5) {\footnotesize$\bullet$};

\node[red] at (1.2,0.2) {$\mathcal{Q}$};
\end{tikzpicture}

\caption{A cartoon depiction of the law of reflection for an arbitrary boundary $ \mathcal{Q} $. The reflecting geodesic in black may hit $\mathcal{Q}$ in red any number at times. At each dot, the incident and reflected angles are equal.}
\label{figs:reflectCartoon}
\end{figure}
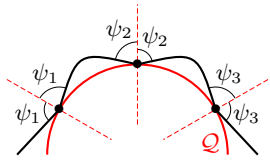

Now, given the ansatz \eqref{dewitt}, the heat equation can be solved by expanding each $ \Omega_n $ as a Taylor series in $\tau$,
\begin{equation}
\Omega_n(X,X';\tau) = \sum_{k=0}^{\infty}a_{n}^{(k)}(X,X')\,\tau^{k},
\label{taylor}
\end{equation}
where $ \smash{a_{n}^{(0)}(X,X)} = 1 $. This is the short-time expansion ($\tau \to 0$). The Seeley-DeWitt coefficients $ \smash{a_{n}^{(k)}} $ are found recursively, as on a manifold without boundary \cite{McAvity_1991}. 

Each boundary condition \eqref{BCs1}--\eqref{BCs2} imposes extra conditions on $ \Omega_n $. When one endpoint is taken to the boundary $( X\rightarrow \mathcal{Q} )$, the geodesic lengths satisfy ``pairing" \cite{McAvity_1991},
\begin{equation}
\sigma_n \rightarrow \sigma_{n+1},\ \ n^{a}\partial_a\sigma_n \rightarrow -n^{a}\partial_a\sigma_{n+1}
\label{pairing}
\end{equation}
for even $ n $. The boundary conditions then imply,
\begin{equation}
(\Omega_n + \Omega_{n+1})\lvert_{\mathcal{Q}} = 0,
\end{equation}
for Dirichlet and,
\begin{equation}
\biggl[(n^{a}\partial_a + \lambda_2) (\Omega_n + \Omega_{n+1})-\frac{n^{a}\partial_a\sigma_n}{2\tau} \,(\Omega_n - \Omega_{n+1})\biggr]\bigg\lvert_{\mathcal{Q}} = 0,
\end{equation}
for Robin \cite{McAvity_1991}. These translate to boundary conditions on the coefficients $ \smash{a_{n}^{(k)}} $. At leading order in \eqref{taylor}, the boundary conditions are satisfied when,
\begin{equation}
a_{n+1}^{(0)} = (-1)^{\kappa}\, a^{(0)}_{n}, \quad X\in \mathcal{Q}
\label{leadingbound}
\end{equation}
where $ \kappa = 0 $ for Robin and $ \kappa = 1 $ for Dirichlet boundary conditions. Notice that $ \lambda_2 $ does not appear in the leading-order coefficients; it only affects $ \smash{a_{n}^{(k)}} $ with $ k\geq 1 $.

The leading order $k = 0$ coefficients are explicitly \cite{McAvity_1991},
\begin{equation}
a_n^{(0)} = (-1)^{\kappa n}\sqrt{\triangle_n},
\label{leadingsol}
\end{equation}
where the van Vleck determinant $\triangle_n$ \cite{poisson_motion_2011,visser_van_1993} of a reflecting geodesic is given by,
\begin{equation}
\triangle_n(X,X') = \frac{1}{\sqrt{g(X) g(X')}} \det\biggl(-\frac{\partial^{2}\sigma_n(X,X')}{\partial X^{a}\,\partial X'^{b}}\biggr),
\end{equation}
and the boundary conditions \eqref{leadingbound} are satisfied due to geodesic pairing \eqref{pairing}.



Now that we have the heat kernel, the propagator is obtained by computing the Laplace transform \eqref{laplace}. Performing the integral order by order and using \eqref{sqrtsigma},
\begin{equation}
G(X,X') = 2\sum_{n,k=0}^{\infty}A^{(k)}_n\,K_{\alpha_k}(mL_n).
\end{equation}
$ K_{\alpha_k} $ is the modified Bessel function of the second kind with $ \alpha_k = k+1-D\slash 2 $. The coefficients are,
\begin{equation}
A^{(k)}_n = \frac{a_n^{(k)}}{(4\pi)^{D\slash 2}}\,\biggl(\frac{L_n}{2m}\biggr)^{\alpha_k}.
\end{equation}
Consider the regime $ mL_n\gg 1 $. We can approximate,
\begin{equation}
2K_{\alpha_k}(mL_n) \sim e^{-mL_n}\sqrt{\frac{2\pi}{m L_n}},
\end{equation}
The $ \smash{a_n^{(k)}} $, being coefficients of the Laplace operator, are independent of $ m $ and depend only on $ \sigma_n $ and possibly its covariant derivatives. Hence,
\begin{equation}
A^{(k)}_{n}\slash A^{(0)}_{n}  = \mathcal{O}(m^{-k}),
\end{equation}
and $ k\geq 1 $ terms are suppressed relative to the $ k=0 $ term when $ m \gg 1 $ (in units of AdS radius $\ell$). Thus to leading-order in the heavy limit, only the short-time terms matter---Robin produces the same leading-order result as Neumann---and we get,
\begin{equation}
G(X,X') \sim \sum_{n=0}^{\infty} (-1)^{\kappa n}\pi  \sqrt{\frac{m^{D-3}\triangle_n}{(2\pi)^{D+1} L_n^{D-1}}}\, e^{-m L_n}.
\label{midstep}
\end{equation}
This expression is valid as long as $ m \gg 1 $ and $ mL_n\gg 1 $. If we also take $ L_n \gg 1 $ (in units of $ \ell $), the expression simplifies further as,\footnote{One can prove that $ \triangle_n = e^{-dL_n + \mathcal{O}(\log{L_n})} $ when $ L_n \rightarrow \infty $. This is subleading in \eqref{midstep} when $ m\gg d\slash 2 $.}
\begin{equation}
G(X,X') \sim \sum_{n=0}^{\infty} (-1)^{\kappa n} e^{-mL_n}.\label{geodesicApproximation}
\end{equation}
This is simply geodesic approximation for the bulk-to-bulk propagator in the presence of the boundary. To reiterate, it is valid for large $m$ and $L_n$. Note $L_n/\ell \gg 1$ is automatically guaranteed when at least one insertion point is near the conformal boundary.

We briefly mention that the approximation \eqref{geodesicApproximation} with reflecting terms can also be derived as the saddle-point approximation of a path integral formula proven in \cite{Ludewig_2017}. In this approach, the reflection data $ \{\hat{x}_k,p_k\} $ should be integrated over and the extremization conditions \eqref{extremeReflect} result from the saddle point. The appearance of van Vleck determinants is also naturally explained \cite{zannias_path-integral_1983}. Note that the formula in \cite{Ludewig_2017} is very rigorous and does not apply to Robin boundary conditions for some subtle reason. It would be interesting to try to understand the relationship between the rigorous path integral and the DeWitt ansatz used in this paper.

When $ X,X' $ are sufficiently close to each other and to the boundary $ \mathcal{Q} $, one expects only a single reflection $ \gamma_1 \equiv \bar{\gamma} $ to exist in addition to the nonreflecting geodesic $ \gamma_0 \equiv \gamma $. Then the ansatz \eqref{dewitt} consists of only two terms,
\begin{equation}
\mathcal{G}(X,X';\tau) = \frac{1}{(4\pi\tau)^{D\slash 2}} \left[e^{-\sigma\slash (2\tau)}\,\Omega +e^{-\bar{\sigma}\slash (2\tau)}\,\bar{\Omega}\right],
\end{equation}
and geodesic approximation yields,
\begin{equation}
G(X,X') \sim e^{-mL}\pm e^{-m\bar{L}}.
\label{G2terms}
\end{equation}
Based on \eqref{pairing}, Dirichlet or Robin boundary conditions are met depending on the relative sign. In AdS with a planar brane, \eqref{G2terms} is true more generally and not just near the brane---when taking one of the insertion points to the conformal boundary, there are no geodesics which reflect more than once (Appendix \ref{appC}).

We are finally ready to give formulas for 1-point and 2-point functions at large $ \Delta $. Starting with the 2-point function \eqref{extr} and defining the renormalized geodesic length between boundary points $\zeta$ and $\zeta'$,
\begin{equation}
L^*(\zeta,\zeta') = \lim_{z,z'\rightarrow 0}[L(X,X') - \log{(zz')}],
\end{equation}
the BCFT 2-point function is given by,
\begin{equation}
\langle \mathcal{O}(\zeta)\mathcal{O}(\zeta') \rangle \sim e^{-\Delta L^*}\pm e^{-\Delta \bar{L}^*}, \quad \Delta\rightarrow \infty.
\end{equation}
For the 1-point function, we focus on Robin boundary conditions. In the expression \eqref{propOP}, the propagator between $X$ and the embedded brane point $Y(\hat{x})$ is,
\begin{equation}
G(X,Y(\hat{x})) \sim 2e^{-\Delta L(X,Y(\hat{x}))}, \quad \Delta \rightarrow \infty,
\end{equation}
where we use the fact that the reflecting geodesic coincides with the nonreflecting one when $ Y(\hat{x}) \in \mathcal{Q} $ \eqref{pairing}. There are no higher reflecting contributions due to the no-go theorem proven in Appendix \ref{appC}.

The integral \eqref{propOP} is approximated by the geodesic with minimal length that corresponds to the point $ Y_b \in \mathcal{Q} $ for which the geodesic hits the brane orthogonally; this length is computed explicitly in Section \ref{sec3A}. Thus at large $ \Delta $, \eqref{propOP} is given by,
\begin{equation}
\expval{\mathcal{O}(\zeta)} = \lambda_1 e^{-\Delta L^{*}(\zeta,Y_b)}.\label{opGeoForm}
\end{equation}
where,
\begin{equation}
L^{*}(\zeta,Y_b) \equiv \lim_{z\rightarrow 0}[L(X,Y_b) - \log{z}],
\end{equation}
is renormalized only in one insertion.

\section{1-Point Function}\label{sec:1point}

We start with the 1-point function. As it is an integral over propagators \eqref{propOP}, we sum over geodesics connecting the insertion point to the brane, termed boundary-to-brane geodesics. We find that the minimal such geodesic gives the leading-order behavior of the full 1-point function (computed in Appendix \ref{appB}) in the large-$\Delta$ limit.

Furthermore, the minimal geodesic only sees the linear coupling $\lambda_1 \Phi$ on the brane---not the quadratic coupling $\lambda_2 \Phi^2/2$. In accordance with the law of reflection (Section \ref{sec2B}), the relevant geodesic hits the brane at a right angle.

Throughout this section, we assume the non-Dirichlet boundary condition. The Dirichlet answer is exactly $0$.

\subsection{Geodesic Approximation at Large $ \Delta $}\label{sec3A}

Denote the insertion point $\zeta$ as $(y,\vec{x})$ and the brane intersection point $Y_b$ as $(z_b,y_b,\vec{x}_b)$, noting $y_b = z_b\cot\theta$. The 1-point function is an integral on $\mathcal{Q}$ \eqref{propOP}, but, as the length is,
\begin{equation}
L = \int \frac{1}{z}\sqrt{dz^2 + dy^2 + d\vec{x}^2},
\end{equation}
the boundary-to-brane geodesics with $\vec{x}_b = \vec{x}$ are local minima of $L$ with respect to $\vec{x}_b$. In the full parameter space, these particular geodesics comprise a ``valley" of the length function dominating the path integral of $e^{-\Delta L}$.

\begin{figure}
\centering
\subfloat[$T \geq 0$]
{
\begin{tikzpicture}[scale=0.65]
\draw[->] (0,0) to (0,2);

\node[] at (2.5,-0.3) {$y$};

\draw[-,thick,black!20!yellow] (2.5,0) arc (0:180:1.25);
\draw[-,thick,black!20!yellow] (2.5,0) arc (0:180:1.5);
\draw[-,thick,black!20!yellow] (2.5,0) arc (0:180:1.75);
\draw[-,thick,black!20!yellow] (2.5,0) arc (0:180:2);

\draw[-,draw=none,fill=white] (-2,0) to (0,0) to (-2,2) to (-2,0);

\draw[-,dashed,black!50] (0,0) to (-2,0);
\draw[->] (0,0) to (3,0);

\node[red] at (0,0) {$\bullet$};
\node[black!20!yellow] at (2.5,0) {$\bullet$};

\draw[-,red,thick] (0,0) to (-2,2);

\draw[<->] (-1.14,1.14) to (-1.14,0);
\node at (-1.34,0.57) {$z_b$};

\node at (-0.2,2) {$z$};
\end{tikzpicture}
}
\subfloat[$T < 0$]
{
\begin{tikzpicture}[scale=0.95]

\node[] at (1.75,-0.2) {$y$};

\draw[-,thick,black!20!yellow] (1.75,0) arc (0:110:0.42);
\draw[-,thick,black!20!yellow,dashed] (1.75,0) arc (0:180:0.42); 

\draw[-,thick,black!20!yellow] (1.75,0) arc (0:90:0.42+0.25);
\draw[-,thick,black!20!yellow] (1.75,0) arc (0:45:0.42+0.5);
\draw[-,thick,black!20!yellow] (1.75,0) arc (0:45:0.42+0.75);

\draw[-,thick,black!20!yellow] (1.75,0) arc (180:150:1/0.67);
\draw[-,thick,black!20!yellow] (1.75,0) arc (180:130:1/0.67-0.25);
\draw[-,thick,black!20!yellow] (1.75,0) arc (180:120:1/0.67-0.5);

\draw[-,thick,black!20!yellow] (1.75,0) arc (180:110:0.81);
\draw[-,thick,black!20!yellow,dashed] (1.75,0) arc (180:0:0.81);

\draw[-,draw=none,fill=white] (0,0) to (3,1) to (3,1) to (0,1) to (0,0);

\draw[->] (0,0) to (3.5,0);
\draw[->] (0,0) to (0,1);

\node[black!20!yellow] at (1.75,0) {$\bullet$};

\draw[-,red,thick] (0,0) to (3,1);

\draw[<->] (1.19,0) to (1.19,1.19/3);
\node at (1.19+0.2,1.19/6) {$z_\ell$};

\draw[<->] (2.3,0) to (2.3,2.3/3);
\node at (2.3+0.2,2.3/6) {$z_u$};

\node at (-0.2,1) {$z$};
\node[red] at (0,0) {$\bullet$};
\end{tikzpicture}
}
\caption{The boundary-to-brane geodesics confined to a slice of AdS$_{d+1}$. Each is labeled by its intersection depth $z_b$, but negative $T$ imposes a finite range on $z_b$.}
\label{figs:opGeodesics}
\end{figure}
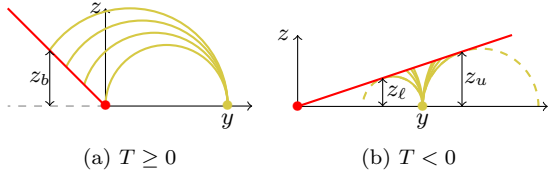

Such geodesics are uniquely determined by $z_b$, but there are two cases (Figure \ref{figs:opGeodesics}). For $T \geq 0$, we have only counterclockwise trajectories from the insertion, with each one corresponding to some $z_b > 0$. For $T < 0$ however, we have both counterclockwise and clockwise trajectories, with only a finite range of $z_b$,
\begin{equation}
z_\ell = y \frac{\cos\theta}{\csc\theta + 1} \leq z_b \leq y \frac{\cos\theta}{\csc\theta - 1} = z_u.\label{boundsNegTen}
\end{equation}
For all of these geodesics, we can use \eqref{met} to integrate up to some cutoff surface $z = \epsilon \to 0$ and compute the length in terms of a dimensionless ratio $\chi_b = z_b/y$,
\begin{equation}
\begin{split}
L_0(\chi_b) =\ &\log y - \log \chi_b\\
&+ \log\left[\left(1 - \chi_b \cot\theta\right)^2 + \chi_b^2\right] - \log\epsilon.\label{length}
\end{split}
\end{equation}
Adding the counterterm $+\log\epsilon$ to renormalize, the minimum length $L_0^*$ found at $\chi_b^* = \sin\theta$ is,
\begin{equation}
L_0^* = \log\left(2y\tan\frac{\theta}{2}\right),
\end{equation}
which we plug into \eqref{opGeoForm}. Thus to leading order in the $\Delta \to \infty$ expansion, the 1-point function has a $\theta$-dependence of the expected form \eqref{opGeneric},
\begin{equation}
\expval{\mathcal{O}(y,\vec{x})} \sim \lambda_1 e^{-\Delta L_0^*} = \frac{\lambda_1}{(2y)^{\Delta}} \cot^\Delta\frac{\theta}{2}.\label{opGeo}
\end{equation}
As an aside, this minimal geodesic hits the brane orthogonally, so as expected it satisfies the law of reflection.

\subsection{Asymptotics of the Finite-$\Delta$ Function}

We now compare \eqref{opGeo} against the $\Delta \to \infty$ limit of the 1-point function found for,
\begin{equation}
V(\Phi) = \lambda_1 \Phi + \frac{1}{2}\lambda_2 \Phi^2.
\end{equation}
All of the details including the definitions of the coefficients, the discrepancy with \cite{Fujita:2011fp}, and the asymptotic analysis are explored in greater depth in Appendix \ref{appB}.

The 1-point function is,
\begin{equation}
\expval{\mathcal{O}(y,\vec{x})} = -\frac{C_{1}(\theta)}{(2y)^\Delta}\left[\frac{2^\Delta \Gamma\left(\frac{\Delta+2}{2}\right)\Gamma\left(\frac{\Delta-d+1}{2}\right)}{\sqrt{\pi}\Gamma\left(\frac{2\Delta - d+2}{2}\right)}\right],\label{opQuad}
\end{equation}
and the coefficient is,
\begin{equation}
C_{1}(\theta)
= \frac{-\lambda_1}{F_1(\theta) - RF_2(\theta) + \lambda_2 [G_1(\theta) - RG_2(\theta)]},\label{coeff1}
\end{equation}
where $R$ depends on $\Delta$ and $F_{1,2}(\theta)$ and $G_{1,2}(\theta)$ depend on $\Delta$ and $\theta$. Observe \eqref{opQuad} for $\lambda_2 = 0$ differs from the 1-point function in \cite{Fujita:2011fp}. In fact, the latter fails to reproduce \eqref{opGeo} as $\Delta \to \infty$ (see Appendix \ref{appB}).

In computing the large-$\Delta$ asymptotics of our expression, as there are a lot of pieces, we individually write the large-$\Delta$ expressions of the quantities comprising \eqref{coeff1}:
\begin{align}
F_1(\theta) \sim G_2(\theta) &\sim \frac{1}{2}\left(\cot^\Delta \frac{\theta}{2} + \tan^\Delta \frac{\theta}{2}\right),\\
F_2(\theta) &\sim \frac{\Delta}{2}\left(\cot^\Delta \frac{\theta}{2} - \tan^\Delta\frac{\theta}{2}\right),\\
G_1(\theta) &\sim \frac{1}{2\Delta}\left(\cot^\Delta\frac{\theta}{2} - \tan^\Delta\frac{\theta}{2}\right),\\
R &\sim \frac{1}{\Delta}.
\end{align}
We thus write the leading-order term in the heavy regime of the $C_1(\theta)$ coefficient as,
\begin{equation}
C_{1}(\theta) \sim \lambda_1 \cot^\Delta\frac{\theta}{2},
\end{equation}
In conjunction with the other prefactors and the approximation $\Delta - d \approx \Delta$, we ultimately find that,
\begin{equation}
\expval{\mathcal{O}(y,\vec{x})} \sim \frac{\lambda_1}{(2y)^\Delta}\cot^\Delta\frac{\theta}{2}.\label{opAsympt}
\end{equation}
The overall dependence on $\theta$ matches \eqref{opGeo}. However, the only coupling which matters in the leading-order term is $\lambda_1$. The quadratic coupling $\lambda_2$ only appears in subleading terms---it weights the finite-$\Delta$ corrections. 

To summarize, the asymptotic heavy 1-point function is encoded by a classical trajectory in the bulk and only ``sees" the term dependent on the linear coupling, while the finite-$\Delta$ corrections are computed by ``quantum" trajectories which are controlled by higher-order couplings.

\subsection{Computing the Boundary Entropy}

In $d = 2$, the 1-point function of a heavy scalar operator can be related to the partition function of a conical defect. This allows us to compute the boundary entropy.

Consider an insertion of a scalar operator $ (\Delta\slash 2, \Delta \slash 2) $ with dimension $ \Delta \sim c \rightarrow \infty $ such that $ \Delta\slash c $ is fixed. For $ \Delta\slash c \in [0,1] $, the insertion creates a conical singularity with an angular deficit $ 2\pi(1-\alpha) $ where \cite{asplund_holographic_2015,fitzpatrick_virasoro_2015},
\begin{equation}
\alpha = \sqrt{1-\frac{12\Delta}{c}}.
\end{equation}
This was proven at the level of Virasoro blocks with no boundary, but we assume that it holds more generally. The 1-point function of the heavy operator is then,
\begin{equation}
\langle \mathcal{O}(y,x)\rangle \sim \frac{Z_n}{(Z_1)^{n}}\bigg\lvert_{n\rightarrow 1\slash \alpha}.
\label{1pointasdefect}
\end{equation}
where $ Z_n $ for $ n\in \mathbb{N} $ is the partition function of a replicated orbifold theory $ \text{BCFT}^{n}\slash \mathbb{Z}_n $ with a twist operator insertion. This approach of computing partition functions of CFTs on $ n $-fold branched covers is familiar from the definitions of twist operators.

\begin{figure}
\centering
\begin{tikzpicture}[scale=1.2]
\draw[->,red, thick] (0,-0.5) to (0,1.5);
\draw[->] (0,0) to (2,0);
\node at (-0.25,1.5) {$x_h$};
\node at (2.2,0) {$y_h$};

\draw[-,dashed,thick,blue,fill=black!5] (1,0.75) circle (0.275);
\node at (1.35,0.4) {$\epsilon_0$};

\draw[->,color=black!50!yellow] (1,0.75) to[bend left] (1.3,1.2);
\node[color=black!50!yellow] at (1.65,1.2) {$(y,x)$};

\node[color=black!20!yellow] at (1,0.75) {$\bullet$};

\node[blue] at (0.5,0.75) {$\ket{a}$};
\node[red] at (-0.25,0.75) {$\ket{b}$};
\end{tikzpicture}

\caption{The regulated orbifolded BCFT$^n/\mathbb{Z}_n$ state with an operator insertion at $(y,x)$. We excise a disk of radius $\epsilon_0$, on whose boundary we impose the state $\ket{a}$.}
\label{figs:replicaBCFT}
\end{figure}
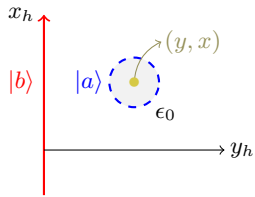

The singularity has to be regulated by cutting out a disk of radius $ \epsilon_0 $ around the insertion point (Figure \ref{figs:replicaBCFT}). On the boundary of this disk, we impose the boundary state $ \lvert a\rangle $. The resulting partition function is then \cite{cardy_entanglement_2016,sully_bcft_2020},
\begin{equation}
\begin{split}
\frac{Z_n}{(Z_1)^{n}} &= (\langle a\lvert 0 \rangle \langle 0 \lvert b \rangle)^{1-n}\biggl(\frac{\epsilon_0}{2y}\biggr)^{d_n},\\
d_n &= \frac{c}{12}\biggl(n-\frac{1}{n}\biggr).
\end{split}
\end{equation}
where $ d_n $ is the scaling dimension of the twist operator and $ \lvert b \rangle $ is the boundary state on the physical boundary $ y=0 $. The first overlap $ \langle a\lvert 0 \rangle $ is not physical and can be absorbed into the regulator \cite{sully_bcft_2020},
\begin{equation}
\langle a\lvert 0 \rangle^{1-n}\epsilon_0^{d_n} \equiv \epsilon^{d_n}.
\label{regulator}
\end{equation}
The second overlap $ \langle 0 \lvert b \rangle $ is physical and gives the boundary entropy by $ S_{bdy} = \log{\langle 0 \lvert b \rangle} $.

The formula \eqref{1pointasdefect} agrees with the 1-point function obtained from geodesics. In our case, we neglect the backreaction of the scalar field to the bulk AdS background: $mG_{N} \rightarrow 0$. On the field theory side, this corresponds to the limit $ \Delta\slash c\rightarrow 0 $ $ (\alpha \rightarrow 1) $ where the conical singularity disappears. In this limit,
\begin{equation}
1-\frac{1}{\alpha} = - \frac{6\Delta}{c} + \mathcal{O}\left(\frac{\Delta}{c}\right)^2,\ \ d_{1\slash \alpha} = \Delta + \mathcal{O}\left(\frac{\Delta}{c}\right)^2,
\end{equation}
so as $\Delta/c \to 0$,
\begin{equation}
\frac{Z_n}{(Z_1)^{n}}\bigg\lvert_{n\rightarrow 1\slash \alpha} \sim \epsilon^{\Delta}\frac{\left[e^{-(6\slash c)S_{bdy}}\right]^{\Delta}}{(2y)^{\Delta}}.
\end{equation}
The regulator $ \epsilon $ defined in \eqref{regulator} can be identified with the bulk IR cutoff $ z = \epsilon $ in Poincar\'e coordinates \cite{sully_bcft_2020}. Absorbing the regulator into the operator and equating this with our 1-point function \eqref{opAsympt}, we find,
\begin{equation}
S_{bdy} \sim -\frac{c}{6}\log{\cot{\frac{\theta}{2}}},\ \ \Delta \to \infty.\label{bdryEntropyCheck}
\end{equation}
So the 1-point function obtained from geodesic approximation indeed reproduces the boundary entropy \eqref{bdryEntropy} in the large-$\Delta$ regime after identifying $c = 3/(2G_N)$.

\section{2-Point Function}\label{sec:2point}

We now study the 2-point function, starting by computing the connected geodesic lengths between insertion points $(y_1,\vec{x})$ and $(y_2,\vec{x})$ on the same transverse slices. The cross-ratio \eqref{crossRatio} becomes,
\begin{equation}
\xi = \frac{(y_2 - y_1)^2}{4y_1 y_2},\label{crossCons}
\end{equation}
and the sum over geodesics yields an expression \eqref{tpGeneric} which is the large-$\Delta$ connected 2-point function. We only find two connected saddles---a nonreflecting geodesic and a (once) reflecting geodesic (Figure \ref{figs:2ptSaddles})---see Appendix \ref{appC}.

Upon writing our expressions in terms of $\xi$, we can then use them for more general insertion points on different transverse slices---this only changes implicit coordinate dependence but not explicit functional dependence on $\xi$.

The resulting 2-point function has two interesting features. The first, a phase transition, is seen from the existence (or lack thereof) of the connected saddles; starting from zero tension $\theta = \pi/2$ and keeping the insertions fixed, as we tune the brane angle down to,
\begin{equation}
\tan\theta = \sqrt{\xi},
\end{equation}
the nonreflecting and reflecting saddles coincide with one another (Figure \ref{figs:factorizationAng}). Tuning it down even further results in the loss of these saddles entirely, so the connected 2-point function becomes $0$ and the full 2-point function becomes factorizable---a product of the 1-point functions.\footnote{This is certainly only a large-$\Delta$ effect. The suppressed contributions coming from other connected, nongeodesic trajectories would prevent factorization.}

The second feature comes from the boundary operator product expansion (BOPE) \cite{McAvity:1995zd,Karch:2017fuh}. By performing the expansion, we find an extra tower of terms labeled by half-integer powers of $\xi$. These correspond to anomalous BCFT boundary operators which emerge only away from the probe limit, i.e. only for nonzero tension. Indeed, such terms cannot arise for zero tension even in the $\Delta$-exact answer from the method of images (Appendix \ref{app:images}).

However, there is still a problem; the resulting anomalous dimensions themselves are independent of $\theta$. The boundary operator spectrum can also be computed holographically by generalizing \cite{aharony_defect_2003} to nonzero tension, but this is expected to yield a $\theta$-dependent spectrum. We discuss this inconsistency, but further work is required.

Throughout this section, we again assume the Robin boundary condition, but the connected 2-point function in the heavy limit reduces to that of Neumann condition.

\subsection{Reflecting and Nonreflecting Trajectories}

\begin{figure}
\centering
\subfloat[$T \geq 0$]
{
\begin{tikzpicture}[scale=0.65]

\draw[->] (0,0) to (0,2);

\node[] at (0.5,-0.3) {$y_1$};
\node[] at (2.5,-0.3) {$y_2$};

\draw[-,thick,black!20!green] (0.5,0) arc (180:0:1);

\draw[-,thick,black!20!yellow] (0.5,0) arc (0:180:0.887426);
\draw[-,thick,black!20!yellow] (2.5,0) arc (0:180:1.74025);

\draw[-,draw=none,fill=white] (-2,0) to (0,0) to (-2,2) to (-2,0);

\draw[-,red,thick] (0,0) to (-2,2);
\draw[->] (-2,0) to (3,0);

\node at (-0.2,2) {$z$};

\draw[draw=none,fill=black!20!yellow] (0.5,-0.07) arc (270:450:0.09);
\draw[draw=none,fill=black!20!yellow] (2.5,-0.07) arc (270:450:0.09);

\draw[draw=none,fill=black!20!green] (0.5,-0.07) arc (270:90:0.09);
\draw[draw=none,fill=black!20!green] (2.5,-0.07) arc (270:90:0.09);

\node[black!20!yellow] at (-0.790569,0.790569) {$\bullet$};

\draw[<->,very thin] (-0.790569,0.790569) to (-0.790569,0.0);

\node at (-0.790569-0.2,0.79/2) {$z_b^*$};

\end{tikzpicture}
}
\subfloat[$T < 0$]
{
\begin{tikzpicture}[scale=0.65]

\node[] at (1.75,-0.3) {$y_1$};
\node[] at (3.37,-0.3) {$y_2$};

\draw[-,thick,black!20!green] (1.75,0) arc (180:0:0.81);

\draw[-,thick,black!20!yellow] (1.75,0) arc (180:180-37.935:1.859);
\draw[-,thick,black!20!yellow] (3.37,0) arc (0:94.080:1.146);

\draw[->] (-0.25,0) to (3.75,0);
\draw[->] (0,0) to (0,2);

\node[black!20!yellow] at (1.75,0) {$\bullet$};

\draw[-,red,thick] (0,0) to (3.75,2);

\draw[draw=none,fill=black!20!yellow] (1.75,-0.07) arc (270:450:0.09);
\draw[draw=none,fill=black!20!yellow] (3.37,-0.07) arc (270:450:0.09);

\draw[draw=none,fill=black!20!green] (1.75,-0.07) arc (270:90:0.09);
\draw[draw=none,fill=black!20!green] (3.37,-0.07) arc (270:90:0.09);

\node[black!20!yellow] at (2.143,1.143) {$\bullet$};

\draw[<->,very thin] (2.143,1.143) to (2.143,0.0);

\node at (2.143+0.3,1.143/3) {$z_b^*$};

\node at (-0.2,2) {$z$};
\end{tikzpicture}}
\caption{The reflecting (yellow) and nonreflecting (green) connected geodesics with insertion points $(y_1,\vec{x})$ and $(y_2,\vec{x})$. As the brane angle decreases, the reflecting trajectory gets ``closer" to the nonreflecting trajectory.}
\label{figs:2ptSaddles}
\end{figure}
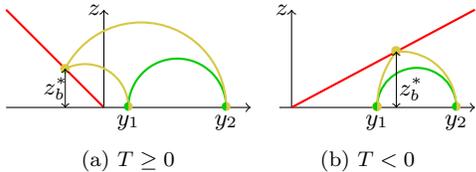

We start with insertions $(y_1,\vec{x})$ and $(y_2,\vec{x})$ ($y_2 > y_1$); just as for the 1-point function, the geodesic trajectories connecting these points to the brane or to one another are on a fixed-$\vec{x}$ slice. As there are only two connected saddles---one nonreflecting and another once-reflecting (Appendix \ref{appC})---we restrict our attention to computing these lengths. We depict both geodesics in Figure \ref{figs:2ptSaddles}.

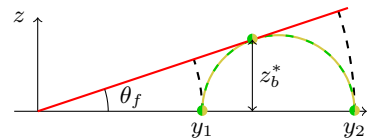
\begin{figure}
\centering
\begin{tikzpicture}[scale=1.25]

\node[] at (1.75,-0.2) {$y_1$};
\node[] at (3.37,-0.2) {$y_2$};

\draw[-,dashed,thick] (1.75,0) arc (0:18.5:1.75);
\draw[-,dashed,thick] (3.37,0) arc (0:18.5:3.37);

\draw[-,thick,black!20!yellow] (1.75,0) arc (180:0:0.81);
\draw[-,thick,black!20!green,dashed,dash pattern= on 3pt off 5pt] (1.75,0) arc (180:0:0.81);

\draw[-,draw=none,fill=white] (0,0) to (3,1) to (3,1) to (0,1) to (0,0);

\draw[->] (-0.25,0) to (3.5,0);
\draw[->] (0,0) to (0,1);

\node[black!20!yellow] at (1.75,0) {$\bullet$};

\draw[-,red,thick] (0,0) to (3.3,1.1);

\draw[draw=none,fill=black!20!yellow] (1.75,-0.047) arc (270:450:0.056);
\draw[draw=none,fill=black!20!yellow] (3.37,-0.047) arc (270:450:0.056);

\draw[draw=none,fill=black!20!green] (1.75,-0.047) arc (270:90:0.056);
\draw[draw=none,fill=black!20!green] (3.37,-0.047) arc (270:90:0.056);

\draw[draw=none,fill=black!20!yellow] (2.28298,0.761151-0.047) arc (270:450:0.056);
\draw[draw=none,fill=black!20!green] (2.28298,0.761151-0.047) arc (270:90:0.056);

\draw[<->] (2.28298,0.761151) to (2.28298,0);
\node at (2.28298+0.2,0.761151/2) {$z_b^*$};

\node at (-0.2,1) {$z$};

\draw[-] (0.75,0) arc (0:18:0.75);
\node at (1,0.15) {$\theta_f$};
\end{tikzpicture}
\caption{The KR braneworld with $\theta$ at the factorization angle $\theta_f$. We only have one connected geodesic, with features of both the reflecting and nonreflecting geodesics. Also shown are the disconnected geodesics.}
\label{figs:factorizationAng}
\end{figure}

Observe however that, with the insertion points fixed, there exists a brane angle $\theta = \theta_f$ that we call the \textit{factorization angle} below which both of these trajectories no longer ``fit" in the braneworld---either connected geodesic is excised by the brane. A geometric calculation (using \eqref{crossCons} and generalizing by conformal symmetry)
yields,
\begin{equation}
\tan\theta_f = \sqrt{\xi}.\label{factorAng}
\end{equation}
At $\theta = \theta_f$, the nonreflecting and reflecting geodesics consolidate and become the same path, giving us just one saddle shown in Figure \ref{figs:factorizationAng}. We for now set $\theta > \theta_f$ and discuss the details of the phase transition in Section \ref{sec3B}.

The nonreflecting trajectory is simply a circle centered at $(y_2+y_1)/2$ with radius $(y_2-y_1)/2$ up to a cutoff $z = \epsilon$. Renormalizing with the counterterm $+2\log\epsilon$, we find,
\begin{equation}
L_{12N} = 2\log(y_2 - y_1).
\end{equation}
The corresponding contribution of this geodesic to the 2-point function in the $\Delta \to \infty$ limit is (in terms of $\xi$),
\begin{equation}
e^{-\Delta L_{12N}} = \frac{1}{(y_2 - y_1)^{2\Delta}} = \frac{1}{(4y_1 y_2)^\Delta} \xi^{-\Delta},\label{contributionN}
\end{equation}
where we have written the exponential in the form \eqref{tpGeneric}.

As for the reflecting trajectory, we follow a similar analysis to Section \ref{sec3A}. We compute a more generic function for the length of a piecewise concatenation of geodesic arcs which hits the brane at a depth $z_b$.\footnote{For $\theta < \pi/2$, the existence of such trajectories is not trivial because of the finite range of depths \eqref{boundsNegTen} for each piece. However, one can prove existence precisely above the factorization angle.}
\begin{equation}
\begin{split}
L_{12R}(z_b)
= &-2\log z_b + \log\left[\left(y_1 - z_b \cot\theta\right)^2 + z_b^2\right]\\
&+ \log\left[\left(y_2 - z_b \cot\theta\right)^2 + z_b^2\right] - 2\log\epsilon.
\end{split}
\end{equation}
There are four saddles, but above the factorization angle only one is ``physical" (i.e. at positive depth),
\begin{equation}
z_b^* = \sqrt{y_1 y_2}\sin\theta,\label{depth}
\end{equation}
Thus the minimal length is,
\begin{align}
L_{12R}^*
&= \log(4y_1 y_2) - 2\log\left[\frac{\sin\theta}{\left(\frac{y_1 + y_2}{2\sqrt{y_1 y_2}}\right) - \cos\theta}\right].\label{minLengthRef}
\end{align}
This calculation does not use the law of reflection to get \eqref{minLengthRef}. However an explicit calculation of the angles for the trajectory corresponding to \eqref{depth} confirms the law holds.

Now from \eqref{crossCons},
\begin{equation}
1 + \xi = \frac{(y_1 + y_2)^2}{4y_1 y_2}.
\end{equation}
We use this to write the contribution of $L_{12R}^*$ to the 2-point function in the form \eqref{tpGeneric},
\begin{equation}
e^{-\Delta L_{12R}^*} = \frac{1}{(4y_1 y_2)^\Delta} \left(\frac{\sin\theta}{\sqrt{1 + \xi} - \cos\theta}\right)^{2\Delta}.\label{contributionR}
\end{equation}
We are at last ready to write the 2-point function to leading order in $\Delta \to \infty$ by combining \eqref{contributionN} and \eqref{contributionR},
\begin{align}
&\expval{\mathcal{O}(y_1,\vec{x}_1)\mathcal{O}(y_2,\vec{x}_2)} \sim \frac{1}{(4y_1 y_2)^\Delta}\mathcal{F}_G(\xi),\label{tpGeo}\\
&\mathcal{F}_G(\xi) = \xi^{-\Delta} \pm \left(\frac{\sin\theta}{\sqrt{1+\xi} - \cos\theta}\right)^{2\Delta}.\label{tpGeoStructure}
\end{align}
The $+$ sign corresponds to Robin boundary conditions while the $-$ sign is for Dirichlet.

We may now switch to arbitrary insertion points $(y_1,\vec{x}_1)$ and $(y_2,\vec{x}_2)$ because we have dropped all explicit dependence on the transverse coordinate $\vec{x}$ in \eqref{contributionN} and \eqref{contributionR}. Specifically, because the 2-point function must be of the form \eqref{tpGeneric} with $\mathcal{F}_G$ being entirely fixed up to the cross-ratio by symmetry \cite{Liendo:2012hy,Mazac:2018biw}, we simply use \eqref{tpGeoStructure}.

Furthermore, for the tensionless case $\theta = \pi/2$ we get,
\begin{equation}
\mathcal{F}_G(\xi) = \xi^{-\Delta} \pm (1 + \xi)^{-\Delta},
\end{equation}
in agreement with the $\Delta$-exact result from the method of images computed in Appendix \ref{app:images}.

However, because the expression is simply part of an expansion around $\Delta \to \infty$ for $\theta \neq \pi/2$, we must note the relative order of the terms. Indeed, at $\theta > \theta_f$,
\begin{equation}
\frac{1}{\xi} > \left(\frac{\sin\theta}{\sqrt{1+\xi} - \cos\theta}\right)^2,
\end{equation}
so the reflecting term is subleading to the nonreflecting term.

As this work was being written, \cite{Reeves:2021sab} presented a Lorentzian analog to our calculation using a null geodesic reflecting off of the brane. They observe a singularity in the 2-point function, which our expression reproduces upon analytic continuation of the cross-ratio. To see how this works, first define the radial coordinate $\rho$ by \cite{Pappadopulo:2012jk,Hogervorst:2013sma},
\begin{equation}
\xi = \frac{(1-\rho)^2}{4\rho},\ \ 1 + \xi = \frac{(1 + \rho)^2}{4\rho}.
\end{equation}
Our reflecting term in \eqref{tpGeoStructure} exhibits a singularity when $1 + \xi = \cos^2\theta$, which in terms of $\rho$ is equivalent to,
\begin{equation}
\rho = e^{\pm 2i\theta}.
\end{equation}

\subsection{Factorization Phase Transition}\label{sec3B}

First, note that the previous discussion is entirely about the connected part of the 2-point function. We may also consider the disconnected part,
\begin{align}
\expval{\mathcal{O}(y_1,\vec{x}_1)\mathcal{O}(y_2,\vec{x}_2)}_{\text{dc}}
&= \expval{\mathcal{O}(y_1,\vec{x}_1)}\expval{\mathcal{O}(y_2,\vec{x}_2)}\nonumber\\
&\sim \frac{\lambda_1^2}{(4y_1 y_2)^\Delta} \cot^{2\Delta}\frac{\theta}{2},
\end{align}
where we use the geodesic result \eqref{opGeo} to write this term to leading order in large $\Delta$. It is constant in $\xi$ and vanishes when there is no scalar coupling to the brane.

A quick calculation shows that, in the large-$\Delta$ expansion, this disconnected part is leading over the reflecting geodesic contribution, but whether or not it is leading or subleading with respect to the nonreflecting contribution depends on $\xi$ and $\theta$. 

As touched upon in the previous section however, there exists a brane angle \eqref{factorAng} below which the connected geodesics undergo a phase transition. Below this factorization angle, only the disconnected part of the 2-point function remains; in other words, the large-$\Delta$ 2-point function factorizes into a product of 1-point functions.

Another way to think about this is by keeping the brane angle $\theta$ fixed. If the brane tension is non-negative, then the phase transition simply never happens. If the tension is negative, then there exists a cross-ratio,
\begin{equation}
\xi_f = \tan^2\theta,
\end{equation}
such that 2-point functions whose insertion points have a cross-ratio $\xi > \xi_f$ factorize in the large-$\Delta$ regime.

This phase transition is reminiscent of others in large-$N$ limits of holographic gauge theories. A particularly well-known example is the confinement-deconfinement transition \cite{Witten:1998zw,Brandhuber:1998bs}---one finds a qualitative change in the Wilson lines when tuning temperature, reflecting a phase transition from Coulombic to free behavior of quarks.

In our case, the inserted heavy operators are ``decorrelated" in the phase transition. Qualitatively, this is akin to chiral symmetry breaking; a similar story was discussed by \cite{Johnson:2008vna} but in terms of flavor branes \cite{Karch:2002sh} ending on horizons, as opposed to geodesics ending on branes. Roughly speaking, we assign global symmetry groups to the two insertion points. If the points are correlated, then the symmetry is spontaneously broken. Once the points are decorrelated, the symmetry is restored.

\subsection{Consistency of the BOPE and Anomalous Defect Operators}

Within the context of the BCFT, we will now study the 2-point function obtained from doing geodesic approximation in the \textit{ambient channel} $ \xi \rightarrow 0 $ and the \textit{defect channel} $ \xi\rightarrow \infty $.\footnote{The BCFT literature \cite{Liendo:2012hy,Mazac:2018biw} often refers to the ``bulk" and ``boundary" channels of the field theory. To avoid confusion with the holographic terminology, in this section we refer to the former by ``ambient" and the latter by ``defect" \cite{aharony_defect_2003,Karch:2017fuh}.} These limits are determined by operator product expansions which depend on BCFT data. While the ambient channel corresponds to taking operator insertions very far from the BCFT boundary or close to one another, the defect channel corresponds to taking operator insertions very close to the BCFT boundary or far from each other.




The ambient channel $ \xi \rightarrow 0 $ expansion of \eqref{tpGeoStructure} is obtained by rewriting,
\begin{equation}
\mathcal{F}_G(\xi) = \xi^{-\Delta} \pm \left(\cot{\frac{\theta}{2}}\right)^{2\Delta}\left(1 + \frac{\sqrt{\xi+1}-1}{1-\cos{\theta}}\right)^{-2\Delta},
\end{equation}
which expands in integer powers of $ \xi $,
\begin{equation}
\mathcal{F}_G(\xi) = \xi^{-\Delta}\pm \Bigl(\cot{\frac{\theta}{2}}\Bigr)^{2\Delta}\sum_{n=0}^{\infty}A_n\, \xi^{n}.
\label{bulkchannel}
\end{equation}
The first three coefficients are
\begin{equation}
A_0 = 1,\ A_1 = \frac{\Delta}{\cos{\theta}-1},\ A_2 = \frac{\Delta(2\Delta+2-\cos{\theta})}{4(\cos{\theta}-1)^{2}}.
\end{equation}
The series \eqref{bulkchannel} can be organized to be an infinite series over conformal blocks of primary operators with dimensions $ \Delta_n = 2\Delta + 2n $ with $ n\in \mathbb{N} $; these are the double-trace primaries $\normord{\mathcal{O}\square^{n}_{\zeta}\mathcal{O}}$. 

We denote the $ n=0 $ primary by $ \mathcal{O}^{2} $, and from the 2-point function we find that,\footnote{Normal ordering subtracts the $ \xi^{-\Delta} $ divergence.}
\begin{equation}
\langle \mathcal{O}^{2}(y,\vec{x}) \rangle = \frac{\pm 1}{(2y)^{2\Delta}} \cot^{2\Delta}\frac{\theta}{2},
\end{equation}
which has the form required by defect conformal symmetry and a coefficient in agreement with a geodesic terminating on the brane. However, in contrast to the 1-point function of $ \mathcal{O} $ \eqref{opAsympt}, no $ \lambda_1 $-prefactor appears.

The coefficients of the conformal blocks can be in principle computed using the methods of \cite{hogervorst_crossing_2017,Hogervorst:2017kbj}, but we have not managed to do this analytically. Nonetheless, from the expansion we see that the OPE coefficients of these operators are modified at nonzero tension $ \theta \neq \pi/2 $, but scaling dimensions remain independent as one would expect for a free theory.

To expand in the defect channel $ \xi\rightarrow \infty $, we write,
\begin{equation}
\mathcal{F}_G(\xi) = \xi^{-\Delta} \pm (\sin{\theta})^{2\Delta} \left(\sqrt{\xi + 1} - \cos{\theta}\right)^{-2\Delta},
\label{brewriting}
\end{equation}
which expands as,
\begin{equation}
\mathcal{F}_G(\xi) = \xi^{-\Delta}\pm (\sin{\theta})^{2\Delta}\sum_{n=0}^{\infty}B_{n/2}\, \xi^{-\Delta-n\slash 2}.
\label{boundaryexp}
\end{equation}
The first three coefficients are,
\begin{equation}
B_0 = 1,\ B_{1/2} = 2\Delta \cos{\theta},\ B_1 = \Delta[(2\Delta+1)\cos^{2}{\theta}-1].
\end{equation}
At zero tension, all of the half-integer coefficients vanish. In this case, the expansion can be reorganized in defect conformal blocks of single-trace defect primaries \cite{liendo_bootstrap_2013},
\begin{align}
&\text{Dirichlet:}&& \widehat{\Delta}_n = \Delta + 2n+1,\\
&\text{Neumann:}&&\widehat{\Delta}_n = \Delta + 2n.
\end{align}

\noindent At nonzero tension however, the appearance of half-integer powers is unusual and requires a new tower of operators to appear in the defect operator spectrum. Since an operator of dimension $ \widehat{\Delta}_n $ appears as a term of order $ \smash{\xi^{-\widehat{\Delta}_n}} $ in the expansion of $ \mathcal{F}(\xi) $ \cite{aharony_defect_2003}, \eqref{boundaryexp} requires the existence of a primary of dimension $ \Delta + 1\slash 2 $. Such a dimension has to be anomalous since engineering dimensions of local operators constructed from $ \mathcal{O} $ cannot produce a fractional power.

There is another problem however. The fact that dimensions of boundary operators extracted from \eqref{boundaryexp} are independent of $ \theta $ appears to be inconsistent. The defect operator spectrum can be computed holographically by generalizing the calculations of \cite{aharony_defect_2003} and the bulk calculations of \cite{Mazac:2018biw} to nonzero tension. The idea is to solve the Klein-Gordon equation using the ansatz,
\begin{equation}
\Phi(X) = \sum_n \psi_n(r)\, \widehat{\Phi}_n(r,\vec{x}),
\end{equation}
where these coordinates are defined in \eqref{btzslicing}. The $n$th mode $ \widehat{\Phi}_n $ solves the Klein-Gordon equation in AdS$ _d $ with mass $ \widehat{\Delta}_n(\widehat{\Delta}_n-d-1) $, where $ \widehat{\Delta}_n $ is the dimension of a boundary operator. This leads to an equation for $ \psi_n $ that can be solved in terms of hypergeometric functions.

Requiring normalizability and imposing boundary conditions at the brane $ \mathcal{Q} $ gives the equation (akin to \eqref{ratio12}),
\begin{equation}
R_n = \frac{\Gamma\bigl(\frac{\Delta-\widehat{\Delta}_n}{2} \bigr)\Gamma\bigl(\frac{\Delta+\widehat{\Delta}_n-d+1}{2} \bigr)}{2\Gamma\bigl(\frac{\Delta-\widehat{\Delta}_n+1}{2} \bigr)\Gamma\bigl(\frac{\Delta+\widehat{\Delta}_n-d+2}{2} \bigr)}
\end{equation}
where $ R_n = R_n(\Delta,\widehat{\Delta}_n,\theta) $ is a known function given in terms of the solutions $ \psi_n $. From this equation, the boundary operator spectrum $ \widehat{\Delta}_n $ can be solved for in principle. They depend on not just $ \Delta $ but also tension through $ \theta $.

One possible cause for the disagreement could be a mismatch between the boundary conditions of the bulk scalar field $\phi$ and the dual BCFT operator $\mathcal{O}$. The mismatch occurs at the $ z = y= 0 $ corner of the holographic bulk and only arises when the brane has angle $\theta \neq \pi/2$, in which case the brane's normal vector in the bulk is not orthogonal to the defect in the BCFT. 

It would thus be interesting to examine if there is an order-of-limits issue when taking $z \to 0$ and $y \to 0$. We took the former limit first, but doing the latter (or more correctly, taking $y \to z\cot\theta$) first would physically mean taking a bulk scalar insertion to the KR brane, then going to the BCFT boundary. In this case, one would obtain a ``brane OPE'' at finite-$ z $ from which the BOPE coefficients could be obtained by taking $ z\rightarrow 0 $. However, it is not clear if taking the limits in this order is consistent.






\section{Conclusions and Future Directions}

We have studied $ d $-dimensional BCFTs using a bottom-up\footnote{We briefly mention that much work on 1-point functions has also been done in top-down models. See \cite{Nagasaki:2012re,Kristjansen:2012tn,Buhl-Mortensen:2015gfd} for a sample of this story.} holographic model involving a massive free scalar field with a Karch-Randall end-of-the-world brane. We have computed $1$-point and $2$-point functions of the dual scalar operator in the large-$ \Delta $ limit when the brane has nonzero tension. This requires a generalization of geodesic approximation to manifolds with boundaries which we have provided.

The $1$-point function is cleanly replicated by our geodesic calculation, and we are also able to use it to reproduce the known boundary entropy.

However, our analysis of the $2$-point function manifests unusual phenomena. We find a factorization phase transition for negative-tension branes, so such BCFT states cannot have large-$\Delta$ correlations beyond a particular cross-ratio. Furthermore, the ambient OPE appears to be completely consistent, but the BOPE gives rise to an anomalous tower of boundary operators with tension-independent scaling dimensions. This appears to be inconsistent with another holographic calculation of the boundary operator spectrum. We propose that the problem is related to the corner at $ z=y=0 $ which is not orthogonal for $ \theta \neq \pi \slash 2 $. We leave the resolution of this issue and other possible questions about the influence of corners in AdS/BCFT to future work.

There are plenty of clear avenues along which to continue this work. We list some below:
\begin{itemize}
\item We study a very simple class of KR branes and thus only have a BCFT on half-space. It would be interesting to study other configurations (such as a configuration dual to a BCFT on a disk) or even a two-brane ``wedge" setup \cite{Akal:2020wfl,Geng:2020fxl,Geng:2021iyq}.

\item Our work is in Euclidean signature, so it would be interesting to recast everything in real time, particularly the 2-point function. As this work was being written up, a real-time study of constraints on holographic BCFT from a bulk causal structure using reflecting null geodesics was performed \cite{Reeves:2021sab}. We found agreement in the singularity structure, but future work could involve consolidating more of the results.

\item We only consider empty, nonbackreacting AdS. One could ask how our methods apply to other solutions, e.g. black holes or backreacting geometries.

\item One could interpret our results in terms of the localized gravitational theory on the brane coupled to the BCFT bath and use geodesic approximation to study correlations between bath radiation and corresponding entanglement islands. Along these lines, one could also study correlations in the context of braneworld cosmologies \cite{Antonini:2019qkt}.

\end{itemize}


\section*{Acknowledgments}

We thank Elena C\'aceres, Mitsutoshi Fujita, Andreas Karch, Esko Keski-Vakkuri, Arnab Kundu, Wyatt Reeves, Moshe Rozali, Petar Simidzija, James Sully, Tadashi Takayanagi, Christopher Waddell, and David Wakeham for reviewing and commenting on the draft. We also thank Elena C\'aceres, Niko Jokela, Andreas Karch, Esko Keski-Vakkuri, Arnab Kundu, Francesco Nitti, and Miika Sarkkinen for useful discussions during the completion of this work.

JK was supported by the Osk. Huttunen Foundation. SS was supported by NSF Grants No. PHY-1820712 and No. PHY-1914679.

\begin{appendix}
\section{Method of Images and Mean Field Theory}\label{app:images}

We can explicitly construct a tensionless KR braneworld with a scalar field as an orbifold theory. Specifically we quotient empty AdS$_{d+1}$ with a scalar by some discrete subgroup of isometries $\Gamma$ \cite{Horowitz:1998xk,Gao:1999er,Shashi:2020mkd}. Generically, if $\Gamma$ has fixed points $\mathcal{Q}_\Gamma$, then these will comprise a defect in the quotient space. Tensionless KR branes occur when $\mathcal{Q}_\Gamma$ is codimension-one to the bulk.\footnote{It is unclear if one can construct all tensionless KR branes this way. An argument in the affirmative would need to demonstrate that, for any AdS$_d$ slice in AdS$_{d+1}$ with $K = 0$, there exists an isometry fixing each point of the slice.}

We can get propagators from a method of images. For bulk-to-boundary propagators in $\text{AdS}_{d+1}/\Gamma$ \cite{Gao:1999er},\footnote{The method of images is a general technique. \cite{KeskiVakkuri:1998nw} uses it to write the bulk-to-boundary propagator in the BTZ black hole \cite{Banados:1992gq}. \cite{Balasubramanian:2003kq} does so in a nonsingular multiboundary AdS$_3$ orbifold. \cite{Arefeva:2015zra,Ageev:2015qbz,Ageev:2017yno,Arefeva:2016wek,Arefeva:2016nic} studies $\mathbb{Z}_n$ orbifolds with conical defects.}
\begin{equation}
\mathcal{K}_{\mathbb{H}/\Gamma}(X,\zeta') = \frac{1}{|\Gamma|} \sum_{\gamma \in \Gamma} \chi(\gamma)\Omega_\gamma(\zeta')^\Delta K_\mathbb{H}(X,\gamma \zeta').
\end{equation}
$X = (z,y,\vec{x})$ is a point in Euclidean AdS$_{d+1}$ ($\cong \mathbb{H}^{d+1}$), $\zeta' = (y',\vec{x}')$ is a boundary point, $\chi$ is some 1-dimensional unitary representation of $\Gamma$, $\Omega_\gamma$ is the Jacobian of the boundary's conformal transformation induced by $\gamma$, and $K_\mathbb{H}$ is the empty AdS$_{d+1}$ bulk-to-boundary propagator, i.e. the Poisson kernel \cite{Freedman:1998tz,DHoker:2002nbb},
\begin{equation}
\mathcal{K}_\mathbb{H}(X,\zeta') = C_\Delta \left[\frac{z}{z^2 + (y-y')^2 + |\vec{x}-\vec{x}'|^2}\right]^\Delta,
\end{equation}
with normalization $C_\Delta = \pi^{-d/2}\Gamma(\Delta)/\Gamma(\Delta - d/2)$.

In the quotient space, the scalar field will generically have different sectors which are defined by how they transform under $\Gamma$. These correspond to different representations $\chi$ which weight the image terms so that the propagator satisfies the appropriate boundary condition on $\mathcal{Q}$. The sum above using a 1-dimensional unitary representation $\chi$ accommodates an untwisted sector ($\chi = 1$) and twisted sectors ($\chi \neq 1$ but $|\chi|= 1$).

To actualize this discussion, take an orbifold by parity,
\begin{equation}
\mathcal{P}: (z,y,\vec{x}) \to (z,-y,\vec{x}).
\end{equation}
The resulting brane is a planar KR probe brane of the sort discussed in the main text ($y = 0$). Such an isometry generates a $\mathbb{Z}_2$ subgroup, so there exists an untwisted sector corresponding to Neumann boundary conditions on the scalar field and a $\mathbb{Z}_2$-twisted sector corresponding to Dirichlet boundary conditions.

This story about propagators carries into the boundary theory on half-space. We still have two sectors---an untwisted (Neumann) sector and a $\mathbb{Z}_2$-twisted (Dirichlet) sector---in which the respective propagators are \cite{Almheiri:2018ijj},
\begin{align}
\begin{split}
\expval{\mathcal{O}(y_1,\vec{x}_1)\mathcal{O}(y_2,\vec{x}_2)} \sim\ &\frac{1}{[(y_2-y_1)^2 + |\vec{x}_2-\vec{x}_1|^2]^\Delta}\\
&+\frac{1}{[(y_2 + y_1)^2 + |\vec{x}_2-\vec{x}_1|^2]^\Delta},
\end{split}\label{nImages}\\
\begin{split}
\expval{\mathcal{O}(y_1,\vec{x}_1)\mathcal{O}(y_2,\vec{x}_2)} \sim\ &\frac{1}{[(y_2-y_1)^2 + |\vec{x}_2-\vec{x}_1|^2]^\Delta}\\
&-\frac{1}{[(y_2 + y_1)^2 + |\vec{x}_2-\vec{x}_1|^2]^\Delta}.
\end{split}\label{dImages}
\end{align}
That being said, we also comment that the above boundary propagators are those of a mean field theory \cite{Mazac:2018biw}. This becomes evident when we write \eqref{nImages} and \eqref{dImages} in terms of the cross-ratio \eqref{crossRatio},
\begin{equation}
\expval{\mathcal{O}(y_1,\vec{x}_1)\mathcal{O}(y_2,\vec{x}_2)} \sim \frac{1}{(4y_1 y_2)^\Delta} \left[\xi^{-\Delta} \pm (\xi + 1)^{-\Delta}\right],
\end{equation}
At this point, note that the method of images appears to be rather robust. Not only do we not make any assumptions about the order of $\Delta$, but we may consider other quotients which provide more nontrivial shapes for the boundary. For instance, we may consider another $\mathbb{Z}_2$ isometry in the bulk: parity composed with inversion,
\begin{equation}
\mathcal{P} \mathcal{I}_a: (z,y,\vec{x}) \to \frac{a^2}{z^2 + y^2 + |\vec{x}|^2}(z,y,\vec{x}).
\end{equation}
Rather than a plane, we have a brane which is a hemisphere. The dual BCFT state thus has as its boundary a $(d-1)$-sphere of radius $a$.

\begin{figure}
\centering
\begin{tikzpicture}[scale=0.7]
\draw[-,thick,black!20] (0.2,-2) to (0.2,2);
\draw[-,thick,black!20] (0.4,-2) to (0.4,2);
\draw[-,thick,black!20] (0.6,-2) to (0.6,2);
\draw[-,thick,black!20] (0.8,-2) to (0.8,2);
\draw[-,thick,black!20] (1,-2) to (1,2);
\draw[-,thick,black!20] (1.2,-2) to (1.2,2);
\draw[-,thick,black!20] (1.4,-2) to (1.4,2);
\draw[-,thick,black!20] (1.6,-2) to (1.6,2);
\draw[-,thick,black!20] (1.8,-2) to (1.8,2);

\draw[->,red, thick] (0,-2) to (0,2);
\draw[->] (0,0) to (2,0);
\node at (-0.35,2) {$\vec{x}_h$};
\node at (2.3,0) {$y_h$};

\draw[thick,->] (2.3,0.5) to[bend left] (4.7,0.5);

\draw[-,thick,black!20] (7+0.2/1.2,0) circle (1.5-0.2/1.2);
\draw[-,thick,black!20] (7+0.4/1.2,0) circle (1.5-0.4/1.2);
\draw[-,thick,black!20] (7+0.6/1.2,0) circle (1.5-0.6/1.2);
\draw[-,thick,black!20] (7+0.8/1.2,0) circle (1.5-0.8/1.2);
\draw[-,thick,black!20] (7+1/1.2,0) circle (1.5-1/1.2);
\draw[-,thick,black!20] (7+1.2/1.2,0) circle (1.5-1.2/1.2);
\draw[-,thick,black!20] (7+1.4/1.2,0) circle (1.5-1.4/1.2);
\draw[-,thick,black!20] (7+1.6/1.2,0) circle (1.5-1.6/1.2);
\draw[-,thick,black!20] (7+1.8/1.2,0) circle (1.5-1.8/1.2);

\draw[-,red,thick] (7,0) circle (1.5);

\draw[->] (5,0) to (9,0);
\draw[->] (7,-2) to (7,2);
\node at (6.65,2) {$\vec{x}_d$};
\node at (9.3,0) {$y_d$};
\end{tikzpicture}

\caption{A visual representation of the conformal transformation mapping half-space to the unit disk (which can then be rescaled). The boundary at $y_h = 0$ is mapped to the boundary at $|\vec{x}_d|^2 + y_d^2 = 1$.}
\label{figs:confHalfToDisk}
\end{figure}
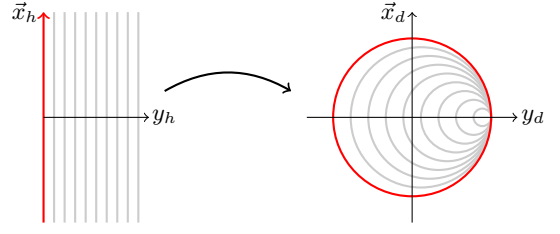

However, digging deeper reveals the limitations to this approach. In the bulk, the extrinsic curvature of even the hemispherical brane is still $0$---the brane is a tensionless probe brane. Additionally, there exists a conformal transformation between the disk and half-plane (Figure \ref{figs:confHalfToDisk}), so studying the BCFT state on one geometry is equivalent to studying the related state on the other geometry.

This suggests the method of images is only good for tensionless branes, which we argue more generally. Let $ \mathcal{M} $ be a smooth manifold with a $ \mathbb{Z}_2$ isometry whose fixed points form a codimension-1 submanifold $ \mathcal{Q} $. Near $ \mathcal{Q} $, we write the metric of $ \mathcal{M} $ in Gaussian normal coordinates,
\begin{equation}
g_{ab}dX^{a}dX^{b} = d\rho^{2} + g_{\mu\nu}(\rho,\hat{x})\,d\hat{x}^{\mu}d\hat{x}^{\nu},
\end{equation}
where $ X^{a} = (\rho,\hat{x}^{\mu}) $ are coordinates on $ \mathcal{M} $ and $\hat{x}^{\mu} $ are coordinates on $ \mathcal{Q} $.
On the former coordinates, a general $\mathbb{Z}_2$ action is written as (suppressing indices for convenience),
\begin{equation}
(\rho,\hat{x}) \mapsto A(\rho,\hat{x}),\ \ A^2 = \textbf{1}.\label{actionA}
\end{equation}
The surface $ \mathcal{Q} $ is located at $ \rho = 0 $ with its induced metric and extrinsic curvature given by,
\begin{equation}
h_{\mu\nu}(\hat{x}) = g_{\mu\nu}(0,\hat{x}), \quad K_{\mu\nu} = \left.\frac{1}{2}\partial_{\rho}g_{\mu\nu}\right|_{\rho = 0}.
\end{equation}
Taylor expanding $ g_{\mu\nu} $ around $ \rho = 0 $ gives,
\begin{equation}
g_{\mu\nu}(\rho,\hat{x}) =  h_{\mu\nu}(\hat{x}) + 2K_{\mu\nu}(\hat{x})\,\rho + \mathcal{O}(\rho^{2}),
\label{taylorg}
\end{equation}
which generalizes to any coordinate system of $ \mathcal{Q} $.

The $ \mathbb{Z}_2 $ action \eqref{actionA} has to keep each point $ (0,\hat{x}) $ fixed. Thus near $ \rho = 0 $, all such actions are reflections,
\begin{equation}
A(\rho,\hat{x}) =  (-\rho + \mathcal{O}(\rho^{2}),\hat{x} + \mathcal{O}(\rho)),
\label{fixedpointaction}
\end{equation}
and higher-order corrections in $\rho$ do not contribute at linear order in \eqref{taylorg}.

An action \eqref{fixedpointaction} is an isometry if and only if the linear term in \eqref{taylorg} vanishes, since we must have,
\begin{equation}
g_{\mu\nu}\left(A(\rho,\hat{x})\right) = g_{\mu\nu}(\rho,\hat{x}).
\end{equation}
Thus for isometries, $ K_{\mu\nu} = 0 $ at each point on $ \mathcal{Q} $. From \eqref{krNeumann}, $\mathcal{Q}$ thus has constant and vanishing tension $ T = 0 $.

In summary, although the orbifolds of AdS$_{d+1}$ gravity coupled to a scalar are a limited class of theories, we can obtain a mean field theory in this way.




\section{$\Delta$-Exact 1-Point Functions in AdS/BCFT}\label{appB}

The starting point of computing the 1-point functions, regardless of the scalar field's coupling to the brane, is to solve \eqref{kgEq} for the background field $\phi_B$ to compute $f_B$ in \eqref{modeBack}. Here we do so for a quadratic brane potential,
\begin{equation}
V(\Phi) = \lambda_1 \Phi + \frac{1}{2}\lambda_2 \Phi^2,
\end{equation}
and the corresponding boundary condition on $\phi_B$,
\begin{equation}
\left.\left(z\sin{\theta}\,\partial_y - z\cos{\theta}\, \partial_z + \lambda_2\right)\phi_B\right|_{\mathcal{Q}} = -\lambda_1.
\end{equation}
We define the coordinates,
\begin{equation}
r = \frac{y}{z},\ \ w = \sqrt{y^2 + z^2},
\label{btzslicing}
\end{equation}
for which the metric \eqref{met} takes the form,
\begin{equation}
ds^2 = \frac{dr^2}{1+r^2} + \frac{1+r^2}{w^2}(dw^2 + d\vec{x}^2).
\end{equation}
This is Euclidean AdS$_{d+1}$ in hyperbolic slicing; the slices are written in lower-dimensional Poincar\'e coordinates. Now the planar KR brane \eqref{planarKR} is located at $r = \cot\theta$ and the $y > 0$ half of the conformal boundary is at $r \to \infty$.

In these coordinates, the equation of motion for the scalar field $\Phi$ takes the form,
\begin{equation}
\begin{split}
0 =&\ (1 + r^2)\partial_r^2 \Phi + (d+1)r\partial_r \Phi\\
&+ \frac{w^2 \partial_w^2 \Phi - (d-2)w\partial_w \Phi + w^2 \partial_{\vec{x}}^2 \Phi}{1 + r^2} - \Delta(\Delta - d)\Phi.
\end{split}
\end{equation}
For the ansatz $\Phi(z,y,\vec{x}) = \phi_B(r)$, this simplifies to,
\begin{equation}
(1 + r^2)\phi_B'' + (d+1)r\phi_B' - \Delta(\Delta - d)\phi_B = 0,
\end{equation}
which has the general solution,
\begin{equation}
\phi_B(r) = C_1 \phi_1(r) + C_2 \phi_2(r),\label{radPhi}
\end{equation}
where $C_{1,2}$ are constants and,
\begin{align}
\phi_1(r) &= r\,_2F_1\left(\frac{\Delta + 1}{2},\frac{d-\Delta + 1}{2};\frac{3}{2};-r^2\right),\\
\phi_2(r) &=\,\!_2F_1\left(\frac{\Delta}{2},\frac{d-\Delta}{2};\frac{1}{2};-r^2\right).
\end{align}
Recall that the background field is normalizable \eqref{modeBack}, so $\phi_B(r) \sim r^{-\Delta}$ as $r \to \infty$. This fixes the (negative) ratio of the coefficients as,
\begin{equation}
R = -\frac{C_2}{C_1} = \frac{\Gamma\left(\frac{\Delta}{2}\right)\Gamma\left(\frac{\Delta - d + 1}{2}\right)}{2\Gamma\left(\frac{\Delta + 1}{2}\right)\Gamma\left(\frac{\Delta - d + 2}{2}\right)}.\label{ratio12}
\end{equation}
Additionally the modified Robin boundary condition at the brane is now,
\begin{equation}
\left.\left[\csc\theta\,\partial_r + \lambda_2\right]\phi_B(r) \right|_{r = \cot\theta} = -\lambda_1,
\end{equation}
which for \eqref{radPhi} becomes,
\begin{equation}
F_1(\theta) - RF_2(\theta) + \lambda_2\left[G_1(\theta) - RG_2(\theta)\right] = -\frac{\lambda_1}{C_1},
\end{equation}
where we have defined the functions,
\begin{align}
F_1(\theta)
&= \csc\theta\,_2F_1\left(\frac{\Delta + 1}{2},\frac{d-\Delta + 1}{2};\frac{1}{2};-\cot^2\theta\right),\\
F_2(\theta)
&= \csc\theta\cot\theta\,\Delta(\Delta - d)\nonumber\\
&\qquad\times\,\!_2F_1\left(\frac{\Delta + 2}{2}, \frac{d-\Delta + 2}{2};\frac{3}{2};-\cot^2\theta\right),\\
G_1(\theta)
&= \cot\theta\,_2F_1\left(\frac{\Delta + 1}{2},\frac{d-\Delta + 1}{2};\frac{3}{2};-\cot^2\theta\right),\\
G_2(\theta)
&=\,_2F_1\left(\frac{\Delta}{2},\frac{d-\Delta}{2};\frac{1}{2};-\cot^2\theta\right).
\end{align}
From the Robin boundary condition, we find that,
\begin{equation}
C_1(\theta) = \frac{-\lambda_1}{F_1(\theta) - RF_2(\theta) + \lambda_2[G_1(\theta) - RG_2(\theta)]}.
\end{equation}
And in Poincar\'e coordinates, asymptotically we get,
\begin{equation}
\phi_B(z,y) = z^\Delta f_B(y) + \cdots,\ \ z \to 0,
\end{equation}
where using Euler's reflection formula yields,
\begin{equation}
f_B(y) = -\frac{\Gamma\left(\frac{\Delta}{2}\right)\Gamma\left(\frac{\Delta - d + 1}{2}\right)}{2\sqrt{\pi}\Gamma\left(\frac{2\Delta - d + 2}{2}\right)}\frac{C_1(\theta)}{y^\Delta}.
\end{equation}
The 1-point function is,
\begin{equation}
\expval{\mathcal{O}(y,\vec{x})} = \Delta f_B(y) = \frac{a_\mathcal{O}(\theta)}{(2y)^\Delta},\label{normAns}
\end{equation}
so the coefficient in \eqref{opGeneric} is,
\begin{equation}
a_\mathcal{O}(\theta) = -C_1(\theta)\frac{2^\Delta \Gamma\left(\frac{\Delta+2}{2}\right)\Gamma\left(\frac{\Delta - d + 1}{2}\right)}{\sqrt{\pi}\Gamma\left(\frac{2\Delta - d + 2}{2}\right)}.\label{exactCoeff}
\end{equation}
When taking $\lambda_2 = 0$, our result differs from the 1-point function of \cite{Fujita:2011fp}; their tension-dependent coefficient is,
\begin{equation}
a_\mathcal{O}^*(\theta) \sim \frac{1}{F_1(\theta)},
\end{equation}
i.e. no $F_2(\theta)$, which happens in our analysis if $R = 0$. For large $\Delta$, we then have (omitting $\theta$-independent factors),
\begin{equation}
\expval{\mathcal{O}} \sim \frac{1}{\tan^{\Delta}\left(\frac{\theta}{2}\right) + \cot^\Delta\left(\frac{\theta}{2}\right)}.
\end{equation}
This behavior is inconsistent with geodesic approximation, hence the need for a nonzero $R$.

\subsection{Tensionless Brane}

At zero tension $\theta = \pi/2$, our result matches the literature. Specifically, we have,
\begin{align}
&F_1\left(\frac{\pi}{2}\right) = G_2\left(\frac{\pi}{2}\right) = 1,\ F_{2}\left(\frac{\pi}{2}\right) = G_{1}\left(\frac{\pi}{2}\right) = 0,\\
&\implies C_1\left(\frac{\pi}{2}\right) = -\frac{\lambda_1}{1-\lambda_2 R}.
\end{align}
The $\lambda_2$ coupling allows us to interpolate between modified Neumann ($\lambda_2 = 0$) and Dirichlet ($\lambda_2 = \infty$) boundary conditions. For the tensionless case at these limits,
\begin{equation}
a_\mathcal{O}\left(\frac{\pi}{2}\right) = \begin{cases}
\lambda_1 \dfrac{2^\Delta \Gamma\left(\frac{\Delta+2}{2}\right)\Gamma\left(\frac{\Delta - d + 1}{2}\right)}{\sqrt{\pi}\Gamma\left(\frac{2\Delta - d + 2}{2}\right)},&\text{modified N},\\
0,&\text{D}.
\end{cases}\label{tensionlessA}
\end{equation}
The tensionless $\lambda_2 = 0$ expression can also be computed from a method of images. Take the Neumann bulk-to-boundary propagator,
\begin{equation}
\begin{split}
&\mathcal{K}(z,y_1,\vec{x}_1;y_2,\vec{x}_2)\\
&= \frac{\Gamma(\Delta)}{\pi^{d/2}\Gamma\left(\frac{2\Delta - d}{2}\right)}
\left[\left(\frac{z}{z^2 + (y_1 - y_2)^2 + |\vec{x}_1 - \vec{x}_2|^2}\right)^{\Delta}\right.\\
&\qquad\qquad\qquad\qquad+\left.\left(\frac{z}{z^2 + (y_1 + y_2)^2 + |\vec{x}_1 - \vec{x}_2|^2}\right)^{\Delta}\right],
\end{split}
\end{equation}
and use \eqref{1pointK},
\begin{equation}
\expval{\mathcal{O}(y,\vec{x})} = \frac{\lambda_1\Delta}{2\Delta - d}\int_{\mathbb{R}^{d-1}} d^{d-1} \vec{x}_b \int_0^\infty \frac{dz_b}{z_b^d} \,\mathcal{K}(z_b,0,\vec{x}_b;y,\vec{x}).
\label{againstbrane}
\end{equation}
Using the integral \cite{Rastelli:2017ecj},
\begin{equation}
\begin{split}
&\int_{\mathbb{R}^{d-1}} d^{d-1} \vec{x}_b \int_0^\infty \frac{dz_b}{z_b^d} \left(\frac{z_b}{z_b^2 + y^2 + |\vec{x}_b - \vec{x}|^2}\right)^\Delta\\
&= \frac{\pi^{(d-1)/2}\Gamma\left(\frac{\Delta}{2}\right)\Gamma\left(\frac{\Delta - d + 1}{2}\right)}{2\Gamma(\Delta)}\frac{1}{y^\Delta},
\end{split}
\end{equation}
the Neumann result \eqref{tensionlessA} is reproduced.

In \cite{DeWolfe:2001pq}, the 1-point function was also computed using \eqref{againstbrane} for $ d=4 $, but without both the overall normalization and the factor of two coming from the image term in the Neumann propagator. These factors are important to match with the alternative calculation \eqref{tensionlessA}.

\subsection{Large-$ \Delta $ Asymptotics}

\begin{figure}
\includegraphics[scale=0.35]{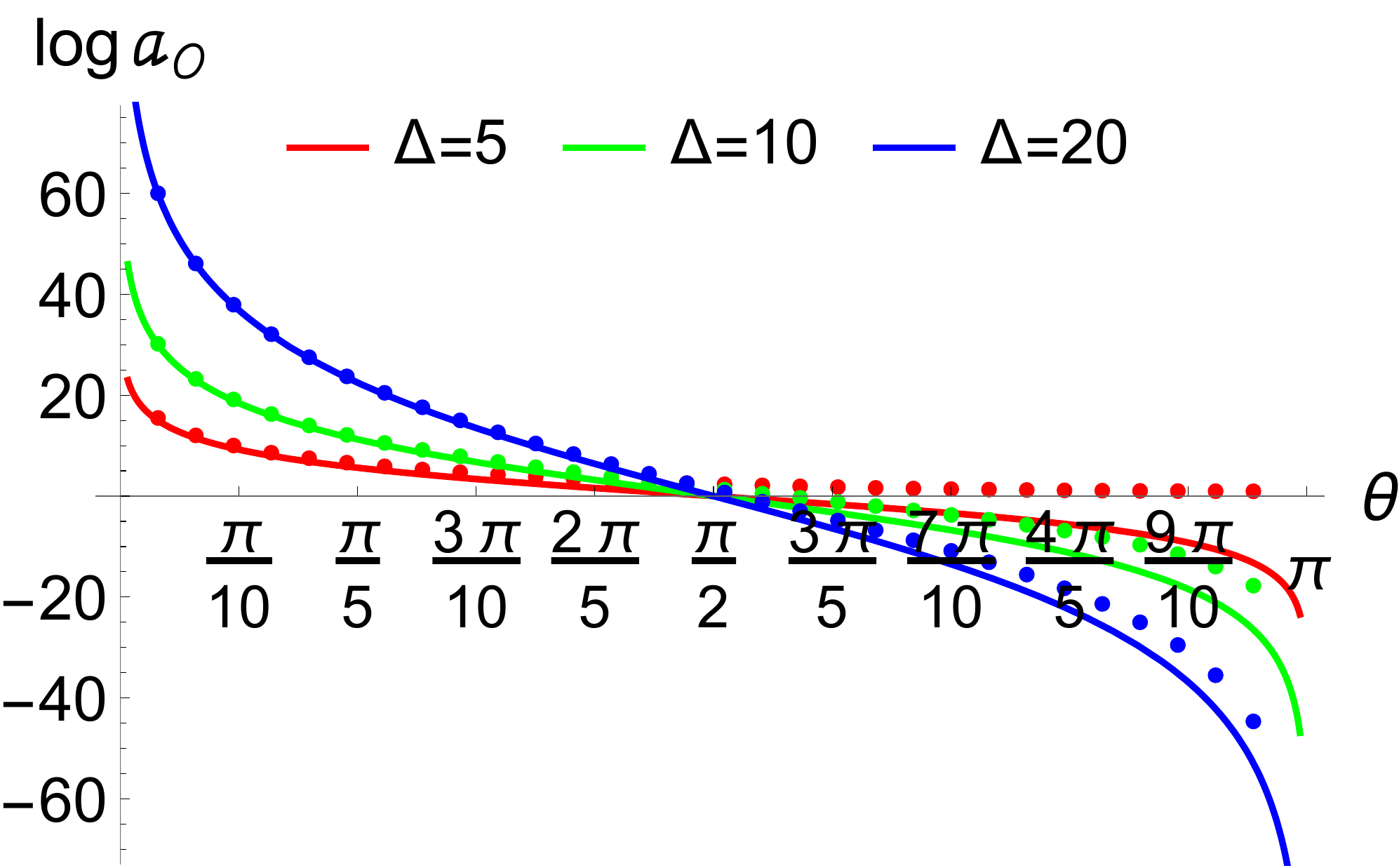}
\caption{Log plot comparing the exact coefficient $a_\mathcal{O}(\theta)$ of the 1-point function \eqref{exactCoeff} (computed numerically as dots) to the asymptotic limit \eqref{largeD} (plotted as lines) at various $\Delta$, setting $d = 3$, $\lambda_1 = 1$, and $\lambda_2 = 2$. The asymptotic expression fails for $\Delta \sim d$ (particularly for $\theta > \pi/2$), but the error rapidly decreases as $\Delta$ increases.}
\label{figs:expansionA}
\end{figure}

From the expansion presented in \cite{Jones:2001hyp}, as $\Delta \to \infty$,
\begin{align}
&_2F_1\left(a + \frac{\Delta}{2},b-\frac{\Delta}{2};\frac{1}{2};\frac{1-z}{2}\right)\nonumber\\
&\qquad\qquad\qquad\qquad\sim \frac{1}{2}\left(e^{\Delta\zeta/2} + e^{-\Delta\zeta/2}\right),\\
&\sqrt{\frac{z-1}{2}}\,_2F_1\left(a + \frac{\Delta}{2},b-\frac{\Delta}{2};\frac{3}{2};\frac{1-z}{2}\right)\nonumber\\
&\qquad\qquad\qquad\qquad\sim \frac{1}{2\Delta}\left(e^{\Delta\zeta/2} - e^{-\Delta\zeta/2}\right),
\end{align}
where $\zeta = \text{cosh}^{-1}z$. We set $(z-1)/2 = \cot^2\theta$ so that,
\begin{equation}
\zeta = \begin{cases}
2\log\cot\dfrac{\theta}{2},&0 < \theta \leq \dfrac{\pi}{2},\vspace{0.2cm}\\
2\log\tan\dfrac{\theta}{2},&\dfrac{\pi}{2} < \theta < \pi.
\end{cases}
\end{equation}
Hence for the hypergeometric functions appearing in the 1-point function, we get,
\begin{align}
F_1(\theta) \sim G_2(\theta) &\sim \frac{1}{2}\left(\cot^\Delta \frac{\theta}{2} + \tan^\Delta \frac{\theta}{2}\right),\\
F_2(\theta) &\sim \frac{\Delta}{2}\left(\cot^\Delta \frac{\theta}{2} - \tan^\Delta\frac{\theta}{2}\right),\\
G_1(\theta) &\sim \frac{1}{2\Delta}\left(\cot^\Delta\frac{\theta}{2} - \tan^\Delta\frac{\theta}{2}\right).
\end{align}
The ratio \eqref{ratio12} goes as $R \sim 1/\Delta$, so we get,
\begin{equation}
a_\mathcal{O}(\theta) \sim \lambda_1 \cot^\Delta\frac{\theta}{2},\ \ \Delta \to \infty,\label{largeD}
\end{equation}
where the dependence on $\lambda_2$ completely disappears at leading order in $\Delta \to \infty$. This term is plotted against the exact $a_\mathcal{O}(\theta)$ in Figure \ref{figs:expansionA} with good agreement---the error is high for $\Delta \sim d$ but decreases as $\Delta$ increases.

\section{Proof of the Law of Reflection}\label{app:extrem}

In this Appendix, we solve the extremization conditions \eqref{extremeReflect} imposed on Synge's world function for a reflecting geodesic. We find that the conditions impose a standard law of reflection at each reflection point.

\subsection{Extremization over $ p_k $}

Recalling the notation of Section \ref{sec2B}, Synge's world function for a reflecting geodesic is,
\begin{equation}
\sigma_n(X,X') = \sum_{k=0}^n \frac{\sigma(X_k,X_{k+1})}{p_{k+1} - p_k}.\label{nRefActapp}
\end{equation}
$\sigma(X_k,X_{k+1})$ is Synge's world function for a segment.

The conditions,
\begin{equation}
\frac{\partial \sigma_n}{\partial p_k} = 0 \quad (k = 1,...,n),
\end{equation}
are equivalent to,
\begin{equation}
\frac{p_{k+1}-p_k}{p_k - p_{k-1}} = \frac{L(X_k,X_{k+1})}{L(X_{k-1},X_{k})} \quad (k = 1,...,n),\label{ratioGeo}
\end{equation}
with the length of a geodesic segment defined as,\footnote{Note that the square root can be moved under the integral in $\sigma$ because the integrand is a constant.}
\begin{equation}
\begin{split}
L(X_{k},X_{k+1})
&= \sqrt{2\sigma(X_{k},X_{k+1})}\\
&= \int_0^1 ds\sqrt{\smash[b]{g_{\mu\nu} \dot{\gamma}_{n,k}^{\mu}\dot{\gamma}_{n,k}^{\nu}}}.\label{lengthDef}
\end{split}
\end{equation}
One can check that \eqref{ratioGeo} are solved by,
\begin{equation}
p_k = \frac{\sum_{i=1}^k L(X_{i-1},X_{i})}{\sum_{i=0}^n L(X_{i},X_{i+1})} \quad (k = 1,...,n).
\end{equation}
Substituting into \eqref{nRefActapp} yields,
\begin{equation}
\sqrt{2\sigma_n(X,X')} = \sum_{k=0}^{n} L(X_{k},X_{k+1}) = L_n(X,X'),
\label{appsqrtsigma}
\end{equation}
where $L_n(X,X')$ is the total length of the geodesic $ \gamma_n $ with $n$ reflections.

\subsection{Extremization over $ \hat{x}_k $}

Extremizing over $ p_k $ first, \eqref{appsqrtsigma} implies that,
\begin{equation}
\frac{\partial \sigma_n}{\partial \hat{x}^{\mu}_k} = 0 \iff \frac{\partial L_n}{\partial \hat{x}^{\mu}_k} = 0.
\label{Lext}
\end{equation}
Defining $L[\gamma_{n,k}]$ as the length functional of $\gamma_{n,k}$, we have,
\begin{equation}
\frac{\partial L_n}{\partial \hat{x}^{\mu}_k} = \frac{\partial L[\gamma_{n,k}]}{\partial \hat{x}^{\mu}_k}+ \frac{\partial L[\gamma_{n,k-1}]}{\partial \hat{x}^{\mu}_k}.
\label{Lnvar}
\end{equation}
with the endpoints of $\gamma_{n,k}$ being,
\begin{equation}
\gamma_{n,k}(0) = X_k, \quad \gamma_{n,k}(1) = X_{k+1}.
\end{equation}
By the geodesic equation for each geodesic segment $ \gamma_{n,k} $, the variation of $L[\gamma_{n,k}]$ is a pure boundary term,
\begin{equation}
\delta L[\gamma_{n,k}] = u^{a}_{k}(1)\,\delta (X_{k+1})_a - u^{a}_{k}(0)\,\delta (X_{k})_a,
\end{equation}
where,
\begin{equation}
u^{a}_{k}(s)  = \frac{\dot{\gamma}_{n,k}^{a}(s)}{\sqrt{\smash[b]{g_{ab}\dot{\gamma}^{a}_{n,k}\dot{\gamma}^{b}_{n,k}}}}. 
\end{equation}
is the unit tangent vector of $\gamma_{n,k}$.

The reflections points are $ X_k = X_k(\hat{x}_k)\in \mathcal{Q} $, so that for the incoming segment,
\begin{equation}
\frac{\partial L[\gamma_{n,k-1}]}{\partial \hat{x}^{\mu}_{k}} = u^{a}_{k-1}(1)\,M_{a\mu}(\hat{x}_k),
\end{equation}
where,
\begin{equation}
M_{\;\;\mu}^{a}(\hat{x}_k) = \frac{\partial X^{a}_k}{\partial \hat{x}^{\mu}_k}.
\end{equation}
Similarly for the outgoing segment:
\begin{equation}
\frac{\partial L[\gamma_{n,k}]}{\partial \hat{x}^{\mu}_k} = -u^{a}_{k}(0)\,M_{a\mu}(\hat{x}_k).
\end{equation}
Using \eqref{Lnvar}, the extremization condition \eqref{Lext} is thus equivalent to,
\begin{equation}
\left[u^{a}_{k-1}(1) - u^{a}_{k}(0)\right]M_{a\mu}(\hat{x}_k) = 0
\label{refcond}
\end{equation}
where $ u^{a}_{k-1}(1) $ is the unit tangent vector of the incoming geodesic segment and $ u^{a}_{k}(0) $ is that of the outgoing segment, both at the reflection point.

\begin{figure}
\centering
\begin{tikzpicture}[scale=1.55]
\draw[-,red,thick] (0,0) arc (0:90:1.5);
\draw[-,dashed] (-0.8,0.7) to (0.5,2);
\node[red] at (-0.3,0.2) {$\mathcal{Q}$};

\draw[->,red,thin] (-0.45,1.05) to (0.2+0.2,1.7+0.2);
\node[red] at (0.3+0.2,1.55+0.2) {$n^a$}; 
\draw[->,red,thin] (-0.45,1.05) to (-1.1-0.2,1.7+0.2);
\node[red] at (-1.25-0.2,1.7+0.2) {$t^a$};

\draw[-,thick] (-0.45,1.05) to[bend left] (1,0.5);
\node at (1.2,0.3) {$\gamma_{n,k-1}$};

\draw[-,thick] (-0.45,1.05) to[bend left] (0,2);
\node at (0.2,2.2) {$\gamma_{n,k}$};

\node at (-0.45,1.05) {$\bullet$};
\node at (-1,1.05) {$X_k(\hat{x}_k)$};

\draw[->,thin] (-0.45,1.05) to (0.55,1.2);
\node at (1.15,1.2) {$-u_{k-1}^a(1)$};

\draw[->,thin] (-0.45,1.05) to (-0.5,2.05);
\node at (-0.5,2.25) {$u_{k}^a(0)$};

\draw[-,very thin] (-0.2,1.3) arc (45:15:0.43);
\node at (0.21,1.3) {$\psi_{k-1}$};

\draw[-,very thin] (-0.3,1.2) arc (45:85:0.27);
\node at (-0.27,1.425) {$\psi_k$};

\end{tikzpicture}

\caption{A cartoon of reflection at a point $X_k(\hat{x}_k)$. The oriented tangent vectors $u_k^a(0)$ and $-u_{k-1}^a(1)$ are decomposed into $t^a$ and $n^a$ components, upon which extremization yields $\psi_{k-1} = \psi_k$.}
\label{figs:reflectProof}
\end{figure}
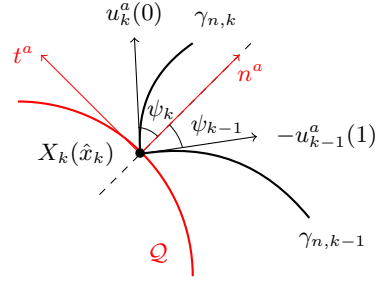

Let $ t^{a} $ and $ n^{a} $ be the unit tangent and normal vectors of $ \mathcal{Q} $ respectively. By definition, the pullback of the normal vector $ n^{a} $ to $ \mathcal{Q} $ vanishes,
\begin{equation}
M_{a\mu}n^{a} = 0.
\end{equation}
Thus by decomposing (Figure \ref{figs:reflectProof}),
\begin{align}
-u_{k-1}^a(1) &= -\sin\psi_{k-1}\,t^a + \cos\psi_{k-1}\,n^a,\\
u_k^a(0) &= \sin\psi_k\,t^a + \cos\psi_k\,n^a,
\end{align}
where $\psi_{k-1}$ is the angle between $-u_{k-1}^a(1)$ and $n^a$ while $\psi_k$ is the angle between $u_k^a(0)$ and $n^a$, \eqref{refcond} becomes,
\begin{equation}
\sin{\psi_{k-1}} - \sin{\psi_{k}} = 0 \implies \psi_{k-1} = \psi_k,
\end{equation}
which is the law of reflection.




\section{No-Go for $n > 1$ Reflections in AdS}\label{appC}

In Section \ref{sec2B}, we discuss an ansatz for the bulk-to-bulk propagator between insertion points $X$ and $X'$ and in the presence of a boundary $\mathcal{Q}$ in which we sum over geodesics obeying the law of reflection. However, this is a formal ansatz---the sum as presented is over any even integer number of reflections by geodesic segments regardless of whether such trajectories exist.

Here, we demonstrate that for a planar KR brane in AdS$_{d+1}$---the geometry used throughout this paper---taking at least one of the insertion points to the conformal boundary disallows more than one reflection. For simplicity, we restrict our attention to a fixed-$\vec{x}$ slice. As our analysis in the main paper also starts on a single $\vec{x}$ slice, the arguments of this appendix are sufficient to prevent geodesics with more than one reflection.

\begin{figure}
\centering
\subfloat[Boundary-to-Brane]
{
\begin{tikzpicture}[scale=0.75]
\draw[->] (0,0) to (0,2.66);

\node[] at (2.5,-0.3) {$y_i$};

\draw[-] (-1.3,1.3) arc (135:45:0.2);
\node at (-1.14,1.65) {$\psi_b$};

\draw[-,thick,black!20!yellow] (2.5,0) arc (0:180:2);

\draw[-,draw=none,fill=white] (-2,0) to (0,0) to (-2,2) to (-2,0);

\draw[-,dashed,black!50] (0,0) to (-2,0);
\draw[->] (0,0) to (3,0);

\node[red] at (0,0) {$\bullet$};
\node[black!20!yellow] at (2.5,0) {$\bullet$};

\draw[-,red,thick] (0,0) to (-2,2);

\draw[<->] (-1.14,1.14) to (-1.14,0);
\node at (-1.34,0.57) {$z_b$};

\node at (-0.2,2.66) {$z$};
\end{tikzpicture}
}\quad
\subfloat[Brane-to-Brane]
{
\begin{tikzpicture}[scale=1]
\draw[->] (0,0) to (0,2);

\draw[-,draw=none,fill=white] (-2,0) to (0,0) to (-2,2) to (-2,0);

\draw[-,dashed,black!50] (0,0) to (-2,0);
\draw[->] (0,0) to (1/2,0);

\node[red] at (0,0) {$\bullet$};

\draw[-,red,thick] (0,0) to (-2,2);

\node at (-0.2,2) {$z$};

\draw[-,thick,purple!50!blue] (-1/2,1/2) arc (18.435:71.5653:1.58114);

\draw[-] (-0.4,0.4) arc (-45:112.5:0.141421);
\node at (-0.25,0.75) {$\psi_{1b}$};

\draw[-] (-1.6,1.6) arc (135:-22.5:0.141421);
\node at (-0.25-1,0.75+1) {$\psi_{2b}$};

\node[white] at (-0.5,-0.225) {$y_i$};

\draw[<->] (-1/2,1/2) to (-1/2,0);
\node at (-0.725,0.25) {$z_{1b}$};

\draw[<->] (-3/2,3/2) to (-3/2,0);
\node at (-1.725,0.75) {$z_{2b}$};
\end{tikzpicture}}
\caption{Boundary-to-brane and brane-to-brane geodesics shown with their characteristic angles and endpoints. These examples are on a particular $\vec{x}$ slice.}
\label{figs:geodesicAngles}
\end{figure}
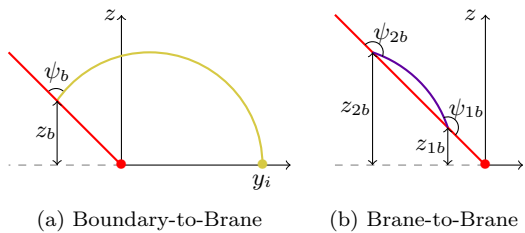

We first introduce brane-to-brane geodesics---semicircles connecting two insertion points along the brane. Such arcs on a particular $\vec{x}$ slice can be determined by fixing the depths of their two endpoints: $z_{1b}$ for the point closer to the boundary and $z_{2b}$ for the point further from the boundary. This, in addition to a boundary-to-brane geodesic from a boundary point $y_i$ to the brane at a depth $z_b$, are shown in Figure \ref{figs:geodesicAngles}.

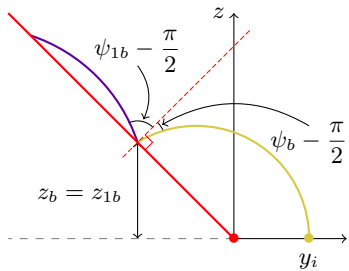
\begin{figure}
\centering
\begin{tikzpicture}[scale=1]
\draw[->] (0,0) to (0,3);

\node[] at (1,-0.3) {$y_i$};

\draw[-] (-1.28,1.28+0.286) arc (90:45:0.286);
\draw[-] (-1.28,1.28+0.286) arc (90:115:0.286);
\node at (-1.28,2.5) {$\psi_{1b}-\dfrac{\pi}{2}$};
\draw[->] (-1.28,2.3) to[bend left] (-1.2,1.28+0.306);

\draw[-] (-1.28+0.2652,1.28+0.2652) arc (45:22.5:0.375);

\node at (1,1.28) {$\psi_b - \dfrac{\pi}{2}$};
\draw[->] (1,1.5) to[bend right] (-1.28+0.3252,1.28+0.2352);

\draw[-,thick,black!20!yellow] (1,0) arc (0:180:1.5);
\draw[-,thick,purple!50!blue] (-1.28,1.28) arc (18.435:71.5653:2.25);

\draw[-,draw=none,fill=white] (-3,0) to (0,0) to (-3,3) to (-3,0);

\draw[<->] (-1.28,1.28) to (-1.28,0);
\node at (-2.05,0.57) {$z_b = z_{1b}$};

\draw[-,red,very thin,dashed,dash pattern=on 2pt off 1pt] (-1.28-0.2,1.28-0.2) to (-1.28+1.5,1.28+1.5);

\draw[-,dashed,black!50] (0,0) to (-3,0);
\draw[->] (0,0) to (1.5,0);

\node[red] at (0,0) {$\bullet$};
\node[black!20!yellow] at (1,0) {$\bullet$};

\draw[-,red,thick] (0,0) to (-3,3);

\draw[-,red] (-1.28+0.1,1.28+0.1) to (-1.28+0.2,1.28) to (-1.28+0.1,1.28-0.1);

\node at (-0.2,3) {$z$};
\end{tikzpicture}
\caption{A piecewise trajectory consisting of a boundary-to-brane geodesic and a brane-to-brane geodesic connected at $z_b = z_{1b}$. Also shown are each piece's angle with respect to the normal. The above picture only makes sense if $\psi_b > \pi/2$, while a piecewise trajectory with $z_b = z_{2b}$ only makes sense if $\psi_b < \pi/2$.}
\label{figs:2reflections}
\end{figure}

If we have $n > 1$ reflections, a brane-to-brane geodesic must be involved. However, these trajectories cannot exist if $\theta \leq \pi/2$---that is for zero and negative tensions---because there is no way to draw a brane-to-brane semicircle centered on the conformal boundary in such cases. Immediately, this rules out $n > 1$ reflections for $\theta \leq \pi/2$.

The argument for positive tensions ($\theta > \pi/2$) however relies on the angles that these trajectories make with the brane, also depicted in Figure \ref{figs:geodesicAngles}. Starting with the boundary-to-brane geodesic, for a particular insertion point $y_i$ we can parametrize the different possible trajectories by the angle $\psi_b$ made with the brane, instead of by the depth $z_b$. This angle monotonically increases with $z_b$, ultimately falling within a calculable range,
\begin{equation}
\theta - \frac{\pi}{2} \leq \psi_b < \frac{3\pi}{2} - \theta.\label{rangeBdry}
\end{equation}
A similar statement holds for the brane-to-brane geodesics---we can parametrize them in terms of the pairs of angles made between the brane and their endpoints. For the $z_{ib}$ endpoint ($i = 1,2$), this angle is denoted as $\psi_{ib}$. However, these angles can be instead written in terms of the dimensionless parameter $\chi_b = z_{2b}/z_{1b} > 1$. Doing so reveals the constraint,
\begin{equation}
\frac{3\pi}{2} - \theta < \psi_{1b} = \psi_{2b} < \pi.\label{rangeBrane}
\end{equation}
We now present an argument by contradiction. Suppose that we do have a geodesic trajectory starting at a boundary point and with $n > 1$ reflections. Then there must exist a boundary-to-brane arc connected to a brane-to-brane arc with the law of reflection satisfied---the former may connect to the latter at either $z_{1b}$ (closer to the boundary) or $z_{2b}$ (further from the boundary). As the arguments for both of these cases are identical, we focus on the former (depicted in Figure \ref{figs:2reflections}).

By equating the incident and reflected angles, the law of reflection gives us the constraint,
\begin{equation}
\psi_b = \psi_{1b}.
\end{equation}
However this contradicts \eqref{rangeBdry} and \eqref{rangeBrane}; the possible values of $\psi_b$ and $\psi_{1b}$ do not overlap.

So, for planar KR branes in AdS$_{d+1}$, there is no way to connect a boundary-to-brane geodesic to a brane-to-brane geodesic while obeying the law of reflection. This rules out any $n > 1$ reflecting geodesic trajectories.
\end{appendix}

\bibliographystyle{apsrev4-2}
\bibliography{multi.bib}

\begin{thebibliography}{81}%
\makeatletter
\providecommand \@ifxundefined [1]{%
 \@ifx{#1\undefined}
}%
\providecommand \@ifnum [1]{%
 \ifnum #1\expandafter \@firstoftwo
 \else \expandafter \@secondoftwo
 \fi
}%
\providecommand \@ifx [1]{%
 \ifx #1\expandafter \@firstoftwo
 \else \expandafter \@secondoftwo
 \fi
}%
\providecommand \natexlab [1]{#1}%
\providecommand \enquote  [1]{``#1''}%
\providecommand \bibnamefont  [1]{#1}%
\providecommand \bibfnamefont [1]{#1}%
\providecommand \citenamefont [1]{#1}%
\providecommand \href@noop [0]{\@secondoftwo}%
\providecommand \href [0]{\begingroup \@sanitize@url \@href}%
\providecommand \@href[1]{\@@startlink{#1}\@@href}%
\providecommand \@@href[1]{\endgroup#1\@@endlink}%
\providecommand \@sanitize@url [0]{\catcode `\\12\catcode `\$12\catcode
  `\&12\catcode `\#12\catcode `\^12\catcode `\_12\catcode `\%12\relax}%
\providecommand \@@startlink[1]{}%
\providecommand \@@endlink[0]{}%
\providecommand \url  [0]{\begingroup\@sanitize@url \@url }%
\providecommand \@url [1]{\endgroup\@href {#1}{\urlprefix }}%
\providecommand \urlprefix  [0]{URL }%
\providecommand \Eprint [0]{\href }%
\providecommand \doibase [0]{https://doi.org/}%
\providecommand \selectlanguage [0]{\@gobble}%
\providecommand \bibinfo  [0]{\@secondoftwo}%
\providecommand \bibfield  [0]{\@secondoftwo}%
\providecommand \translation [1]{[#1]}%
\providecommand \BibitemOpen [0]{}%
\providecommand \bibitemStop [0]{}%
\providecommand \bibitemNoStop [0]{.\EOS\space}%
\providecommand \EOS [0]{\spacefactor3000\relax}%
\providecommand \BibitemShut  [1]{\csname bibitem#1\endcsname}%
\let\auto@bib@innerbib\@empty
\bibitem [{\citenamefont {Randall}\ and\ \citenamefont
  {Sundrum}(1999)}]{Randall:1999vf}%
  \BibitemOpen
  \bibfield  {author} {\bibinfo {author} {\bibfnamefont {L.}~\bibnamefont
  {Randall}}\ and\ \bibinfo {author} {\bibfnamefont {R.}~\bibnamefont
  {Sundrum}},\ }\href {https://doi.org/10.1103/PhysRevLett.83.4690} {\bibfield
  {journal} {\bibinfo  {journal} {Phys. Rev. Lett.}\ }\textbf {\bibinfo
  {volume} {83}},\ \bibinfo {pages} {4690} (\bibinfo {year} {1999})},\ \Eprint
  {https://arxiv.org/abs/hep-th/9906064} {arXiv:hep-th/9906064} \BibitemShut
  {NoStop}%
\bibitem [{\citenamefont {Karch}\ and\ \citenamefont
  {Randall}(2001{\natexlab{a}})}]{Karch:2000ct}%
  \BibitemOpen
  \bibfield  {author} {\bibinfo {author} {\bibfnamefont {A.}~\bibnamefont
  {Karch}}\ and\ \bibinfo {author} {\bibfnamefont {L.}~\bibnamefont
  {Randall}},\ }\href {https://doi.org/10.1088/1126-6708/2001/05/008}
  {\bibfield  {journal} {\bibinfo  {journal} {JHEP}\ }\textbf {\bibinfo
  {volume} {05}},\ \bibinfo {pages} {008}},\ \Eprint
  {https://arxiv.org/abs/hep-th/0011156} {arXiv:hep-th/0011156} \BibitemShut
  {NoStop}%
\bibitem [{\citenamefont {Karch}\ and\ \citenamefont
  {Randall}(2001{\natexlab{b}})}]{Karch:2000gx}%
  \BibitemOpen
  \bibfield  {author} {\bibinfo {author} {\bibfnamefont {A.}~\bibnamefont
  {Karch}}\ and\ \bibinfo {author} {\bibfnamefont {L.}~\bibnamefont
  {Randall}},\ }\href {https://doi.org/10.1088/1126-6708/2001/06/063}
  {\bibfield  {journal} {\bibinfo  {journal} {JHEP}\ }\textbf {\bibinfo
  {volume} {06}},\ \bibinfo {pages} {063}},\ \Eprint
  {https://arxiv.org/abs/hep-th/0105132} {arXiv:hep-th/0105132} \BibitemShut
  {NoStop}%
\bibitem [{\citenamefont {Geng}\ and\ \citenamefont
  {Karch}(2020)}]{Geng:2020qvw}%
  \BibitemOpen
  \bibfield  {author} {\bibinfo {author} {\bibfnamefont {H.}~\bibnamefont
  {Geng}}\ and\ \bibinfo {author} {\bibfnamefont {A.}~\bibnamefont {Karch}},\
  }\href {https://doi.org/10.1007/JHEP09(2020)121} {\bibfield  {journal}
  {\bibinfo  {journal} {JHEP}\ }\textbf {\bibinfo {volume} {09}},\ \bibinfo
  {pages} {121}},\ \Eprint {https://arxiv.org/abs/2006.02438} {arXiv:2006.02438
  [hep-th]} \BibitemShut {NoStop}%
\bibitem [{\citenamefont {Neuenfeld}(2021)}]{Neuenfeld:2021wbl}%
  \BibitemOpen
  \bibfield  {author} {\bibinfo {author} {\bibfnamefont {D.}~\bibnamefont
  {Neuenfeld}},\ }\href@noop {} {\  (\bibinfo {year} {2021})},\ \Eprint
  {https://arxiv.org/abs/2104.02801} {arXiv:2104.02801 [hep-th]} \BibitemShut
  {NoStop}%
\bibitem [{\citenamefont {McAvity}\ and\ \citenamefont
  {Osborn}(1995)}]{McAvity:1995zd}%
  \BibitemOpen
  \bibfield  {author} {\bibinfo {author} {\bibfnamefont {D.~M.}\ \bibnamefont
  {McAvity}}\ and\ \bibinfo {author} {\bibfnamefont {H.}~\bibnamefont
  {Osborn}},\ }\href {https://doi.org/10.1016/0550-3213(95)00476-9} {\bibfield
  {journal} {\bibinfo  {journal} {Nucl. Phys. B}\ }\textbf {\bibinfo {volume}
  {455}},\ \bibinfo {pages} {522} (\bibinfo {year} {1995})},\ \Eprint
  {https://arxiv.org/abs/cond-mat/9505127} {arXiv:cond-mat/9505127}
  \BibitemShut {NoStop}%
\bibitem [{\citenamefont {Cardy}(2004)}]{Cardy:2004hm}%
  \BibitemOpen
  \bibfield  {author} {\bibinfo {author} {\bibfnamefont {J.~L.}\ \bibnamefont
  {Cardy}},\ }\href@noop {} {\  (\bibinfo {year} {2004})},\ \Eprint
  {https://arxiv.org/abs/hep-th/0411189} {arXiv:hep-th/0411189} \BibitemShut
  {NoStop}%
\bibitem [{\citenamefont {Penington}(2020)}]{2020JHEP...09..002P}%
  \BibitemOpen
  \bibfield  {author} {\bibinfo {author} {\bibfnamefont {G.}~\bibnamefont
  {Penington}},\ }\href {https://doi.org/10.1007/JHEP09(2020)002} {\bibfield
  {journal} {\bibinfo  {journal} {JHEP}\ }\textbf {\bibinfo {volume} {09}},\
  \bibinfo {pages} {002}},\ \Eprint {https://arxiv.org/abs/1905.08255}
  {arXiv:1905.08255 [hep-th]} \BibitemShut {NoStop}%
\bibitem [{\citenamefont {Almheiri}\ \emph
  {et~al.}(2020{\natexlab{a}})\citenamefont {Almheiri}, \citenamefont
  {Mahajan}, \citenamefont {Maldacena},\ and\ \citenamefont
  {Zhao}}]{Almheiri:2019hni}%
  \BibitemOpen
  \bibfield  {author} {\bibinfo {author} {\bibfnamefont {A.}~\bibnamefont
  {Almheiri}}, \bibinfo {author} {\bibfnamefont {R.}~\bibnamefont {Mahajan}},
  \bibinfo {author} {\bibfnamefont {J.}~\bibnamefont {Maldacena}},\ and\
  \bibinfo {author} {\bibfnamefont {Y.}~\bibnamefont {Zhao}},\ }\href
  {https://doi.org/10.1007/JHEP03(2020)149} {\bibfield  {journal} {\bibinfo
  {journal} {JHEP}\ }\textbf {\bibinfo {volume} {03}},\ \bibinfo {pages}
  {149}},\ \Eprint {https://arxiv.org/abs/1908.10996} {arXiv:1908.10996
  [hep-th]} \BibitemShut {NoStop}%
\bibitem [{\citenamefont {Almheiri}\ \emph
  {et~al.}(2020{\natexlab{b}})\citenamefont {Almheiri}, \citenamefont
  {Mahajan},\ and\ \citenamefont {Santos}}]{Almheiri:2019psy}%
  \BibitemOpen
  \bibfield  {author} {\bibinfo {author} {\bibfnamefont {A.}~\bibnamefont
  {Almheiri}}, \bibinfo {author} {\bibfnamefont {R.}~\bibnamefont {Mahajan}},\
  and\ \bibinfo {author} {\bibfnamefont {J.~E.}\ \bibnamefont {Santos}},\
  }\href {https://doi.org/10.21468/SciPostPhys.9.1.001} {\bibfield  {journal}
  {\bibinfo  {journal} {SciPost Phys.}\ }\textbf {\bibinfo {volume} {9}},\
  \bibinfo {pages} {001} (\bibinfo {year} {2020}{\natexlab{b}})},\ \Eprint
  {https://arxiv.org/abs/1911.09666} {arXiv:1911.09666 [hep-th]} \BibitemShut
  {NoStop}%
\bibitem [{\citenamefont {Chen}\ \emph {et~al.}(2020)\citenamefont {Chen},
  \citenamefont {Myers}, \citenamefont {Neuenfeld}, \citenamefont {Reyes},\
  and\ \citenamefont {Sandor}}]{Chen:2020uac}%
  \BibitemOpen
  \bibfield  {author} {\bibinfo {author} {\bibfnamefont {H.~Z.}\ \bibnamefont
  {Chen}}, \bibinfo {author} {\bibfnamefont {R.~C.}\ \bibnamefont {Myers}},
  \bibinfo {author} {\bibfnamefont {D.}~\bibnamefont {Neuenfeld}}, \bibinfo
  {author} {\bibfnamefont {I.~A.}\ \bibnamefont {Reyes}},\ and\ \bibinfo
  {author} {\bibfnamefont {J.}~\bibnamefont {Sandor}},\ }\href
  {https://doi.org/10.1007/JHEP10(2020)166} {\bibfield  {journal} {\bibinfo
  {journal} {JHEP}\ }\textbf {\bibinfo {volume} {10}},\ \bibinfo {pages}
  {166}},\ \Eprint {https://arxiv.org/abs/2006.04851} {arXiv:2006.04851
  [hep-th]} \BibitemShut {NoStop}%
\bibitem [{\citenamefont {Almheiri}\ \emph
  {et~al.}(2020{\natexlab{c}})\citenamefont {Almheiri}, \citenamefont
  {Hartman}, \citenamefont {Maldacena}, \citenamefont {Shaghoulian},\ and\
  \citenamefont {Tajdini}}]{Almheiri:2020cfm}%
  \BibitemOpen
  \bibfield  {author} {\bibinfo {author} {\bibfnamefont {A.}~\bibnamefont
  {Almheiri}}, \bibinfo {author} {\bibfnamefont {T.}~\bibnamefont {Hartman}},
  \bibinfo {author} {\bibfnamefont {J.}~\bibnamefont {Maldacena}}, \bibinfo
  {author} {\bibfnamefont {E.}~\bibnamefont {Shaghoulian}},\ and\ \bibinfo
  {author} {\bibfnamefont {A.}~\bibnamefont {Tajdini}},\ }\href@noop {} {\
  (\bibinfo {year} {2020}{\natexlab{c}})},\ \Eprint
  {https://arxiv.org/abs/2006.06872} {arXiv:2006.06872 [hep-th]} \BibitemShut
  {NoStop}%
\bibitem [{\citenamefont {Balasubramanian}\ and\ \citenamefont
  {Ross}(2000)}]{Balasubramanian:1999zv}%
  \BibitemOpen
  \bibfield  {author} {\bibinfo {author} {\bibfnamefont {V.}~\bibnamefont
  {Balasubramanian}}\ and\ \bibinfo {author} {\bibfnamefont {S.~F.}\
  \bibnamefont {Ross}},\ }\href {https://doi.org/10.1103/PhysRevD.61.044007}
  {\bibfield  {journal} {\bibinfo  {journal} {Phys. Rev. D}\ }\textbf {\bibinfo
  {volume} {61}},\ \bibinfo {pages} {044007} (\bibinfo {year} {2000})},\
  \Eprint {https://arxiv.org/abs/hep-th/9906226} {arXiv:hep-th/9906226}
  \BibitemShut {NoStop}%
\bibitem [{\citenamefont {Balasubramanian}\ \emph {et~al.}(2013)\citenamefont
  {Balasubramanian}, \citenamefont {Bernamonti}, \citenamefont {Craps},
  \citenamefont {Ker\"anen}, \citenamefont {Keski-Vakkuri}, \citenamefont
  {M\"uller}, \citenamefont {Thorlacius},\ and\ \citenamefont
  {Vanhoof}}]{Balasubramanian:2012tu}%
  \BibitemOpen
  \bibfield  {author} {\bibinfo {author} {\bibfnamefont {V.}~\bibnamefont
  {Balasubramanian}}, \bibinfo {author} {\bibfnamefont {A.}~\bibnamefont
  {Bernamonti}}, \bibinfo {author} {\bibfnamefont {B.}~\bibnamefont {Craps}},
  \bibinfo {author} {\bibfnamefont {V.}~\bibnamefont {Ker\"anen}}, \bibinfo
  {author} {\bibfnamefont {E.}~\bibnamefont {Keski-Vakkuri}}, \bibinfo {author}
  {\bibfnamefont {B.}~\bibnamefont {M\"uller}}, \bibinfo {author}
  {\bibfnamefont {L.}~\bibnamefont {Thorlacius}},\ and\ \bibinfo {author}
  {\bibfnamefont {J.}~\bibnamefont {Vanhoof}},\ }\href
  {https://doi.org/10.1007/JHEP04(2013)069} {\bibfield  {journal} {\bibinfo
  {journal} {JHEP}\ }\textbf {\bibinfo {volume} {04}},\ \bibinfo {pages}
  {069}},\ \Eprint {https://arxiv.org/abs/1212.6066} {arXiv:1212.6066 [hep-th]}
  \BibitemShut {NoStop}%
\bibitem [{\citenamefont {Takayanagi}\ and\ \citenamefont
  {Uetoko}(2021)}]{Takayanagi:2020njm}%
  \BibitemOpen
  \bibfield  {author} {\bibinfo {author} {\bibfnamefont {T.}~\bibnamefont
  {Takayanagi}}\ and\ \bibinfo {author} {\bibfnamefont {T.}~\bibnamefont
  {Uetoko}},\ }\href {https://doi.org/10.1007/JHEP04(2021)193} {\bibfield
  {journal} {\bibinfo  {journal} {JHEP}\ }\textbf {\bibinfo {volume} {04}},\
  \bibinfo {pages} {193}},\ \Eprint {https://arxiv.org/abs/2011.02513}
  {arXiv:2011.02513 [hep-th]} \BibitemShut {NoStop}%
\bibitem [{\citenamefont {Takayanagi}(2011)}]{Takayanagi:2011zk}%
  \BibitemOpen
  \bibfield  {author} {\bibinfo {author} {\bibfnamefont {T.}~\bibnamefont
  {Takayanagi}},\ }\href {https://doi.org/10.1103/PhysRevLett.107.101602}
  {\bibfield  {journal} {\bibinfo  {journal} {Phys. Rev. Lett.}\ }\textbf
  {\bibinfo {volume} {107}},\ \bibinfo {pages} {101602} (\bibinfo {year}
  {2011})},\ \Eprint {https://arxiv.org/abs/1105.5165} {arXiv:1105.5165
  [hep-th]} \BibitemShut {NoStop}%
\bibitem [{\citenamefont {Fujita}\ \emph {et~al.}(2011)\citenamefont {Fujita},
  \citenamefont {Takayanagi},\ and\ \citenamefont {Tonni}}]{Fujita:2011fp}%
  \BibitemOpen
  \bibfield  {author} {\bibinfo {author} {\bibfnamefont {M.}~\bibnamefont
  {Fujita}}, \bibinfo {author} {\bibfnamefont {T.}~\bibnamefont {Takayanagi}},\
  and\ \bibinfo {author} {\bibfnamefont {E.}~\bibnamefont {Tonni}},\ }\href
  {https://doi.org/10.1007/JHEP11(2011)043} {\bibfield  {journal} {\bibinfo
  {journal} {JHEP}\ }\textbf {\bibinfo {volume} {11}},\ \bibinfo {pages}
  {043}},\ \Eprint {https://arxiv.org/abs/1108.5152} {arXiv:1108.5152 [hep-th]}
  \BibitemShut {NoStop}%
\bibitem [{\citenamefont {Ryu}\ and\ \citenamefont
  {Takayanagi}(2006)}]{Ryu:2006bv}%
  \BibitemOpen
  \bibfield  {author} {\bibinfo {author} {\bibfnamefont {S.}~\bibnamefont
  {Ryu}}\ and\ \bibinfo {author} {\bibfnamefont {T.}~\bibnamefont
  {Takayanagi}},\ }\href {https://doi.org/10.1103/PhysRevLett.96.181602}
  {\bibfield  {journal} {\bibinfo  {journal} {Phys. Rev. Lett.}\ }\textbf
  {\bibinfo {volume} {96}},\ \bibinfo {pages} {181602} (\bibinfo {year}
  {2006})},\ \Eprint {https://arxiv.org/abs/hep-th/0603001}
  {arXiv:hep-th/0603001} \BibitemShut {NoStop}%
\bibitem [{\citenamefont {Hubeny}\ \emph {et~al.}(2007)\citenamefont {Hubeny},
  \citenamefont {Rangamani},\ and\ \citenamefont {Takayanagi}}]{Hubeny:2007xt}%
  \BibitemOpen
  \bibfield  {author} {\bibinfo {author} {\bibfnamefont {V.~E.}\ \bibnamefont
  {Hubeny}}, \bibinfo {author} {\bibfnamefont {M.}~\bibnamefont {Rangamani}},\
  and\ \bibinfo {author} {\bibfnamefont {T.}~\bibnamefont {Takayanagi}},\
  }\href {https://doi.org/10.1088/1126-6708/2007/07/062} {\bibfield  {journal}
  {\bibinfo  {journal} {JHEP}\ }\textbf {\bibinfo {volume} {07}},\ \bibinfo
  {pages} {062}},\ \Eprint {https://arxiv.org/abs/0705.0016} {arXiv:0705.0016
  [hep-th]} \BibitemShut {NoStop}%
\bibitem [{\citenamefont {McAvity}\ and\ \citenamefont
  {Osborn}(1991{\natexlab{a}})}]{McAvity_1991}%
  \BibitemOpen
  \bibfield  {author} {\bibinfo {author} {\bibfnamefont {D.~M.}\ \bibnamefont
  {McAvity}}\ and\ \bibinfo {author} {\bibfnamefont {H.}~\bibnamefont
  {Osborn}},\ }\href {https://doi.org/10.1088/0264-9381/8/4/008} {\bibfield
  {journal} {\bibinfo  {journal} {Classical and Quantum Gravity}\ }\textbf
  {\bibinfo {volume} {8}},\ \bibinfo {pages} {603} (\bibinfo {year}
  {1991}{\natexlab{a}})}\BibitemShut {NoStop}%
\bibitem [{\citenamefont {McAvity}\ and\ \citenamefont
  {Osborn}(1991{\natexlab{b}})}]{mcavity_asymptotic_1991}%
  \BibitemOpen
  \bibfield  {author} {\bibinfo {author} {\bibfnamefont {D.~M.}\ \bibnamefont
  {McAvity}}\ and\ \bibinfo {author} {\bibfnamefont {H.}~\bibnamefont
  {Osborn}},\ }\href@noop {} {\bibfield  {journal} {\bibinfo  {journal}
  {Classical and Quantum Gravity}\ }\textbf {\bibinfo {volume} {8}},\ \bibinfo
  {pages} {1445} (\bibinfo {year} {1991}{\natexlab{b}})}\BibitemShut {NoStop}%
\bibitem [{\citenamefont {McAvity}\ and\ \citenamefont
  {Osborn}(1993)}]{mcavity_quantum_1993}%
  \BibitemOpen
  \bibfield  {author} {\bibinfo {author} {\bibfnamefont {D.~M.}\ \bibnamefont
  {McAvity}}\ and\ \bibinfo {author} {\bibfnamefont {H.}~\bibnamefont
  {Osborn}},\ }\href@noop {} {\bibfield  {journal} {\bibinfo  {journal}
  {Nuclear Physics B}\ }\textbf {\bibinfo {volume} {394}},\ \bibinfo {pages}
  {728} (\bibinfo {year} {1993})},\ \Eprint
  {https://arxiv.org/abs/cond-mat/9206009} {arXiv:cond-mat/9206009}
  \BibitemShut {NoStop}%
\bibitem [{\citenamefont {Alishahiha}\ and\ \citenamefont
  {Fareghbal}(2011)}]{Alishahiha:2011rg}%
  \BibitemOpen
  \bibfield  {author} {\bibinfo {author} {\bibfnamefont {M.}~\bibnamefont
  {Alishahiha}}\ and\ \bibinfo {author} {\bibfnamefont {R.}~\bibnamefont
  {Fareghbal}},\ }\href {https://doi.org/10.1103/PhysRevD.84.106002} {\bibfield
   {journal} {\bibinfo  {journal} {Phys. Rev. D}\ }\textbf {\bibinfo {volume}
  {84}},\ \bibinfo {pages} {106002} (\bibinfo {year} {2011})},\ \Eprint
  {https://arxiv.org/abs/1108.5607} {arXiv:1108.5607 [hep-th]} \BibitemShut
  {NoStop}%
\bibitem [{\citenamefont {Almheiri}\ \emph {et~al.}(2018)\citenamefont
  {Almheiri}, \citenamefont {Mousatov},\ and\ \citenamefont
  {Shyani}}]{Almheiri:2018ijj}%
  \BibitemOpen
  \bibfield  {author} {\bibinfo {author} {\bibfnamefont {A.}~\bibnamefont
  {Almheiri}}, \bibinfo {author} {\bibfnamefont {A.}~\bibnamefont {Mousatov}},\
  and\ \bibinfo {author} {\bibfnamefont {M.}~\bibnamefont {Shyani}},\
  }\href@noop {} {\  (\bibinfo {year} {2018})},\ \Eprint
  {https://arxiv.org/abs/1803.04434} {arXiv:1803.04434 [hep-th]} \BibitemShut
  {NoStop}%
\bibitem [{\citenamefont {Witten}(1998)}]{Witten:1998zw}%
  \BibitemOpen
  \bibfield  {author} {\bibinfo {author} {\bibfnamefont {E.}~\bibnamefont
  {Witten}},\ }\href {https://doi.org/10.4310/ATMP.1998.v2.n3.a3} {\bibfield
  {journal} {\bibinfo  {journal} {Adv. Theor. Math. Phys.}\ }\textbf {\bibinfo
  {volume} {2}},\ \bibinfo {pages} {505} (\bibinfo {year} {1998})},\ \Eprint
  {https://arxiv.org/abs/hep-th/9803131} {arXiv:hep-th/9803131} \BibitemShut
  {NoStop}%
\bibitem [{\citenamefont {Brandhuber}\ \emph {et~al.}(1998)\citenamefont
  {Brandhuber}, \citenamefont {Itzhaki}, \citenamefont {Sonnenschein},\ and\
  \citenamefont {Yankielowicz}}]{Brandhuber:1998bs}%
  \BibitemOpen
  \bibfield  {author} {\bibinfo {author} {\bibfnamefont {A.}~\bibnamefont
  {Brandhuber}}, \bibinfo {author} {\bibfnamefont {N.}~\bibnamefont {Itzhaki}},
  \bibinfo {author} {\bibfnamefont {J.}~\bibnamefont {Sonnenschein}},\ and\
  \bibinfo {author} {\bibfnamefont {S.}~\bibnamefont {Yankielowicz}},\ }\href
  {https://doi.org/10.1016/S0370-2693(98)00730-8} {\bibfield  {journal}
  {\bibinfo  {journal} {Phys. Lett. B}\ }\textbf {\bibinfo {volume} {434}},\
  \bibinfo {pages} {36} (\bibinfo {year} {1998})},\ \Eprint
  {https://arxiv.org/abs/hep-th/9803137} {arXiv:hep-th/9803137} \BibitemShut
  {NoStop}%
\bibitem [{\citenamefont {Albash}\ \emph {et~al.}(2008)\citenamefont {Albash},
  \citenamefont {Filev}, \citenamefont {Johnson},\ and\ \citenamefont
  {Kundu}}]{Albash:2006bs}%
  \BibitemOpen
  \bibfield  {author} {\bibinfo {author} {\bibfnamefont {T.}~\bibnamefont
  {Albash}}, \bibinfo {author} {\bibfnamefont {V.~G.}\ \bibnamefont {Filev}},
  \bibinfo {author} {\bibfnamefont {C.~V.}\ \bibnamefont {Johnson}},\ and\
  \bibinfo {author} {\bibfnamefont {A.}~\bibnamefont {Kundu}},\ }\href
  {https://doi.org/10.1088/1126-6708/2008/12/033} {\bibfield  {journal}
  {\bibinfo  {journal} {JHEP}\ }\textbf {\bibinfo {volume} {12}},\ \bibinfo
  {pages} {033}},\ \Eprint {https://arxiv.org/abs/hep-th/0605175}
  {arXiv:hep-th/0605175} \BibitemShut {NoStop}%
\bibitem [{\citenamefont {Johnson}\ and\ \citenamefont
  {Kundu}(2008)}]{Johnson:2008vna}%
  \BibitemOpen
  \bibfield  {author} {\bibinfo {author} {\bibfnamefont {C.~V.}\ \bibnamefont
  {Johnson}}\ and\ \bibinfo {author} {\bibfnamefont {A.}~\bibnamefont
  {Kundu}},\ }\href {https://doi.org/10.1088/1126-6708/2008/12/053} {\bibfield
  {journal} {\bibinfo  {journal} {JHEP}\ }\textbf {\bibinfo {volume} {12}},\
  \bibinfo {pages} {053}},\ \Eprint {https://arxiv.org/abs/0803.0038}
  {arXiv:0803.0038 [hep-th]} \BibitemShut {NoStop}%
\bibitem [{\citenamefont {Karch}\ \emph {et~al.}(2017)\citenamefont {Karch},
  \citenamefont {Sully}, \citenamefont {Uhlemann},\ and\ \citenamefont
  {Walker}}]{Karch:2017fuh}%
  \BibitemOpen
  \bibfield  {author} {\bibinfo {author} {\bibfnamefont {A.}~\bibnamefont
  {Karch}}, \bibinfo {author} {\bibfnamefont {J.}~\bibnamefont {Sully}},
  \bibinfo {author} {\bibfnamefont {C.~F.}\ \bibnamefont {Uhlemann}},\ and\
  \bibinfo {author} {\bibfnamefont {D.~G.~E.}\ \bibnamefont {Walker}},\ }\href
  {https://doi.org/10.1007/JHEP08(2017)039} {\bibfield  {journal} {\bibinfo
  {journal} {JHEP}\ }\textbf {\bibinfo {volume} {08}},\ \bibinfo {pages}
  {039}},\ \Eprint {https://arxiv.org/abs/1703.02990} {arXiv:1703.02990
  [hep-th]} \BibitemShut {NoStop}%
\bibitem [{\citenamefont {Affleck}\ and\ \citenamefont
  {Ludwig}(1991)}]{Affleck:1991tk}%
  \BibitemOpen
  \bibfield  {author} {\bibinfo {author} {\bibfnamefont {I.}~\bibnamefont
  {Affleck}}\ and\ \bibinfo {author} {\bibfnamefont {A.~W.~W.}\ \bibnamefont
  {Ludwig}},\ }\href {https://doi.org/10.1103/PhysRevLett.67.161} {\bibfield
  {journal} {\bibinfo  {journal} {Phys. Rev. Lett.}\ }\textbf {\bibinfo
  {volume} {67}},\ \bibinfo {pages} {161} (\bibinfo {year} {1991})}\BibitemShut
  {NoStop}%
\bibitem [{\citenamefont {Azeyanagi}\ \emph {et~al.}(2008)\citenamefont
  {Azeyanagi}, \citenamefont {Karch}, \citenamefont {Takayanagi},\ and\
  \citenamefont {Thompson}}]{Azeyanagi:2007qj}%
  \BibitemOpen
  \bibfield  {author} {\bibinfo {author} {\bibfnamefont {T.}~\bibnamefont
  {Azeyanagi}}, \bibinfo {author} {\bibfnamefont {A.}~\bibnamefont {Karch}},
  \bibinfo {author} {\bibfnamefont {T.}~\bibnamefont {Takayanagi}},\ and\
  \bibinfo {author} {\bibfnamefont {E.~G.}\ \bibnamefont {Thompson}},\ }\href
  {https://doi.org/10.1088/1126-6708/2008/03/054} {\bibfield  {journal}
  {\bibinfo  {journal} {JHEP}\ }\textbf {\bibinfo {volume} {03}},\ \bibinfo
  {pages} {054}},\ \Eprint {https://arxiv.org/abs/0712.1850} {arXiv:0712.1850
  [hep-th]} \BibitemShut {NoStop}%
\bibitem [{\citenamefont {Karch}\ \emph {et~al.}(2021)\citenamefont {Karch},
  \citenamefont {Luo},\ and\ \citenamefont {Sun}}]{Karch:2020flx}%
  \BibitemOpen
  \bibfield  {author} {\bibinfo {author} {\bibfnamefont {A.}~\bibnamefont
  {Karch}}, \bibinfo {author} {\bibfnamefont {Z.-X.}\ \bibnamefont {Luo}},\
  and\ \bibinfo {author} {\bibfnamefont {H.-Y.}\ \bibnamefont {Sun}},\ }\href
  {https://doi.org/10.1007/JHEP04(2021)018} {\bibfield  {journal} {\bibinfo
  {journal} {JHEP}\ }\textbf {\bibinfo {volume} {04}},\ \bibinfo {pages}
  {018}},\ \Eprint {https://arxiv.org/abs/2012.02067} {arXiv:2012.02067
  [hep-th]} \BibitemShut {NoStop}%
\bibitem [{\citenamefont {Freedman}\ \emph
  {et~al.}(1999{\natexlab{a}})\citenamefont {Freedman}, \citenamefont {Gubser},
  \citenamefont {Pilch},\ and\ \citenamefont {Warner}}]{Freedman:1999gp}%
  \BibitemOpen
  \bibfield  {author} {\bibinfo {author} {\bibfnamefont {D.~Z.}\ \bibnamefont
  {Freedman}}, \bibinfo {author} {\bibfnamefont {S.~S.}\ \bibnamefont
  {Gubser}}, \bibinfo {author} {\bibfnamefont {K.}~\bibnamefont {Pilch}},\ and\
  \bibinfo {author} {\bibfnamefont {N.~P.}\ \bibnamefont {Warner}},\ }\href
  {https://doi.org/10.4310/ATMP.1999.v3.n2.a7} {\bibfield  {journal} {\bibinfo
  {journal} {Adv. Theor. Math. Phys.}\ }\textbf {\bibinfo {volume} {3}},\
  \bibinfo {pages} {363} (\bibinfo {year} {1999}{\natexlab{a}})},\ \Eprint
  {https://arxiv.org/abs/hep-th/9904017} {arXiv:hep-th/9904017} \BibitemShut
  {NoStop}%
\bibitem [{\citenamefont {Aharony}\ \emph {et~al.}(2000)\citenamefont
  {Aharony}, \citenamefont {Gubser}, \citenamefont {Maldacena}, \citenamefont
  {Ooguri},\ and\ \citenamefont {Oz}}]{Aharony:1999ti}%
  \BibitemOpen
  \bibfield  {author} {\bibinfo {author} {\bibfnamefont {O.}~\bibnamefont
  {Aharony}}, \bibinfo {author} {\bibfnamefont {S.~S.}\ \bibnamefont {Gubser}},
  \bibinfo {author} {\bibfnamefont {J.~M.}\ \bibnamefont {Maldacena}}, \bibinfo
  {author} {\bibfnamefont {H.}~\bibnamefont {Ooguri}},\ and\ \bibinfo {author}
  {\bibfnamefont {Y.}~\bibnamefont {Oz}},\ }\href
  {https://doi.org/10.1016/S0370-1573(99)00083-6} {\bibfield  {journal}
  {\bibinfo  {journal} {Phys. Rept.}\ }\textbf {\bibinfo {volume} {323}},\
  \bibinfo {pages} {183} (\bibinfo {year} {2000})},\ \Eprint
  {https://arxiv.org/abs/hep-th/9905111} {arXiv:hep-th/9905111} \BibitemShut
  {NoStop}%
\bibitem [{\citenamefont {Klebanov}\ and\ \citenamefont
  {Witten}(1999)}]{Klebanov:1999tb}%
  \BibitemOpen
  \bibfield  {author} {\bibinfo {author} {\bibfnamefont {I.~R.}\ \bibnamefont
  {Klebanov}}\ and\ \bibinfo {author} {\bibfnamefont {E.}~\bibnamefont
  {Witten}},\ }\href {https://doi.org/10.1016/S0550-3213(99)00387-9} {\bibfield
   {journal} {\bibinfo  {journal} {Nucl. Phys. B}\ }\textbf {\bibinfo {volume}
  {556}},\ \bibinfo {pages} {89} (\bibinfo {year} {1999})},\ \Eprint
  {https://arxiv.org/abs/hep-th/9905104} {arXiv:hep-th/9905104} \BibitemShut
  {NoStop}%
\bibitem [{\citenamefont {Cardy}(1984)}]{Cardy:1984bb}%
  \BibitemOpen
  \bibfield  {author} {\bibinfo {author} {\bibfnamefont {J.~L.}\ \bibnamefont
  {Cardy}},\ }\href {https://doi.org/10.1016/0550-3213(84)90241-4} {\bibfield
  {journal} {\bibinfo  {journal} {Nucl. Phys. B}\ }\textbf {\bibinfo {volume}
  {240}},\ \bibinfo {pages} {514} (\bibinfo {year} {1984})}\BibitemShut
  {NoStop}%
\bibitem [{\citenamefont {DeWolfe}\ \emph {et~al.}(2002)\citenamefont
  {DeWolfe}, \citenamefont {Freedman},\ and\ \citenamefont
  {Ooguri}}]{DeWolfe:2001pq}%
  \BibitemOpen
  \bibfield  {author} {\bibinfo {author} {\bibfnamefont {O.}~\bibnamefont
  {DeWolfe}}, \bibinfo {author} {\bibfnamefont {D.~Z.}\ \bibnamefont
  {Freedman}},\ and\ \bibinfo {author} {\bibfnamefont {H.}~\bibnamefont
  {Ooguri}},\ }\href {https://doi.org/10.1103/PhysRevD.66.025009} {\bibfield
  {journal} {\bibinfo  {journal} {Phys. Rev. D}\ }\textbf {\bibinfo {volume}
  {66}},\ \bibinfo {pages} {025009} (\bibinfo {year} {2002})},\ \Eprint
  {https://arxiv.org/abs/hep-th/0111135} {arXiv:hep-th/0111135} \BibitemShut
  {NoStop}%
\bibitem [{\citenamefont {Liendo}\ \emph
  {et~al.}(2013{\natexlab{a}})\citenamefont {Liendo}, \citenamefont
  {Rastelli},\ and\ \citenamefont {van Rees}}]{Liendo:2012hy}%
  \BibitemOpen
  \bibfield  {author} {\bibinfo {author} {\bibfnamefont {P.}~\bibnamefont
  {Liendo}}, \bibinfo {author} {\bibfnamefont {L.}~\bibnamefont {Rastelli}},\
  and\ \bibinfo {author} {\bibfnamefont {B.~C.}\ \bibnamefont {van Rees}},\
  }\href {https://doi.org/10.1007/JHEP07(2013)113} {\bibfield  {journal}
  {\bibinfo  {journal} {JHEP}\ }\textbf {\bibinfo {volume} {07}},\ \bibinfo
  {pages} {113}},\ \Eprint {https://arxiv.org/abs/1210.4258} {arXiv:1210.4258
  [hep-th]} \BibitemShut {NoStop}%
\bibitem [{\citenamefont {Maz\'a\v{c}}\ \emph {et~al.}(2019)\citenamefont
  {Maz\'a\v{c}}, \citenamefont {Rastelli},\ and\ \citenamefont
  {Zhou}}]{Mazac:2018biw}%
  \BibitemOpen
  \bibfield  {author} {\bibinfo {author} {\bibfnamefont {D.}~\bibnamefont
  {Maz\'a\v{c}}}, \bibinfo {author} {\bibfnamefont {L.}~\bibnamefont
  {Rastelli}},\ and\ \bibinfo {author} {\bibfnamefont {X.}~\bibnamefont
  {Zhou}},\ }\href {https://doi.org/10.1007/JHEP12(2019)004} {\bibfield
  {journal} {\bibinfo  {journal} {JHEP}\ }\textbf {\bibinfo {volume} {12}},\
  \bibinfo {pages} {004}},\ \Eprint {https://arxiv.org/abs/1812.09314}
  {arXiv:1812.09314 [hep-th]} \BibitemShut {NoStop}%
\bibitem [{\citenamefont {McAvity}\ and\ \citenamefont
  {Osborn}(1992)}]{mcavity_heat_1992}%
  \BibitemOpen
  \bibfield  {author} {\bibinfo {author} {\bibfnamefont {D.~M.}\ \bibnamefont
  {McAvity}}\ and\ \bibinfo {author} {\bibfnamefont {H.}~\bibnamefont
  {Osborn}},\ }\href@noop {} {\bibfield  {journal} {\bibinfo  {journal}
  {Journal of Physics A: Mathematical and General}\ }\textbf {\bibinfo {volume}
  {25}},\ \bibinfo {pages} {3287} (\bibinfo {year} {1992})}\BibitemShut
  {NoStop}%
\bibitem [{\citenamefont {McAvity}(1993)}]{mcavity_surface_1993}%
  \BibitemOpen
  \bibfield  {author} {\bibinfo {author} {\bibfnamefont {D.~M.}\ \bibnamefont
  {McAvity}},\ }\href@noop {} {\bibfield  {journal} {\bibinfo  {journal}
  {Journal of Physics A: Mathematical and General}\ }\textbf {\bibinfo {volume}
  {26}},\ \bibinfo {pages} {823} (\bibinfo {year} {1993})}\BibitemShut
  {NoStop}%
\bibitem [{\citenamefont {Giddings}(1999)}]{giddings_boundary_1999}%
  \BibitemOpen
  \bibfield  {author} {\bibinfo {author} {\bibfnamefont {S.~B.}\ \bibnamefont
  {Giddings}},\ }\href@noop {} {\bibfield  {journal} {\bibinfo  {journal}
  {Physical Review Letters}\ }\textbf {\bibinfo {volume} {83}},\ \bibinfo
  {pages} {2707} (\bibinfo {year} {1999})},\ \Eprint
  {https://arxiv.org/abs/hep-th/9903048} {arXiv:hep-th/9903048} \BibitemShut
  {NoStop}%
\bibitem [{\citenamefont {Harlow}\ and\ \citenamefont
  {Stanford}(2011)}]{harlow_operator_2011}%
  \BibitemOpen
  \bibfield  {author} {\bibinfo {author} {\bibfnamefont {D.}~\bibnamefont
  {Harlow}}\ and\ \bibinfo {author} {\bibfnamefont {D.}~\bibnamefont
  {Stanford}},\ }\href@noop {} {\bibfield  {journal} {\bibinfo  {journal}
  {arXiv:1104.2621 [hep-th]}\ } (\bibinfo {year} {2011})},\ \Eprint
  {https://arxiv.org/abs/1104.2621} {arXiv:1104.2621 [hep-th]} \BibitemShut
  {NoStop}%
\bibitem [{\citenamefont {Poisson}\ \emph {et~al.}(2011)\citenamefont
  {Poisson}, \citenamefont {Pound},\ and\ \citenamefont
  {Vega}}]{poisson_motion_2011}%
  \BibitemOpen
  \bibfield  {author} {\bibinfo {author} {\bibfnamefont {E.}~\bibnamefont
  {Poisson}}, \bibinfo {author} {\bibfnamefont {A.}~\bibnamefont {Pound}},\
  and\ \bibinfo {author} {\bibfnamefont {I.}~\bibnamefont {Vega}},\ }\href@noop
  {} {\bibfield  {journal} {\bibinfo  {journal} {Living Reviews in Relativity}\
  }\textbf {\bibinfo {volume} {14}},\ \bibinfo {pages} {7} (\bibinfo {year}
  {2011})}\BibitemShut {NoStop}%
\bibitem [{\citenamefont {Visser}(1993)}]{visser_van_1993}%
  \BibitemOpen
  \bibfield  {author} {\bibinfo {author} {\bibfnamefont {M.}~\bibnamefont
  {Visser}},\ }\href@noop {} {\bibfield  {journal} {\bibinfo  {journal}
  {Physical Review D}\ }\textbf {\bibinfo {volume} {47}},\ \bibinfo {pages}
  {2395} (\bibinfo {year} {1993})},\ \Eprint
  {https://arxiv.org/abs/hep-th/9303020} {arXiv:hep-th/9303020} \BibitemShut
  {NoStop}%
\bibitem [{\citenamefont {Ludewig}(2017)}]{Ludewig_2017}%
  \BibitemOpen
  \bibfield  {author} {\bibinfo {author} {\bibfnamefont {M.}~\bibnamefont
  {Ludewig}},\ }\href {https://doi.org/10.1007/s00220-017-2915-9} {\bibfield
  {journal} {\bibinfo  {journal} {Communications in Mathematical Physics}\
  }\textbf {\bibinfo {volume} {354}},\ \bibinfo {pages} {621} (\bibinfo {year}
  {2017})}\BibitemShut {NoStop}%
\bibitem [{\citenamefont {Zannias}(1983)}]{zannias_path-integral_1983}%
  \BibitemOpen
  \bibfield  {author} {\bibinfo {author} {\bibfnamefont {T.}~\bibnamefont
  {Zannias}},\ }\href@noop {} {\bibfield  {journal} {\bibinfo  {journal}
  {Physical Review D}\ }\textbf {\bibinfo {volume} {27}},\ \bibinfo {pages}
  {1386} (\bibinfo {year} {1983})}\BibitemShut {NoStop}%
\bibitem [{\citenamefont {Asplund}\ \emph {et~al.}(2015)\citenamefont
  {Asplund}, \citenamefont {Bernamonti}, \citenamefont {Galli},\ and\
  \citenamefont {Hartman}}]{asplund_holographic_2015}%
  \BibitemOpen
  \bibfield  {author} {\bibinfo {author} {\bibfnamefont {C.~T.}\ \bibnamefont
  {Asplund}}, \bibinfo {author} {\bibfnamefont {A.}~\bibnamefont {Bernamonti}},
  \bibinfo {author} {\bibfnamefont {F.}~\bibnamefont {Galli}},\ and\ \bibinfo
  {author} {\bibfnamefont {T.}~\bibnamefont {Hartman}},\ }\href@noop {}
  {\bibfield  {journal} {\bibinfo  {journal} {Journal of High Energy Physics}\
  }\textbf {\bibinfo {volume} {2015}},\ \bibinfo {pages} {171} (\bibinfo {year}
  {2015})},\ \Eprint {https://arxiv.org/abs/1410.1392} {arXiv:1410.1392}
  \BibitemShut {NoStop}%
\bibitem [{\citenamefont {Fitzpatrick}\ \emph {et~al.}(2015)\citenamefont
  {Fitzpatrick}, \citenamefont {Kaplan},\ and\ \citenamefont
  {Walters}}]{fitzpatrick_virasoro_2015}%
  \BibitemOpen
  \bibfield  {author} {\bibinfo {author} {\bibfnamefont {A.~L.}\ \bibnamefont
  {Fitzpatrick}}, \bibinfo {author} {\bibfnamefont {J.}~\bibnamefont
  {Kaplan}},\ and\ \bibinfo {author} {\bibfnamefont {M.~T.}\ \bibnamefont
  {Walters}},\ }\href@noop {} {\bibfield  {journal} {\bibinfo  {journal}
  {Journal of High Energy Physics}\ }\textbf {\bibinfo {volume} {2015}},\
  \bibinfo {pages} {200} (\bibinfo {year} {2015})},\ \Eprint
  {https://arxiv.org/abs/1501.05315} {arXiv:1501.05315} \BibitemShut {NoStop}%
\bibitem [{\citenamefont {Cardy}\ and\ \citenamefont
  {Tonni}(2016)}]{cardy_entanglement_2016}%
  \BibitemOpen
  \bibfield  {author} {\bibinfo {author} {\bibfnamefont {J.}~\bibnamefont
  {Cardy}}\ and\ \bibinfo {author} {\bibfnamefont {E.}~\bibnamefont {Tonni}},\
  }\href@noop {} {\bibfield  {journal} {\bibinfo  {journal} {Journal of
  Statistical Mechanics: Theory and Experiment}\ }\textbf {\bibinfo {volume}
  {2016}},\ \bibinfo {pages} {123103} (\bibinfo {year} {2016})},\ \Eprint
  {https://arxiv.org/abs/1608.01283} {arXiv:1608.01283} \BibitemShut {NoStop}%
\bibitem [{\citenamefont {Sully}\ \emph {et~al.}(2020)\citenamefont {Sully},
  \citenamefont {Van~Raamsdonk},\ and\ \citenamefont
  {Wakeham}}]{sully_bcft_2020}%
  \BibitemOpen
  \bibfield  {author} {\bibinfo {author} {\bibfnamefont {J.}~\bibnamefont
  {Sully}}, \bibinfo {author} {\bibfnamefont {M.}~\bibnamefont
  {Van~Raamsdonk}},\ and\ \bibinfo {author} {\bibfnamefont {D.}~\bibnamefont
  {Wakeham}},\ }\href@noop {} {\bibfield  {journal} {\bibinfo  {journal}
  {arXiv:2004.13088 [hep-th]}\ } (\bibinfo {year} {2020})},\ \Eprint
  {https://arxiv.org/abs/2004.13088} {arXiv:2004.13088 [hep-th]} \BibitemShut
  {NoStop}%
\bibitem [{\citenamefont {Aharony}\ \emph {et~al.}(2003)\citenamefont
  {Aharony}, \citenamefont {DeWolfe}, \citenamefont {Freedman},\ and\
  \citenamefont {Karch}}]{aharony_defect_2003}%
  \BibitemOpen
  \bibfield  {author} {\bibinfo {author} {\bibfnamefont {O.}~\bibnamefont
  {Aharony}}, \bibinfo {author} {\bibfnamefont {O.}~\bibnamefont {DeWolfe}},
  \bibinfo {author} {\bibfnamefont {D.~Z.}\ \bibnamefont {Freedman}},\ and\
  \bibinfo {author} {\bibfnamefont {A.}~\bibnamefont {Karch}},\ }\href@noop {}
  {\bibfield  {journal} {\bibinfo  {journal} {Journal of High Energy Physics}\
  }\textbf {\bibinfo {volume} {2003}},\ \bibinfo {pages} {030} (\bibinfo {year}
  {2003})},\ \Eprint {https://arxiv.org/abs/hep-th/0303249}
  {arXiv:hep-th/0303249} \BibitemShut {NoStop}%
\bibitem [{\citenamefont {Reeves}\ \emph {et~al.}(2021)\citenamefont {Reeves},
  \citenamefont {Rozali}, \citenamefont {Simidzija}, \citenamefont {Sully},
  \citenamefont {Waddell},\ and\ \citenamefont {Wakeham}}]{Reeves:2021sab}%
  \BibitemOpen
  \bibfield  {author} {\bibinfo {author} {\bibfnamefont {W.}~\bibnamefont
  {Reeves}}, \bibinfo {author} {\bibfnamefont {M.}~\bibnamefont {Rozali}},
  \bibinfo {author} {\bibfnamefont {P.}~\bibnamefont {Simidzija}}, \bibinfo
  {author} {\bibfnamefont {J.}~\bibnamefont {Sully}}, \bibinfo {author}
  {\bibfnamefont {C.}~\bibnamefont {Waddell}},\ and\ \bibinfo {author}
  {\bibfnamefont {D.}~\bibnamefont {Wakeham}},\ }\href@noop {} {\  (\bibinfo
  {year} {2021})},\ \Eprint {https://arxiv.org/abs/2108.10345}
  {arXiv:2108.10345 [hep-th]} \BibitemShut {NoStop}%
\bibitem [{\citenamefont {Pappadopulo}\ \emph {et~al.}(2012)\citenamefont
  {Pappadopulo}, \citenamefont {Rychkov}, \citenamefont {Espin},\ and\
  \citenamefont {Rattazzi}}]{Pappadopulo:2012jk}%
  \BibitemOpen
  \bibfield  {author} {\bibinfo {author} {\bibfnamefont {D.}~\bibnamefont
  {Pappadopulo}}, \bibinfo {author} {\bibfnamefont {S.}~\bibnamefont
  {Rychkov}}, \bibinfo {author} {\bibfnamefont {J.}~\bibnamefont {Espin}},\
  and\ \bibinfo {author} {\bibfnamefont {R.}~\bibnamefont {Rattazzi}},\ }\href
  {https://doi.org/10.1103/PhysRevD.86.105043} {\bibfield  {journal} {\bibinfo
  {journal} {Phys. Rev. D}\ }\textbf {\bibinfo {volume} {86}},\ \bibinfo
  {pages} {105043} (\bibinfo {year} {2012})},\ \Eprint
  {https://arxiv.org/abs/1208.6449} {arXiv:1208.6449 [hep-th]} \BibitemShut
  {NoStop}%
\bibitem [{\citenamefont {Hogervorst}\ and\ \citenamefont
  {Rychkov}(2013)}]{Hogervorst:2013sma}%
  \BibitemOpen
  \bibfield  {author} {\bibinfo {author} {\bibfnamefont {M.}~\bibnamefont
  {Hogervorst}}\ and\ \bibinfo {author} {\bibfnamefont {S.}~\bibnamefont
  {Rychkov}},\ }\href {https://doi.org/10.1103/PhysRevD.87.106004} {\bibfield
  {journal} {\bibinfo  {journal} {Phys. Rev. D}\ }\textbf {\bibinfo {volume}
  {87}},\ \bibinfo {pages} {106004} (\bibinfo {year} {2013})},\ \Eprint
  {https://arxiv.org/abs/1303.1111} {arXiv:1303.1111 [hep-th]} \BibitemShut
  {NoStop}%
\bibitem [{\citenamefont {Karch}\ and\ \citenamefont
  {Katz}(2002)}]{Karch:2002sh}%
  \BibitemOpen
  \bibfield  {author} {\bibinfo {author} {\bibfnamefont {A.}~\bibnamefont
  {Karch}}\ and\ \bibinfo {author} {\bibfnamefont {E.}~\bibnamefont {Katz}},\
  }\href {https://doi.org/10.1088/1126-6708/2002/06/043} {\bibfield  {journal}
  {\bibinfo  {journal} {JHEP}\ }\textbf {\bibinfo {volume} {06}},\ \bibinfo
  {pages} {043}},\ \Eprint {https://arxiv.org/abs/hep-th/0205236}
  {arXiv:hep-th/0205236} \BibitemShut {NoStop}%
\bibitem [{\citenamefont {Hogervorst}\ and\ \citenamefont {{van
  Rees}}(2017)}]{hogervorst_crossing_2017}%
  \BibitemOpen
  \bibfield  {author} {\bibinfo {author} {\bibfnamefont {M.}~\bibnamefont
  {Hogervorst}}\ and\ \bibinfo {author} {\bibfnamefont {B.~C.}\ \bibnamefont
  {{van Rees}}},\ }\href@noop {} {\bibfield  {journal} {\bibinfo  {journal}
  {Journal of High Energy Physics}\ }\textbf {\bibinfo {volume} {2017}},\
  \bibinfo {pages} {193} (\bibinfo {year} {2017})},\ \Eprint
  {https://arxiv.org/abs/1702.08471} {arXiv:1702.08471} \BibitemShut {NoStop}%
\bibitem [{\citenamefont {Hogervorst}(2017)}]{Hogervorst:2017kbj}%
  \BibitemOpen
  \bibfield  {author} {\bibinfo {author} {\bibfnamefont {M.}~\bibnamefont
  {Hogervorst}},\ }\href@noop {} {\  (\bibinfo {year} {2017})},\ \Eprint
  {https://arxiv.org/abs/1703.08159} {arXiv:1703.08159 [hep-th]} \BibitemShut
  {NoStop}%
\bibitem [{\citenamefont {Liendo}\ \emph
  {et~al.}(2013{\natexlab{b}})\citenamefont {Liendo}, \citenamefont
  {Rastelli},\ and\ \citenamefont {{van Rees}}}]{liendo_bootstrap_2013}%
  \BibitemOpen
  \bibfield  {author} {\bibinfo {author} {\bibfnamefont {P.}~\bibnamefont
  {Liendo}}, \bibinfo {author} {\bibfnamefont {L.}~\bibnamefont {Rastelli}},\
  and\ \bibinfo {author} {\bibfnamefont {B.~C.}\ \bibnamefont {{van Rees}}},\
  }\href@noop {} {\bibfield  {journal} {\bibinfo  {journal} {Journal of High
  Energy Physics}\ }\textbf {\bibinfo {volume} {2013}},\ \bibinfo {pages} {113}
  (\bibinfo {year} {2013}{\natexlab{b}})},\ \Eprint
  {https://arxiv.org/abs/1210.4258} {arXiv:1210.4258} \BibitemShut {NoStop}%
\bibitem [{\citenamefont {Nagasaki}\ and\ \citenamefont
  {Yamaguchi}(2012)}]{Nagasaki:2012re}%
  \BibitemOpen
  \bibfield  {author} {\bibinfo {author} {\bibfnamefont {K.}~\bibnamefont
  {Nagasaki}}\ and\ \bibinfo {author} {\bibfnamefont {S.}~\bibnamefont
  {Yamaguchi}},\ }\href {https://doi.org/10.1103/PhysRevD.86.086004} {\bibfield
   {journal} {\bibinfo  {journal} {Phys. Rev. D}\ }\textbf {\bibinfo {volume}
  {86}},\ \bibinfo {pages} {086004} (\bibinfo {year} {2012})},\ \Eprint
  {https://arxiv.org/abs/1205.1674} {arXiv:1205.1674 [hep-th]} \BibitemShut
  {NoStop}%
\bibitem [{\citenamefont {Kristjansen}\ \emph {et~al.}(2013)\citenamefont
  {Kristjansen}, \citenamefont {Semenoff},\ and\ \citenamefont
  {Young}}]{Kristjansen:2012tn}%
  \BibitemOpen
  \bibfield  {author} {\bibinfo {author} {\bibfnamefont {C.}~\bibnamefont
  {Kristjansen}}, \bibinfo {author} {\bibfnamefont {G.~W.}\ \bibnamefont
  {Semenoff}},\ and\ \bibinfo {author} {\bibfnamefont {D.}~\bibnamefont
  {Young}},\ }\href {https://doi.org/10.1007/JHEP01(2013)117} {\bibfield
  {journal} {\bibinfo  {journal} {JHEP}\ }\textbf {\bibinfo {volume} {01}},\
  \bibinfo {pages} {117}},\ \Eprint {https://arxiv.org/abs/1210.7015}
  {arXiv:1210.7015 [hep-th]} \BibitemShut {NoStop}%
\bibitem [{\citenamefont {Buhl-Mortensen}\ \emph {et~al.}(2016)\citenamefont
  {Buhl-Mortensen}, \citenamefont {de~Leeuw}, \citenamefont {Kristjansen},\
  and\ \citenamefont {Zarembo}}]{Buhl-Mortensen:2015gfd}%
  \BibitemOpen
  \bibfield  {author} {\bibinfo {author} {\bibfnamefont {I.}~\bibnamefont
  {Buhl-Mortensen}}, \bibinfo {author} {\bibfnamefont {M.}~\bibnamefont
  {de~Leeuw}}, \bibinfo {author} {\bibfnamefont {C.}~\bibnamefont
  {Kristjansen}},\ and\ \bibinfo {author} {\bibfnamefont {K.}~\bibnamefont
  {Zarembo}},\ }\href {https://doi.org/10.1007/JHEP02(2016)052} {\bibfield
  {journal} {\bibinfo  {journal} {JHEP}\ }\textbf {\bibinfo {volume} {02}},\
  \bibinfo {pages} {052}},\ \Eprint {https://arxiv.org/abs/1512.02532}
  {arXiv:1512.02532 [hep-th]} \BibitemShut {NoStop}%
\bibitem [{\citenamefont {Akal}\ \emph {et~al.}(2020)\citenamefont {Akal},
  \citenamefont {Kusuki}, \citenamefont {Takayanagi},\ and\ \citenamefont
  {Wei}}]{Akal:2020wfl}%
  \BibitemOpen
  \bibfield  {author} {\bibinfo {author} {\bibfnamefont {I.}~\bibnamefont
  {Akal}}, \bibinfo {author} {\bibfnamefont {Y.}~\bibnamefont {Kusuki}},
  \bibinfo {author} {\bibfnamefont {T.}~\bibnamefont {Takayanagi}},\ and\
  \bibinfo {author} {\bibfnamefont {Z.}~\bibnamefont {Wei}},\ }\href
  {https://doi.org/10.1103/PhysRevD.102.126007} {\bibfield  {journal} {\bibinfo
   {journal} {Phys. Rev. D}\ }\textbf {\bibinfo {volume} {102}},\ \bibinfo
  {pages} {126007} (\bibinfo {year} {2020})},\ \Eprint
  {https://arxiv.org/abs/2007.06800} {arXiv:2007.06800 [hep-th]} \BibitemShut
  {NoStop}%
\bibitem [{\citenamefont {Geng}\ \emph
  {et~al.}(2021{\natexlab{a}})\citenamefont {Geng}, \citenamefont {Karch},
  \citenamefont {Perez-Pardavila}, \citenamefont {Raju}, \citenamefont
  {Randall}, \citenamefont {Riojas},\ and\ \citenamefont
  {Shashi}}]{Geng:2020fxl}%
  \BibitemOpen
  \bibfield  {author} {\bibinfo {author} {\bibfnamefont {H.}~\bibnamefont
  {Geng}}, \bibinfo {author} {\bibfnamefont {A.}~\bibnamefont {Karch}},
  \bibinfo {author} {\bibfnamefont {C.}~\bibnamefont {Perez-Pardavila}},
  \bibinfo {author} {\bibfnamefont {S.}~\bibnamefont {Raju}}, \bibinfo {author}
  {\bibfnamefont {L.}~\bibnamefont {Randall}}, \bibinfo {author} {\bibfnamefont
  {M.}~\bibnamefont {Riojas}},\ and\ \bibinfo {author} {\bibfnamefont
  {S.}~\bibnamefont {Shashi}},\ }\href
  {https://doi.org/10.21468/SciPostPhys.10.5.103} {\bibfield  {journal}
  {\bibinfo  {journal} {SciPost Phys.}\ }\textbf {\bibinfo {volume} {10}},\
  \bibinfo {pages} {103} (\bibinfo {year} {2021}{\natexlab{a}})},\ \Eprint
  {https://arxiv.org/abs/2012.04671} {arXiv:2012.04671 [hep-th]} \BibitemShut
  {NoStop}%
\bibitem [{\citenamefont {Geng}\ \emph
  {et~al.}(2021{\natexlab{b}})\citenamefont {Geng}, \citenamefont {L\"ust},
  \citenamefont {Mishra},\ and\ \citenamefont {Wakeham}}]{Geng:2021iyq}%
  \BibitemOpen
  \bibfield  {author} {\bibinfo {author} {\bibfnamefont {H.}~\bibnamefont
  {Geng}}, \bibinfo {author} {\bibfnamefont {S.}~\bibnamefont {L\"ust}},
  \bibinfo {author} {\bibfnamefont {R.~K.}\ \bibnamefont {Mishra}},\ and\
  \bibinfo {author} {\bibfnamefont {D.}~\bibnamefont {Wakeham}},\ }\href
  {https://doi.org/10.1007/JHEP08(2021)003} {\bibfield  {journal} {\bibinfo
  {journal} {JHEP}\ }\textbf {\bibinfo {volume} {08}},\ \bibinfo {pages}
  {003}},\ \Eprint {https://arxiv.org/abs/2104.07039} {arXiv:2104.07039
  [hep-th]} \BibitemShut {NoStop}%
\bibitem [{\citenamefont {Antonini}\ and\ \citenamefont
  {Swingle}(2020)}]{Antonini:2019qkt}%
  \BibitemOpen
  \bibfield  {author} {\bibinfo {author} {\bibfnamefont {S.}~\bibnamefont
  {Antonini}}\ and\ \bibinfo {author} {\bibfnamefont {B.}~\bibnamefont
  {Swingle}},\ }\href {https://doi.org/10.1038/s41567-020-0909-6} {\bibfield
  {journal} {\bibinfo  {journal} {Nature Phys.}\ }\textbf {\bibinfo {volume}
  {16}},\ \bibinfo {pages} {881} (\bibinfo {year} {2020})},\ \Eprint
  {https://arxiv.org/abs/1907.06667} {arXiv:1907.06667 [hep-th]} \BibitemShut
  {NoStop}%
\bibitem [{\citenamefont {Horowitz}\ and\ \citenamefont
  {Marolf}(1998)}]{Horowitz:1998xk}%
  \BibitemOpen
  \bibfield  {author} {\bibinfo {author} {\bibfnamefont {G.~T.}\ \bibnamefont
  {Horowitz}}\ and\ \bibinfo {author} {\bibfnamefont {D.}~\bibnamefont
  {Marolf}},\ }\href {https://doi.org/10.1088/1126-6708/1998/07/014} {\bibfield
   {journal} {\bibinfo  {journal} {JHEP}\ }\textbf {\bibinfo {volume} {07}},\
  \bibinfo {pages} {014}},\ \Eprint {https://arxiv.org/abs/hep-th/9805207}
  {arXiv:hep-th/9805207} \BibitemShut {NoStop}%
\bibitem [{\citenamefont {Gao}(1999)}]{Gao:1999er}%
  \BibitemOpen
  \bibfield  {author} {\bibinfo {author} {\bibfnamefont {Y.-h.}\ \bibnamefont
  {Gao}},\ }\href {https://doi.org/10.1088/1126-6708/1999/04/005} {\bibfield
  {journal} {\bibinfo  {journal} {JHEP}\ }\textbf {\bibinfo {volume} {04}},\
  \bibinfo {pages} {005}},\ \Eprint {https://arxiv.org/abs/hep-th/9903080}
  {arXiv:hep-th/9903080} \BibitemShut {NoStop}%
\bibitem [{\citenamefont {Shashi}(2020)}]{Shashi:2020mkd}%
  \BibitemOpen
  \bibfield  {author} {\bibinfo {author} {\bibfnamefont {S.}~\bibnamefont
  {Shashi}},\ }\href@noop {} {\  (\bibinfo {year} {2020})},\ \Eprint
  {https://arxiv.org/abs/2005.10244} {arXiv:2005.10244 [hep-th]} \BibitemShut
  {NoStop}%
\bibitem [{\citenamefont {Keski-Vakkuri}(1999)}]{KeskiVakkuri:1998nw}%
  \BibitemOpen
  \bibfield  {author} {\bibinfo {author} {\bibfnamefont {E.}~\bibnamefont
  {Keski-Vakkuri}},\ }\href {https://doi.org/10.1103/PhysRevD.59.104001}
  {\bibfield  {journal} {\bibinfo  {journal} {Phys. Rev. D}\ }\textbf {\bibinfo
  {volume} {59}},\ \bibinfo {pages} {104001} (\bibinfo {year} {1999})},\
  \Eprint {https://arxiv.org/abs/hep-th/9808037} {arXiv:hep-th/9808037}
  \BibitemShut {NoStop}%
\bibitem [{\citenamefont {Banados}\ \emph {et~al.}(1993)\citenamefont
  {Banados}, \citenamefont {Henneaux}, \citenamefont {Teitelboim},\ and\
  \citenamefont {Zanelli}}]{Banados:1992gq}%
  \BibitemOpen
  \bibfield  {author} {\bibinfo {author} {\bibfnamefont {M.}~\bibnamefont
  {Banados}}, \bibinfo {author} {\bibfnamefont {M.}~\bibnamefont {Henneaux}},
  \bibinfo {author} {\bibfnamefont {C.}~\bibnamefont {Teitelboim}},\ and\
  \bibinfo {author} {\bibfnamefont {J.}~\bibnamefont {Zanelli}},\ }\href
  {https://doi.org/10.1103/PhysRevD.48.1506} {\bibfield  {journal} {\bibinfo
  {journal} {Phys. Rev. D}\ }\textbf {\bibinfo {volume} {48}},\ \bibinfo
  {pages} {1506} (\bibinfo {year} {1993})},\ \bibinfo {note} {[Erratum:
  Phys.Rev.D 88, 069902 (2013)]},\ \Eprint
  {https://arxiv.org/abs/gr-qc/9302012} {arXiv:gr-qc/9302012} \BibitemShut
  {NoStop}%
\bibitem [{\citenamefont {Balasubramanian}\ \emph {et~al.}(2004)\citenamefont
  {Balasubramanian}, \citenamefont {Naqvi},\ and\ \citenamefont
  {Simon}}]{Balasubramanian:2003kq}%
  \BibitemOpen
  \bibfield  {author} {\bibinfo {author} {\bibfnamefont {V.}~\bibnamefont
  {Balasubramanian}}, \bibinfo {author} {\bibfnamefont {A.}~\bibnamefont
  {Naqvi}},\ and\ \bibinfo {author} {\bibfnamefont {J.}~\bibnamefont {Simon}},\
  }\href {https://doi.org/10.1088/1126-6708/2004/08/023} {\bibfield  {journal}
  {\bibinfo  {journal} {JHEP}\ }\textbf {\bibinfo {volume} {08}},\ \bibinfo
  {pages} {023}},\ \Eprint {https://arxiv.org/abs/hep-th/0311237}
  {arXiv:hep-th/0311237} \BibitemShut {NoStop}%
\bibitem [{\citenamefont {Arefeva}\ and\ \citenamefont
  {Bagrov}(2015)}]{Arefeva:2015zra}%
  \BibitemOpen
  \bibfield  {author} {\bibinfo {author} {\bibfnamefont {I.}~\bibnamefont
  {Arefeva}}\ and\ \bibinfo {author} {\bibfnamefont {A.}~\bibnamefont
  {Bagrov}},\ }\href {https://doi.org/10.1007/s11232-015-0242-x} {\bibfield
  {journal} {\bibinfo  {journal} {Theor. Math. Phys.}\ }\textbf {\bibinfo
  {volume} {182}},\ \bibinfo {pages} {1} (\bibinfo {year} {2015})}\BibitemShut
  {NoStop}%
\bibitem [{\citenamefont {Ageev}\ \emph {et~al.}(2016)\citenamefont {Ageev},
  \citenamefont {Aref'eva},\ and\ \citenamefont
  {Tikhanovskaya}}]{Ageev:2015qbz}%
  \BibitemOpen
  \bibfield  {author} {\bibinfo {author} {\bibfnamefont {D.}~\bibnamefont
  {Ageev}}, \bibinfo {author} {\bibfnamefont {I.~Y.}\ \bibnamefont
  {Aref'eva}},\ and\ \bibinfo {author} {\bibfnamefont {M.}~\bibnamefont
  {Tikhanovskaya}},\ }\href {https://doi.org/10.1134/S0040577916070060}
  {\bibfield  {journal} {\bibinfo  {journal} {Theor. Math. Phys.}\ }\textbf
  {\bibinfo {volume} {188}},\ \bibinfo {pages} {1038} (\bibinfo {year}
  {2016})},\ \Eprint {https://arxiv.org/abs/1512.03362} {arXiv:1512.03362
  [hep-th]} \BibitemShut {NoStop}%
\bibitem [{\citenamefont {Ageev}\ and\ \citenamefont
  {Aref'eva}(2016)}]{Ageev:2017yno}%
  \BibitemOpen
  \bibfield  {author} {\bibinfo {author} {\bibfnamefont {D.}~\bibnamefont
  {Ageev}}\ and\ \bibinfo {author} {\bibfnamefont {I.~Y.}\ \bibnamefont
  {Aref'eva}},\ }\href {https://doi.org/10.1134/S0040577916120072} {\bibfield
  {journal} {\bibinfo  {journal} {Theor. Math. Phys.}\ }\textbf {\bibinfo
  {volume} {189}},\ \bibinfo {pages} {1742} (\bibinfo {year} {2016})},\ \Eprint
  {https://arxiv.org/abs/1512.03363} {arXiv:1512.03363 [hep-th]} \BibitemShut
  {NoStop}%
\bibitem [{\citenamefont {Aref'eva}\ and\ \citenamefont
  {Khramtsov}(2016)}]{Arefeva:2016wek}%
  \BibitemOpen
  \bibfield  {author} {\bibinfo {author} {\bibfnamefont {I.~Y.}\ \bibnamefont
  {Aref'eva}}\ and\ \bibinfo {author} {\bibfnamefont {M.~A.}\ \bibnamefont
  {Khramtsov}},\ }\href {https://doi.org/10.1007/JHEP04(2016)121} {\bibfield
  {journal} {\bibinfo  {journal} {JHEP}\ }\textbf {\bibinfo {volume} {04}},\
  \bibinfo {pages} {121}},\ \Eprint {https://arxiv.org/abs/1601.02008}
  {arXiv:1601.02008 [hep-th]} \BibitemShut {NoStop}%
\bibitem [{\citenamefont {Aref'eva}\ \emph {et~al.}(2016)\citenamefont
  {Aref'eva}, \citenamefont {Khramtsov},\ and\ \citenamefont
  {Tikhanovskaya}}]{Arefeva:2016nic}%
  \BibitemOpen
  \bibfield  {author} {\bibinfo {author} {\bibfnamefont {I.}~\bibnamefont
  {Aref'eva}}, \bibinfo {author} {\bibfnamefont {M.}~\bibnamefont
  {Khramtsov}},\ and\ \bibinfo {author} {\bibfnamefont {M.}~\bibnamefont
  {Tikhanovskaya}},\ }\href {https://doi.org/10.1134/S0040577916110106}
  {\bibfield  {journal} {\bibinfo  {journal} {Theor. Math. Phys.}\ }\textbf
  {\bibinfo {volume} {189}},\ \bibinfo {pages} {1660} (\bibinfo {year}
  {2016})},\ \Eprint {https://arxiv.org/abs/1604.08905} {arXiv:1604.08905
  [hep-th]} \BibitemShut {NoStop}%
\bibitem [{\citenamefont {Freedman}\ \emph
  {et~al.}(1999{\natexlab{b}})\citenamefont {Freedman}, \citenamefont {Mathur},
  \citenamefont {Matusis},\ and\ \citenamefont {Rastelli}}]{Freedman:1998tz}%
  \BibitemOpen
  \bibfield  {author} {\bibinfo {author} {\bibfnamefont {D.~Z.}\ \bibnamefont
  {Freedman}}, \bibinfo {author} {\bibfnamefont {S.~D.}\ \bibnamefont
  {Mathur}}, \bibinfo {author} {\bibfnamefont {A.}~\bibnamefont {Matusis}},\
  and\ \bibinfo {author} {\bibfnamefont {L.}~\bibnamefont {Rastelli}},\ }\href
  {https://doi.org/10.1016/S0550-3213(99)00053-X} {\bibfield  {journal}
  {\bibinfo  {journal} {Nucl. Phys. B}\ }\textbf {\bibinfo {volume} {546}},\
  \bibinfo {pages} {96} (\bibinfo {year} {1999}{\natexlab{b}})},\ \Eprint
  {https://arxiv.org/abs/hep-th/9804058} {arXiv:hep-th/9804058} \BibitemShut
  {NoStop}%
\bibitem [{\citenamefont {D'Hoker}\ and\ \citenamefont
  {Freedman}(2002)}]{DHoker:2002nbb}%
  \BibitemOpen
  \bibfield  {author} {\bibinfo {author} {\bibfnamefont {E.}~\bibnamefont
  {D'Hoker}}\ and\ \bibinfo {author} {\bibfnamefont {D.~Z.}\ \bibnamefont
  {Freedman}},\ }in\ \href@noop {} {\emph {\bibinfo {booktitle} {{TASI 2001:
  Strings, Branes and EXTRA Dimensions}}}}\ (\bibinfo {year} {2002})\ pp.\
  \bibinfo {pages} {3--158},\ \Eprint {https://arxiv.org/abs/hep-th/0201253}
  {arXiv:hep-th/0201253} \BibitemShut {NoStop}%
\bibitem [{\citenamefont {Rastelli}\ and\ \citenamefont
  {Zhou}(2017)}]{Rastelli:2017ecj}%
  \BibitemOpen
  \bibfield  {author} {\bibinfo {author} {\bibfnamefont {L.}~\bibnamefont
  {Rastelli}}\ and\ \bibinfo {author} {\bibfnamefont {X.}~\bibnamefont
  {Zhou}},\ }\href {https://doi.org/10.1007/JHEP10(2017)146} {\bibfield
  {journal} {\bibinfo  {journal} {JHEP}\ }\textbf {\bibinfo {volume} {10}},\
  \bibinfo {pages} {146}},\ \Eprint {https://arxiv.org/abs/1705.05362}
  {arXiv:1705.05362 [hep-th]} \BibitemShut {NoStop}%
\bibitem [{\citenamefont {Jones}(2001)}]{Jones:2001hyp}%
  \BibitemOpen
  \bibfield  {author} {\bibinfo {author} {\bibfnamefont {D.~S.}\ \bibnamefont
  {Jones}},\ }\href {https://doi.org/https://doi.org/10.1002/mma.208}
  {\bibfield  {journal} {\bibinfo  {journal} {Mathematical Methods in the
  Applied Sciences}\ }\textbf {\bibinfo {volume} {24}},\ \bibinfo {pages} {369}
  (\bibinfo {year} {2001})}\BibitemShut {NoStop}%
\end{thebibliography}%
\end{document}